%% file: manu.tex
\documentclass{article}
\input{header.inc}
\date{}
\author{Jun Li\thanks{
    Department of Computer Science, Australian Artificial Intelligence Institute, 
    Faculty of Engineering and Information Technology, University of Technology Sydney, 
    Ultimo, NSW 2007, Australia
}\\
Email: jun.li@uts.edu.au
}
\title{A Spin Glass Characterization of Neural Networks}
\begin{document}
\maketitle
\input{arxiv_secs/sec1_introduction.tex}

\input{arxiv_secs/sec3_method2.tex}

\input{arxiv_secs/sec4_experiments.tex}

\input{arxiv_secs/sec2_related.tex}

\input{arxiv_secs/sec5_con.tex}
\newpage
\input{arxiv_secs/sec_appendix.tex}
\newpage
\begingroup
\bibliographystyle{plain}
\bibliography{manu}
\endgroup
\end{document}

%% file: header.inc
\usepackage{arxiv}
\usepackage{amsmath}
\usepackage{amsfonts}
\usepackage{amssymb}
\usepackage{amsthm}
\usepackage{amsmath}
\usepackage{graphicx}
\usepackage{refcount}
\usepackage{xspace}
\usepackage{ifoddpage}
\usepackage[ruled,vlined]{algorithm2e}
\usepackage{xcolor}         
\usepackage{algpseudocode}
\usepackage{enumitem}
\usepackage{url}

\newcommand{\qchavg}[1]{\langle \langle #1 \rangle \rangle}
\newcommand{\ve}[1]{\boldsymbol{#1}}
\newcommand{\nbr}[1]{\mathcal{N}_{#1}}
\newcommand{\twover}[2]{{#2}} 

%% file: arxiv_secs/sec1_introduction.tex
\begin{abstract}
This work presents a statistical mechanics characterization of neural
networks, motivated by the replica symmetry breaking (RSB) phenomenon in
spin glasses.
A Hopfield-type spin glass model is constructed from a given
feedforward neural network (FNN).
Overlaps between simulated replica samples serve as a characteristic
descriptor of the FNN.
%
The connection between the spin-glass description and commonly studied
properties of the FNN—such as data fitting, capacity, generalization, and
robustness—has been investigated and empirically demonstrated.
Unlike prior analytical studies that focus on model ensembles, this method
provides a computable descriptor for individual network instances,
which reveals nontrivial structural properties that
are not captured by conventional metrics such as loss or accuracy.
Preliminary results suggests its potential for practical applications such as
model inspection, safety verification, and detection of hidden vulnerabilities.
\end{abstract}

\section{Introduction}
\label{sec:introduction}
Given the full specification of a computational model--its architecture and
parameters -- but without further contextual information, can one tell whether or
how the model has attained the status of ``being intelligent'' through fitting
to a purposeful task?

Essentially, the question is ``what is intelligence?''  
In this ambitious form, the question allows little fruitful investigation
outside philosophical debate.
A similar challenge was faced by the query of ``what is life?'' 
A seminal line of thought was proposed by Schr\"{o}dinger in his 1944
lectures \cite{Schrodinger1944}. He framed life as a thermodynamic phenomenon, a
system resisting thermodynamic equilibrium through structured replication and
energy dissipation.
A similar argument can be made: ``purposeful'' computational models lie far from
the equilibrium distribution of the model family with the same architecture but
randomly distributed parameters.
%
The comparison is suggestive—can we examine computational systems for signatures
of intelligence using statistical mechanical tools?

The statistical mechanics perspective has been taken by early efforts that treat
self-adaptive systems for pattern recognition as spin glasses
\cite{Amit1985,Hopfield1982,Mezard2009,Engel2001}.  
For example, the capacity of model families has been studied via the phase-space
volume of data-consistent model ensembles~\cite{Gardner1988}, and the number of
patterns storable as dynamical attractors~\cite{Amit1985}.
A variational learning framework has been established using generalized
rate-distortion theory~\cite{Tishby1999}.

Recent research efforts in neural networks have mainly focused on architectures
and learning algorithms for feedforward neural networks (FNNs)~\cite{Lecun2015}.
The impressive success of FNNs~\cite{Brown2020,Kirillov2023,Rombach2022} makes
it desirable to investigate these large-scale ``intelligent'' computational
systems from first principles.
The following observations have motivated the present study of large-scale FNNs
from a statistical mechanics perspective.  
First, key macroscopic properties, such as generalization and capacity, can be
derived from the fundamental quantity of free energy (entropy)
\cite{Zdeborova2016}, exhibiting {\em self-averaging}.
This implies that for large systems, studying individual instances becomes
equivalent to analyzing ensemble averages.
Second, statistical mechanical tools provide analytical methods for
characterizing ensemble-level properties.
This formalism enables the analysis of individual instances via their ensemble
representations.

This work introduces a tool inspired by the replica method and replica symmetry
breaking (RSB) for characterizing the thermodynamic signatures of neural
networks.
A given neural network $\mathcal F$ is mapped to an Ising-type Hamiltonian
$\mathcal H$~\cite{Mezard1987} as an instance of a spin glass model.
Multiple Gibbs samples (replicas) are generated according to Hopfield net
thermodynamics \cite{Hopfield1982}.  
The overlap (similarity) between replica samples reflects the structure of the
Gibbs measure in configuration space~\cite{Talagrand2011}, which varies with
temperature and exhibits the RSB phenomenon in the low-temperature regime.
The resultant ``replica-overlap-temperature'' profile of $\mathcal{H}$ serves as
a characteristic description of the original neural network $\mathcal{F}$.

This question is important not only for theoretical insight but also for
practical concerns in designing and deploying learning systems:
How can we tell if a system can fit the data of a specific task~\cite{Xiao2018}?
How flexible is a system in adapting to new domains?~\cite{Wang2021}
How can we tell whether an ongoing optimization process is seeking a better
solution, or entering the regime of overfitting?~\cite{Minegishi2025}
%
%
In a more concrete scenario, a pre-trained large-scale model may be released to
the public and claimed to perform well on specific tasks. 
How can users verify whether it also hides unpublicized sensitivities? For
example, could it (intentionally or not) be made to respond to specific inputs
in predefined ways?
Alternatively, in the seemingly innocuous step of randomly initializing a neural
network, how can one tell whether the random number generator is safe?
Is it possible to plant a pattern (i) without altering commonly monitored
parameter statistics, and (ii) that can survive subsequent training?

The main contributions of this work include: (i) a proposal to investigate
neural network properties from the perspective of statistical mechanics, (ii) a
computational procedure to implement the theoretical characterization, and (iii)
empirical verification and exploration of the associated utilities and
limitations.


%% file: arxiv_secs/sec3_method2.tex

\section{Method}
\label{sec:method}
\newcommand{\Zs}{{\mathrm{Z}}}
\newcommand{\Zb}{Z_\beta(\ve{J})}
\renewcommand{\frac}{\tfrac}
\subsection{Ising model from feedforward neural networks}
The central idea is to derive an Ising model from a feedforward neural network
(FNN). 
The statistical mechanics properties of the Ising model is used to
characterize the FNN.
The FNN computational model makes a {\em directed acyclic graph}. In a common
multi-layer perceptron (MLP), the neurons are connected in a layered
structure, where the neurons in the $l$-th layer are computed as
\begin{align}
x^l_i = \phi \big(\sum\nolimits_{j=1}^{n_{l-1}} W_{i,j}^{l} x_{j}^{l-1}\big) \label{eq:fnn}
\end{align}
where $W_{i,j}^{l}$ are the inter-layer connection weights and $\phi$ is 
the activation.
The model \eqref{eq:fnn} omits the bias terms; a similar treatment also applies
to \eqref{eq:ising} below.  This simplification has little
effect on the discussion of the proposed statistical mechanics trick.

An Ising model describes interacting binary variables called {\em spins}
\cite{Mezard1987}. 
For a system of $N$ spins with $\ve{\sigma} = [\sigma_1, \dots,
\sigma_N]$, $\sigma_i \in \{-1,1\}$, the Hamiltonian is  
\begin{align}
H(\ve{\sigma}; \ve{J}) & = - \sum\nolimits_{1\leq i < j \leq N}
    J_{i,j} \sigma_{i} \sigma_{j} \label{eq:ising} 
\end{align}
where $J_{i,j}$ is the coupling strength between spins $i$ and $j$.
The model \eqref{eq:ising} makes an {\em undirected graph}, spins being nodes
and couplings edges. 
Given \eqref{eq:ising}, the Boltzmann distribution is
\begin{align}
p_\beta(\ve{\sigma}; \ve{J}) & =
\frac{1}{\Zb}
\exp(-\beta {H} (\ve{\sigma}; \ve{J})) \label{eq:boltzmann} \\
Z_{\beta}(\ve{J}) & =\sum\nolimits_{ \{ \ve{\sigma}\} } 
\exp(-\beta H(\ve{\sigma}; \ve J)) \label{eq:Z}
\end{align}
where $\Zb$ is the partition function and $\beta$ the inverse temperature.
Consider a spin system evolving under the thermodynamics,
\begin{align}
p_{\beta, t+1}(\sigma_i=\pm 1;\ve{J}) &= 
\frac{1}{\Zb} \exp\Big(-\beta H^i_t(\pm 1)\Big), \quad \text{for } i=1,\dots,N 
\label{eq:spinprob} \\
H^i_t(s) & = - s \cdot \sum\nolimits_{j \in \nbr{i}} J_{i,j} \sigma_{j, t}
\label{eq:spinH} 
\end{align}
where $\nbr{i}$ is the set of spins that are coupled with $\sigma_i$. The
subscripts $t$ and $t+1$ are nominal to indicate the ``old'' and ``new'' states
in the evolution. 
The equation set \eqref{eq:spinprob} describes the probability of $\sigma_i$ be
found at a state, given the momentary configuration of the rest of the spins via
local fields $\sum_j J_{i,j} \sigma_j$.
The spin stochastic dynamics generate a time-dependent distribution of system
states, which asymptotically relaxes to the equilibrium Boltzmann
distribution\cite{Glauber1963}.

Spin systems served as the computational model of neural networks,
implementing an associative memory by Hopfield \cite{Hopfield1982}.
Henceforth, the term {\em Hopfield network (HNN)} is identified with
the Ising formalism in
\eqref{eq:boltzmann} and \eqref{eq:spinprob}, referring to graph structure,
coupling parameters $\ve J$, and stochastic dynamics where context permits.

It is ready to introduce the technical setup of this work:
{\bf one architecture, two computational models}.
Given a FNN, let a HNN share the same set of neurons, and the same synaptic
connections by aligning \eqref{eq:fnn} and \eqref{eq:ising}.
In an FNN, a neuron in layer $l$ (except for the input/output layers) is
connected to the neurons in the previous layer $l-1$ and the next layer $l+1$. 
In the conversion to HNN, the directions of the connections are removed, and the
connected neurons are collected into the set $\nbr{i}$ with the coupling
strengths $J_{i,j}$ cloned from the FNN weights $W_{\cdot}^{\{l-1, l\}}$.
Intuitively, a ``twin'' HNN is constructed from a FNN, which shares the
topology and parameters.  
But the dynamics is made symmetric and asynchronous. 
See Supplementary Material (SM) for implementation details and schematic
diagrams.

Note that in existing spin-glass-based studies of neural networks, e.g.
\cite{Choromanska2015}, spins typically correspond to weights.  However, this
work maps spins to neurons.  The thermodynamic behavior of the spin system thus
reflects the collective activation patterns.

\begin{figure}
\begin{centering}
\includegraphics[width=16cm]{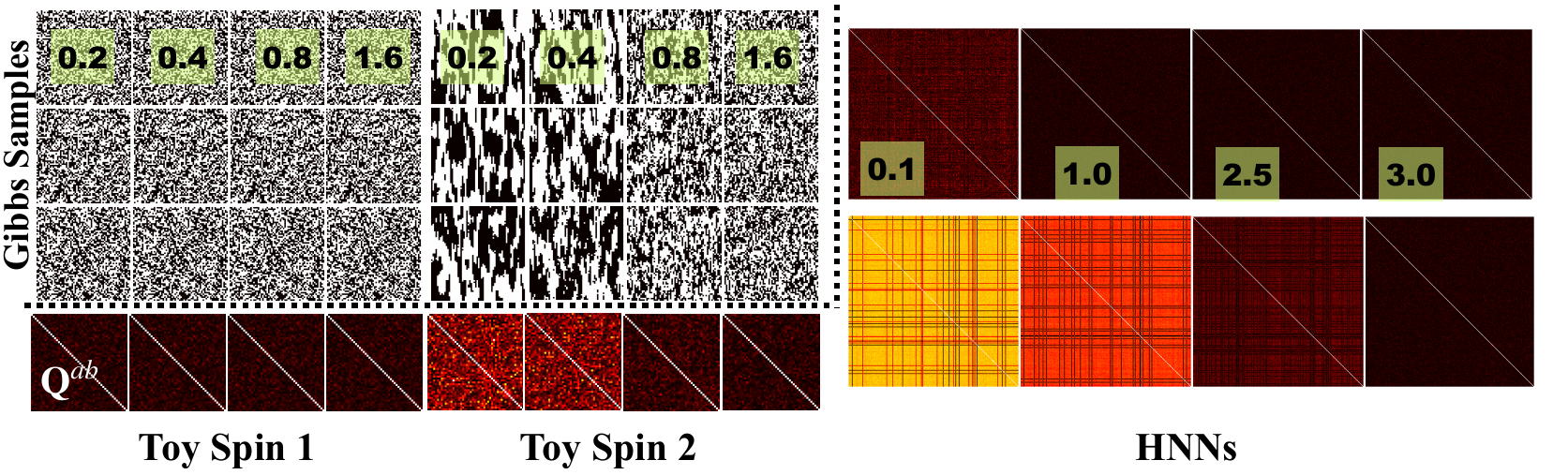}
\par
\end{centering}
\caption{Gibbs samples and the replica overlap matrices.  Left: Gibbs samples
(monochrome plots) of two types of spin systems at different temperatures
(columns), and absolute replica overlaps (heatmap plots).
Right: replica overlaps of random (top) and trained (bottom) neural networks.
\label{fig:vis_overlap}}
\end{figure}

\subsection{Replica overlaps and HNN statistics}
In statistical mechanical systems, macroscopic observables derive from $\log
\Zb$, where the partition function $\Zb$ is defined in
\eqref{eq:Z}.  
In this subsection, $\Zs \equiv \Zb$ is adopted for simple notion, while
noting its dependence on $\ve J$ and $\beta$.
Direct computation of $\Zs$ for specific couplings realizations $\ve{J}$
needs to sum over $2^N$ spin configurations and is generally
intractable.
When the couplings $\ve{J}$ are random variables, the {\em
ensemble-averaged} quantity is considered,
\begin{align}
\qchavg{\log \Zs} & = \int_{\ve{J}} 
\log \Zs \prod\nolimits_{i,j} P(J_{i,j}) dJ_{i,j} 
\label{eq:logZavg}
\end{align}
where $\qchavg{\cdot}$ denotes the average over $\ve J$, which is considered as
{\em fixed} during the time scale of the spin dynamics \eqref{eq:spinprob} and
called {\em quenched disorder}.

The replica trick \cite{Mezard2009,Talagrand2011} is a method to reformulate
$\qchavg{\log \Zs}$ in terms of the partition function of $n$ 
non-interacting replicas of the system using the identity 
$\qchavg{\log \Zs} = \lim_{n\to 0} \frac{\qchavg{\Zs^n} - 1}{n}$.
The quenched average of $\Zs^n$ can be expressed using an effective potential
$F(\ve Q)$, 
\begin{align}
  \qchavg{\Zs^n} = \exp \left( - N F(\ve Q) \right) \label{eq:Zreplica}
\end{align}
where the replica-overlap matrix $\ve{Q}$ with elements
\begin{align}
    q^{ab} & = \frac{1}{N} \sum\nolimits_{i=1}^{N} \sigma_{i}^{(a)} \sigma_{i}^{(b)} \label{eq:Qab} 
\end{align}
quantifies configuration similarity between replicas $a$ and $b$.
%
In statistical mechanics, the {\em replica symmetry (RS) ansatz} assumes
identical off-diagonal elements $q^{ab} = q$ ($a \neq b$). This holds in the
high-temperature phase where weak replica correlations permit mean-field
treatment. 
When the temperature is low, more metastable states start to appear, 
causing {\em replica symmetry breaking (RSB)}. The overlap matrix $\ve{Q}$ then
develops hierarchical structures with $q^{ab}$ values reflecting state
organization. 
See SM for a brief introduction of the background.

To motivate, the following two observations are in order:
(i) For macroscopic properties that are self-averaging, e.g. $\log Z$,
empirical computation from one instance of HNN system is {\em typical} with
overwhelming probability.
(ii) Through the link \eqref{eq:Zreplica}, the overlap matrix $\ve
Q$ reflects the structure of the Gibbs distribution of the HNN of interest.
Notice that the relation in (ii) contains an implicit dependency on temperature
through \eqref{eq:Zreplica} $\rightarrow$ \eqref{eq:Z}.

Fig.~\ref{fig:vis_overlap} illustrates the Gibbs samples and the replica
overlap matrices of a few spin system samples. The left two block of plots show 
the Gibbs samples and the corresponding replica overlap matrices of two types of
simple grid of $64 \times 64=4,096$ spins. ``Toy Spin 1'' is of SK type
\cite{Sherrington1975}, where $J_{i,j}$ follows a Gaussian distribution for all
pairs $1\leq i\neq j \leq 4,096$.  ``Toy Spin 2'' has the same spin grid, but 
the $J_{i,j}$ are non-zero only when $i$ and $j$ are neighbors in the grid, e.g.
{\small \tt i=116(row 2, col 52)} is coupled with {\small \tt j $\in\{$52, 115, 117, 178$\}$}. 
The non-zero couplings follow a Gaussian with $\sigma^2=0.25$, and the
means of the vertical/horizontal-neighbor couplings are $\mu_v=1.0$ (stronger) and
$\mu_h=0.1$ (weaker), respectively.
In the figure, more structural characteristics emarge in the samples of
the short-range spin system in low temperatures. The features are qualitatively
consistent with the setup of stronger vertical couplings. 
The absolute overlap values are shown at different temperatures (denoted as
``$Q^{ab}$''). The structural changes of the Gibbs distribution can
be observed.

The overlap plots on right are obtained from the Gibbs samples of
two HNNs derived from 2 FNNs: (i) randomly initialized and (ii) trained 
on the {\em default task} (See next Section for task details). 
The replica overlaps show a distinction between the trained and
untrained networks: when the temperature drops, more structural features 
emerge in the trained networks.

Given an FNN, Algorithm~\ref{alg:replica} computes a replica overlap–temperature
curve as the statistical mechanics characterization, referred to as the
``$Q^{ab}$ curve''. \hfill $\ $
\begin{algorithm}[H]
{\small
\SetAlgoNoEnd
\LinesNumbered
\caption{Replica Overlap ($Q^{ab}$) Curves for FNN}
\label{alg:replica}
\KwIn{FNN $\mathcal{F}_{\ve{W}}$ with weights $\ve{W}$, 
    number of replica samples $n$, 
    temperature range $\mathcal T$}
\KwOut{Curve of average absolute overlap, 
$\{Q^{ab}_\beta\}, \beta^{-1}\in \mathcal T$}
Construct the HNN $\mathcal{H}_{\ve{J}}$ from the FNN $\mathcal{F}_{\ve{W}}$ \\
\ForEach{temperature $\beta^{-1}$ in $\mathcal T$}{
    Sample $\{\ve{\sigma}^{(a)}\}_{a=1}^n$ from the Gibbs distribution 
        $p_\beta(\ve{\sigma}; \ve{J})$ as \eqref{eq:spinprob}. \label{alg:replica:ln:gibbs}\\
    Compute the $n \times n$ {\em empirical} replica overlap matrix $\ve{Q}_\beta$ as \eqref{eq:Qab} \\
    Calculate the average absolute of off-diagonal elements:
        $Q^{ab}_\beta = \frac{1}{n(n-1)} \sum_{a \neq b} |q^{ab}_\beta|$.
}
}
\end{algorithm}

%% file: arxiv_secs/sec4_experiments.tex
\section{Experiments}
\label{sec:experiment}
In the experiments, an ``{\tt Input Encoder-{\color{blue} MLP}-Readout}''
structure is used, where the characterization is on the {\color{blue} MLP}
block. E.g., following the practice in \cite{Zhang2021}, 
pixels in the MNIST dataset \cite{Lecun1998} are pre-processed into
$10$-dimensional PCA features \cite{Jolliffe2002}, a simple task of classifying
``0/1'' is constructed.  The tested networks consist of fully connected layers
{\tt \{10-256-256-256-2\}}, where the MLP consists of the middle layers of
$3\times 256$ neurons. 
The setup is used as the {\em default task}, which is favored
for its simplicity considering the large number of training sessions.  When
different tasks and models are tested, a similar setup is used with suitable
adaptation of the input encoding and outputs.

A few notes are helpful to interpret the results
and figures:
\textbf{(i)~Gibbs sampling} implements asynchronous spin dynamics from
\eqref{eq:spinprob}, for which a full parallel implementation is cumbersome. 
Algorithm \ref{alg:replica} is implemented using PyTorch \cite{Paszke2019}
with $n=1,\!000$ replica samples.
The potential energy and sampling of spins are performed in groups based on the
FNN layer structure.  
The influence on the $Q^{ab}$ curves is small.
%
\textbf{(ii)~Shaded regions} represent variance across curves of 10 models from
identical task specifications, except Fig.~\ref{fig:curve_invar}(a) showing
variance of Gibbs sampling of a single model.
\textbf{(iii)~Color codes} as in Fig.~\ref{fig:curve_fitness_epoch_task} (a)
carry the semantics of ``how much training a model has received''. 
Alternate contexts use visually discriminative palettes.


\subsection{$Q^{ab}$ curves as effective and consistent characterization of neural networks}
\label{subsec:experi:1}
\paragraph*{Distinctive $Q^{ab}$ curves and variation by Gibbs samples and model ensemble}
\begin{figure}[t] 
\centering
\begin{minipage}{0.45\textwidth}
    \centering
    \includegraphics[width=\textwidth]{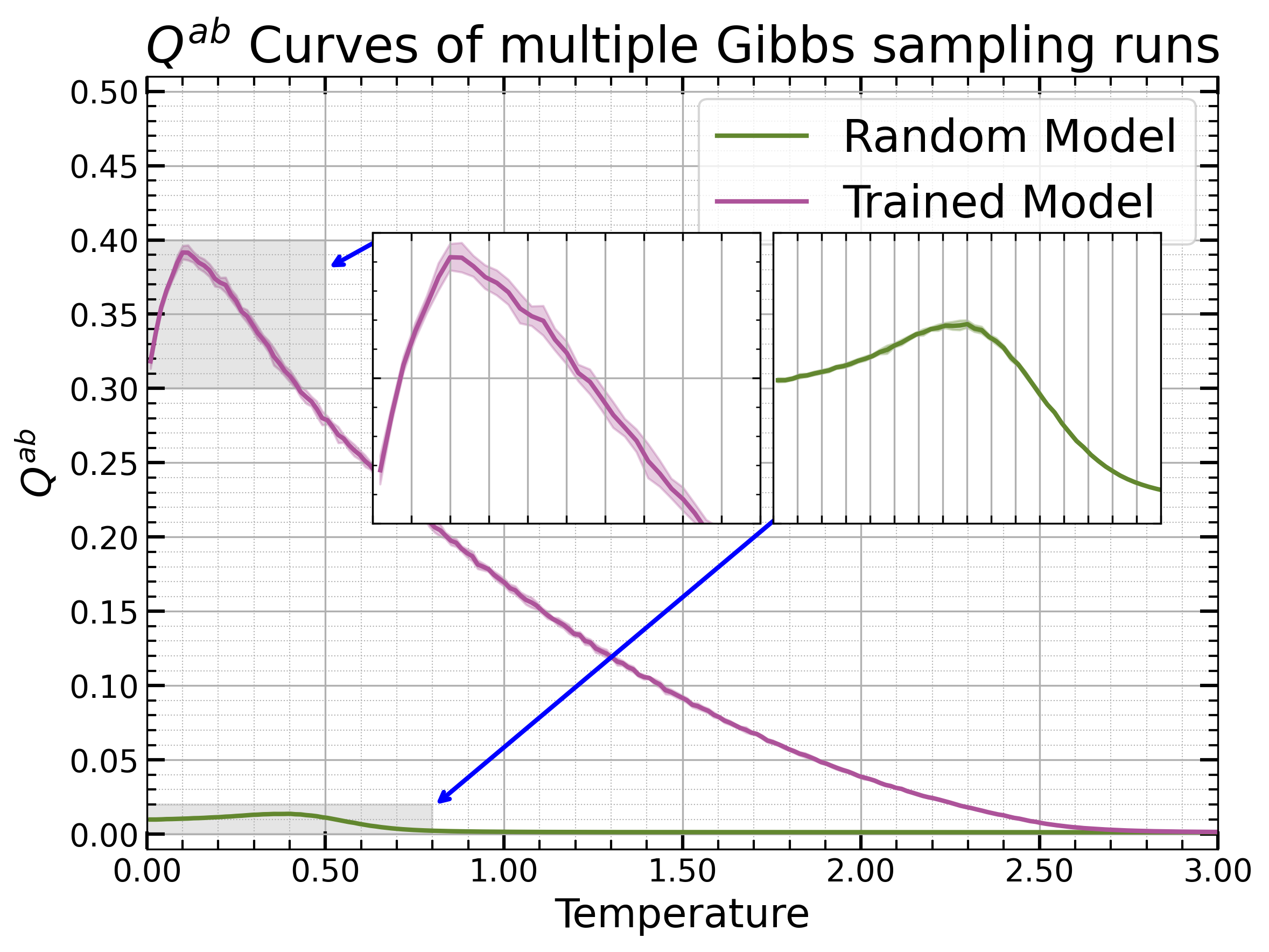}\\
    \textnormal{(a)}
\end{minipage}
\hfill
\begin{minipage}{0.45\textwidth}
    \centering
    \includegraphics[width=\textwidth]{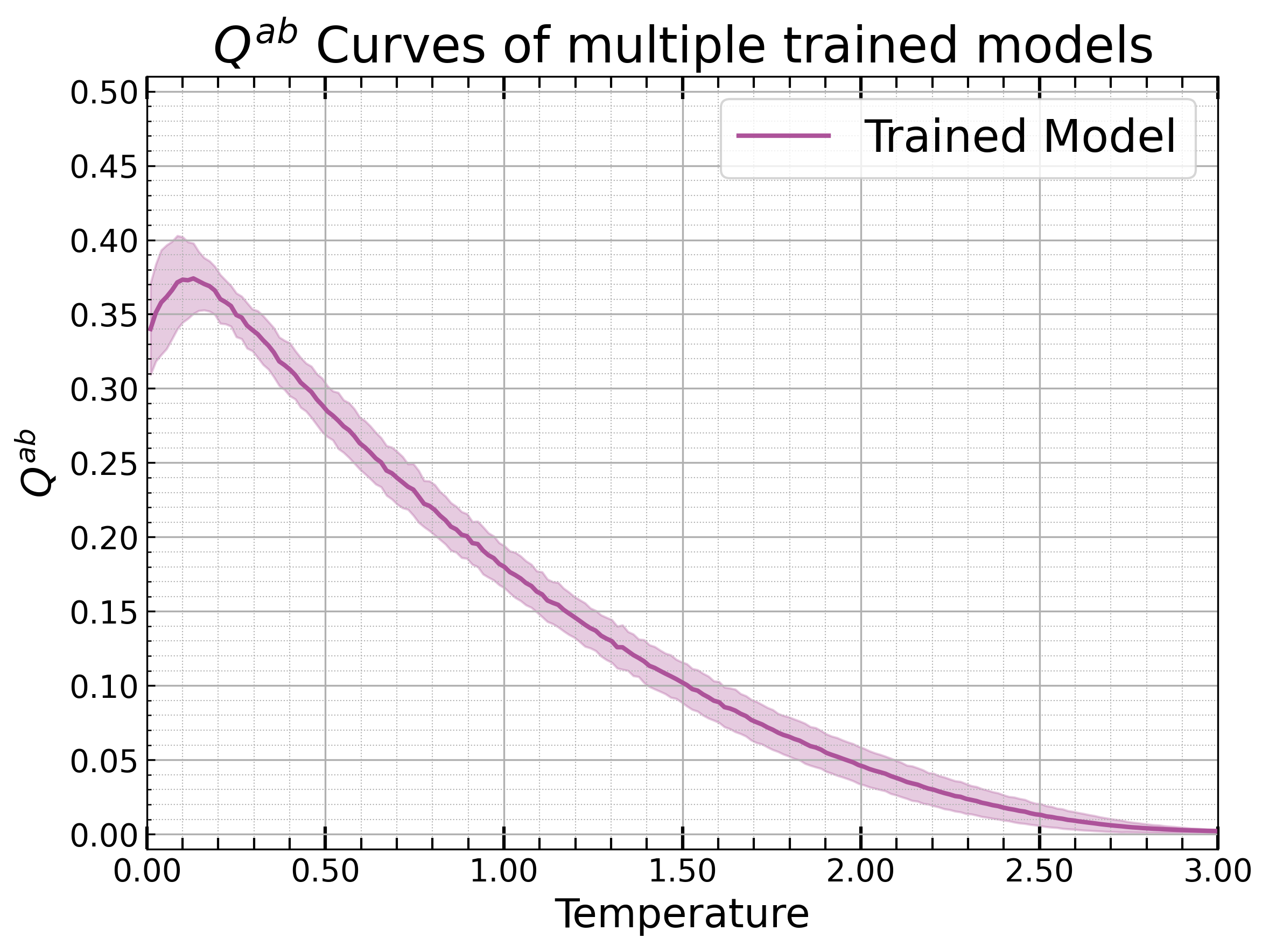}\\
    \textnormal{(b)}
\end{minipage}
\caption{$Q^{ab}$ curves with variations (shaded). 
{\small {\bf (a)} Comparison of random and trained models; Variation of {\em Gibbs sampling 
of same model} (zoom-in insets)
{\bf (b)} Trained models; Variation of {\em multiple models}.
}
\label{fig:curve_invar}}
\end{figure}

The experiment compares $Q^{ab}$ curves between random and trained FNNs
(Fig.~\ref{fig:curve_invar}). This extends observations from
Fig.~\ref{fig:vis_overlap} to quantify replica overlap differentiation across a
range of temperatures.
The distinction is detectable in Fig. \ref{fig:curve_invar}(a)
when the termperature cools down to $\beta^{-1}=T=2.5$, and the
difference is significant for small $T$. 
The high average overlapping at low termperature indicates that metastable
states start emerging in the Gibbs distribution of the trained model. 
As mentioned above, the shaded areas in subplot (a) represent the variation
due to repeated computing the $Q^{ab}$ curves from the same model.
The variation is small and can be examined in the zoom-in insets.
Subplot (b) shows the curves obtained from 10 {\em different models} trained on
the same task. The variation is greater
than that in (a), while the characterization remains effective.
In all the following experiments, the variations refer to the ensemble of models
as in (b).

\paragraph*{Distinctive tasks} 
\begin{figure}[t]
\centering
\begin{minipage}{0.45\textwidth}
    \centering
    \includegraphics[width=\textwidth]{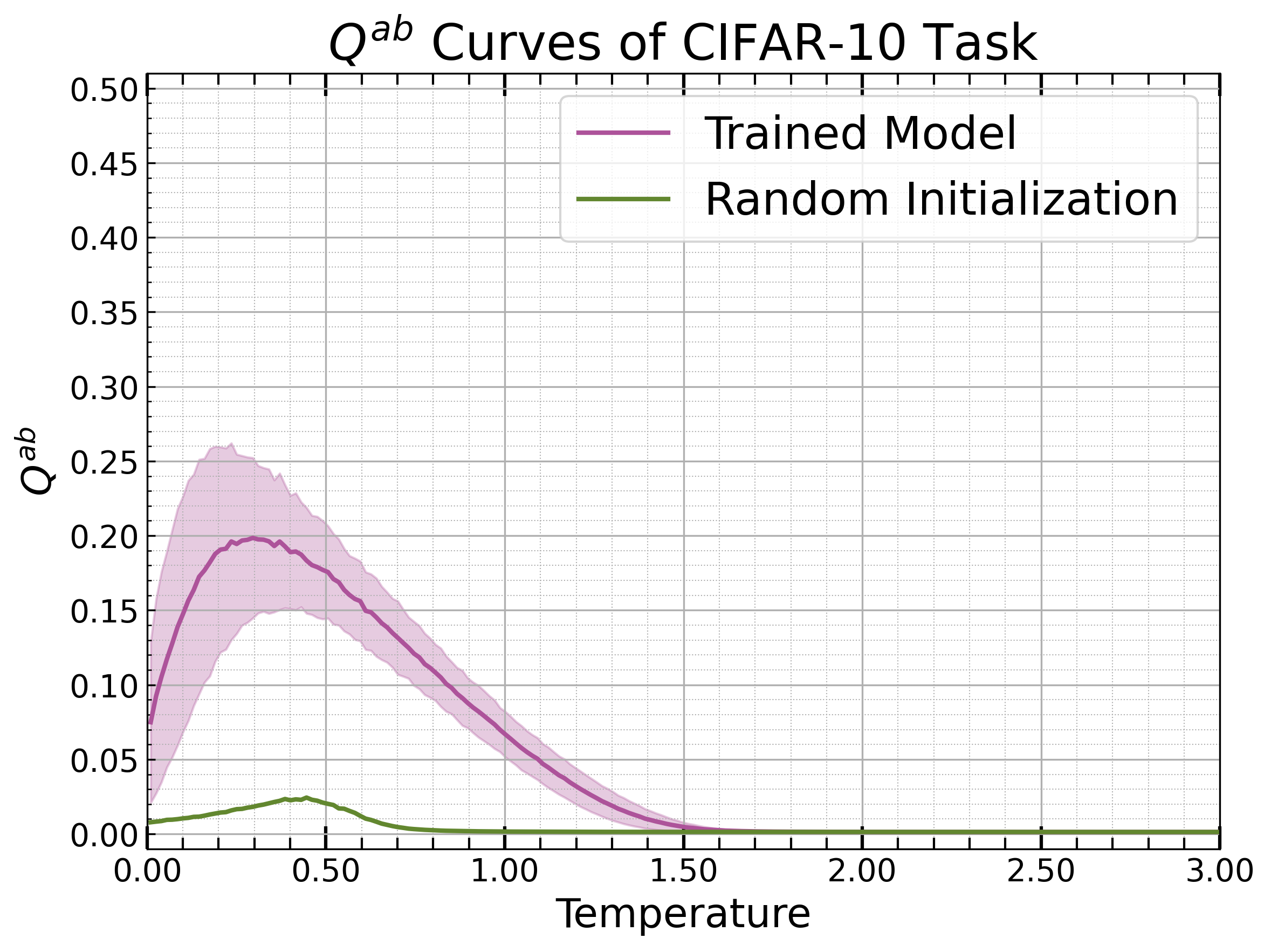} \\
    \textnormal{(a)}
\end{minipage}
\hfill
\begin{minipage}{0.45\textwidth}
    \centering
    \includegraphics[width=\textwidth]{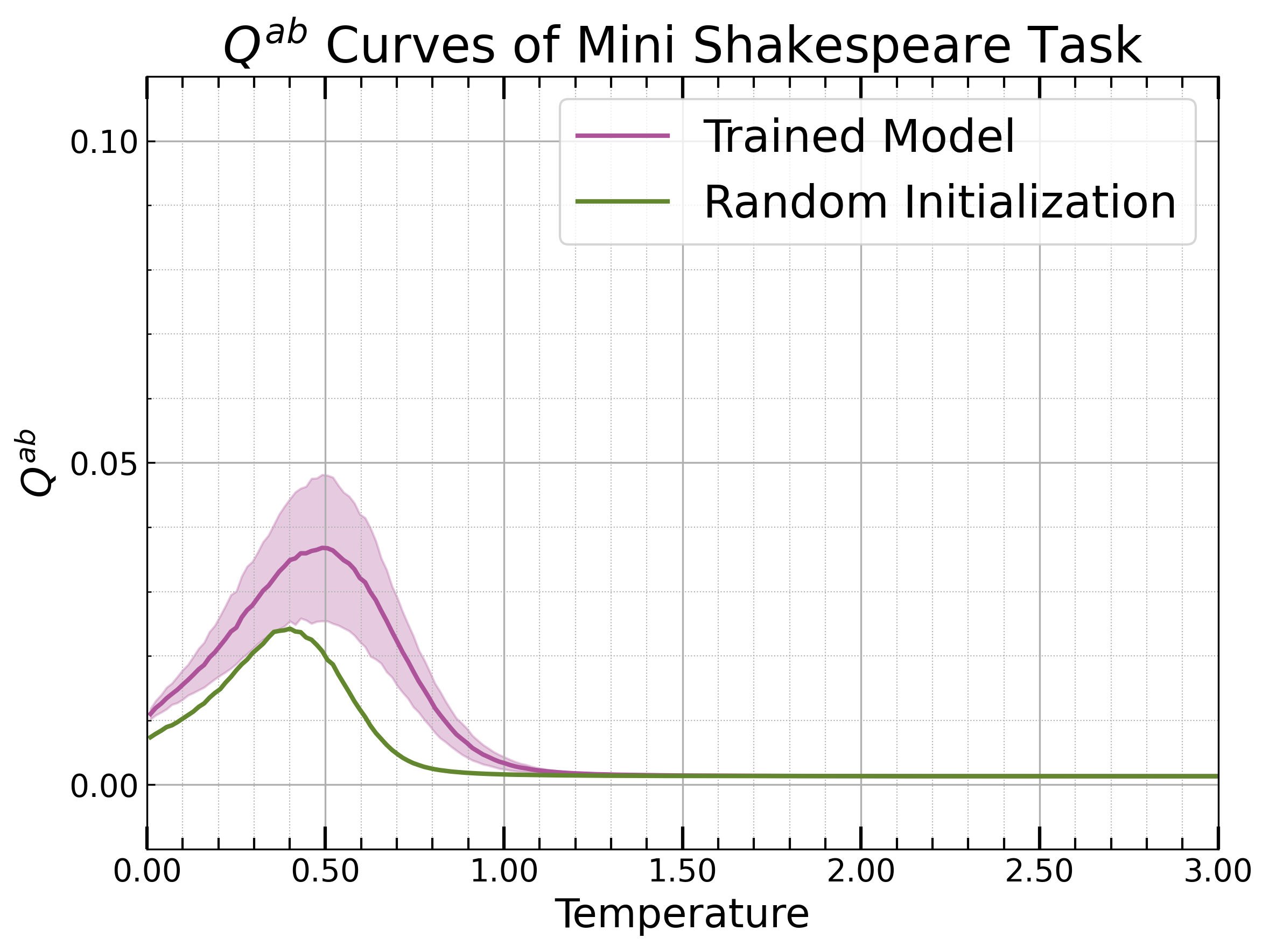}\\
    \textnormal{(b)}
\end{minipage}
\caption{$Q^{ab}$ curves of different tasks. {\bf (a)} Image classification
{\bf (b)} Text generation
\label{fig:curve_task}}
\end{figure}

The distinction revealed by the $Q^{ab}$ curves is observed across different 
learning tasks. 
Fig. \ref{fig:curve_task} shows the comparison  of
random/trained FNNs in two additional tasks:
(i) image classification of CIFAR-10
\cite{Krizhevsky2009}, with a convolutional input encoder and (ii) a text
generater adopted from \cite{Karpathy2020} (Mini-Shakespeare) with a
transformer input encoder. 
The networks are $3 \times 256 $ MLP blocks embedded in the two pipelines.
See SM for details.

\subsection{$Q^{ab}$ curves and model fitness to data}
A hypothesis suggested by the preceding results is that fitting to data
introduces modes in the derived HNNs, which manifest in the $Q^{ab}$ curves.
The following experiments further test the connection.

\begin{figure}
    \centering
    \begin{minipage}{0.43\textwidth}
        \centering
        \includegraphics[width=\textwidth]{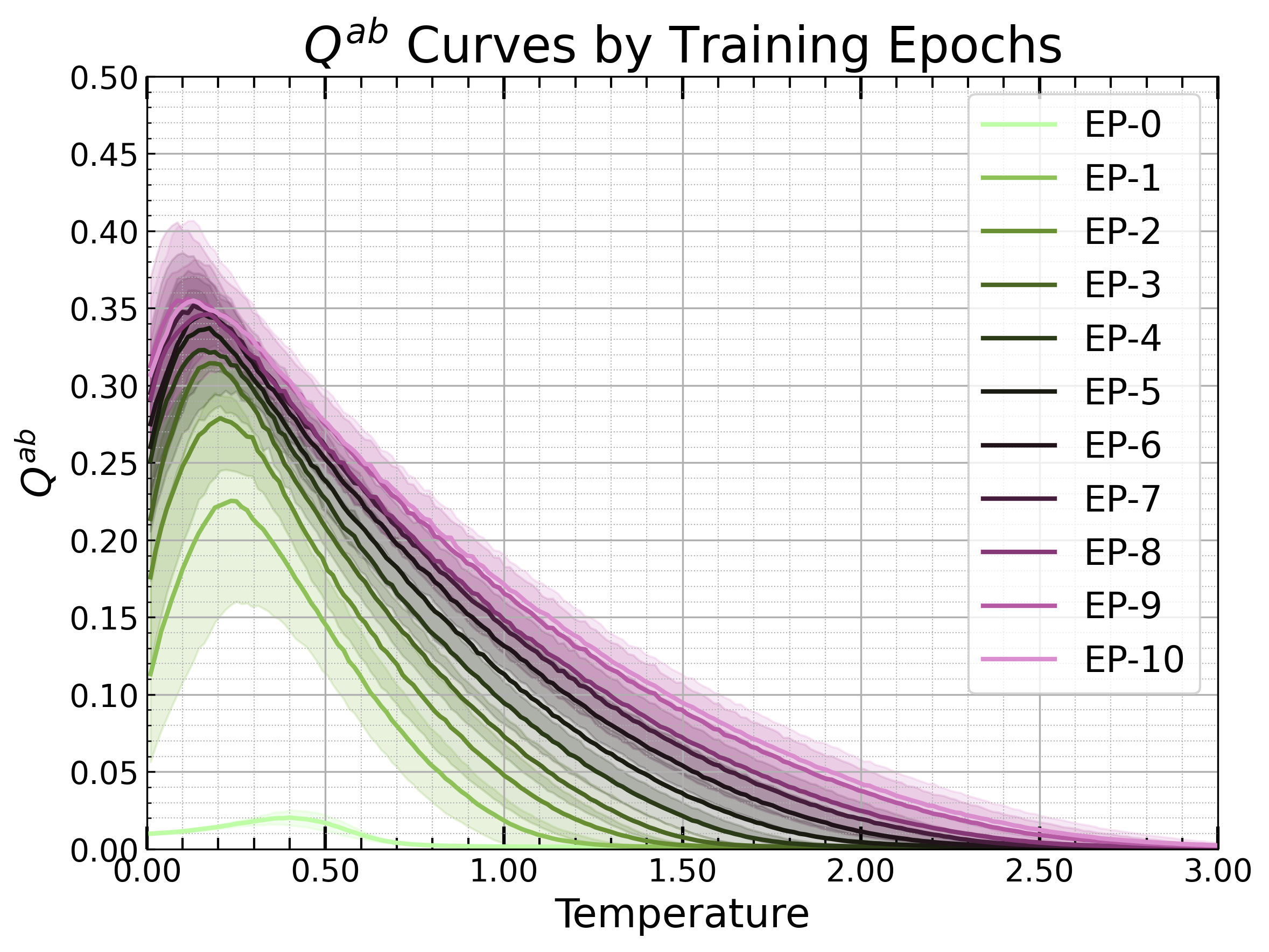} \\
        \textnormal{(a)}
    \end{minipage}
    \hfill
    \begin{minipage}{0.45\textwidth}
        \centering
        \includegraphics[width=\textwidth]{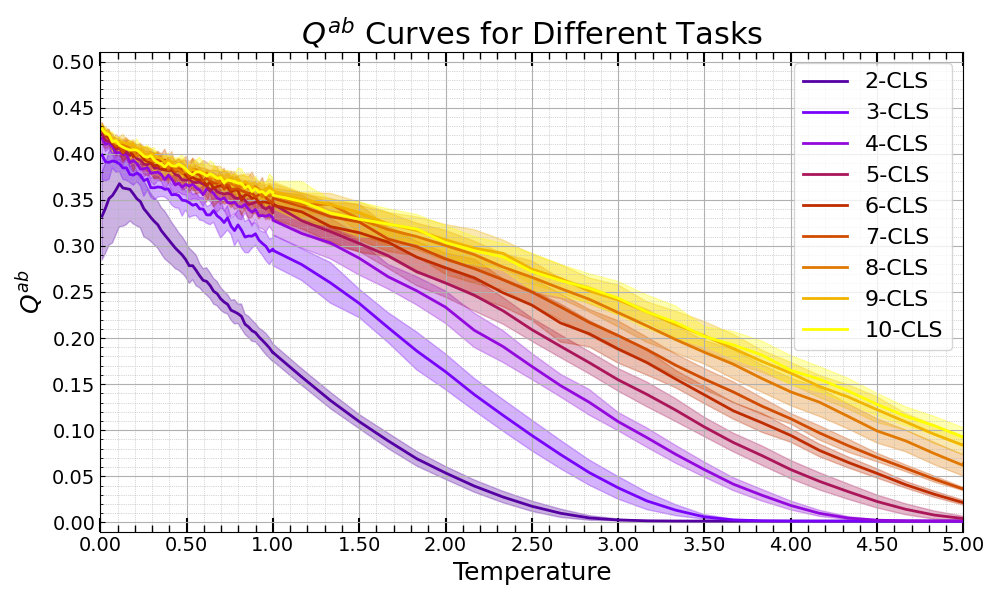} \\
        \textnormal{(b)}
    \end{minipage}
    \caption{$Q^{ab}$ curves and model training. {\bf (a)} different epochs 
    {\bf (b)} tasks of $\{2\dots 10\}$ class targets}
    \label{fig:curve_fitness_epoch_task}
\end{figure}    

\paragraph*{Fitness and task}
Fig. \ref{fig:curve_fitness_epoch_task} shows the $Q^{ab}$ curves of networks 
trained under different procedures. 
Subplot~(a) shows the curves of networks trained for different numbers of epochs
on the {\em default task}.  
Subplot~(b) shows curves from models trained for 10 epochs, with classification
targets varying from 2 to 10 classes: $\{0, 1\}$, $\{0, 1, 2\}$, etc.
The plots demonstrate how increased fitness—via training duration or task
complexity—affects low-temperature replica overlaps.

\paragraph*{Training conditions}
\begin{figure}
\centering
\begin{minipage}{0.45\textwidth}
    \centering
    \includegraphics[width=\textwidth]{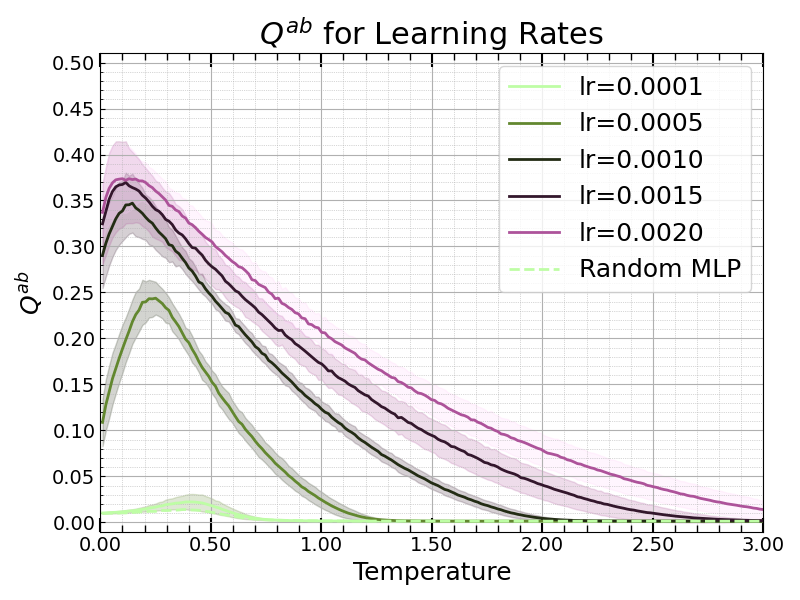} 
    \textnormal{(a)}
\end{minipage}
\hfill
\begin{minipage}{0.45\textwidth}
    \centering
    \includegraphics[width=\textwidth]{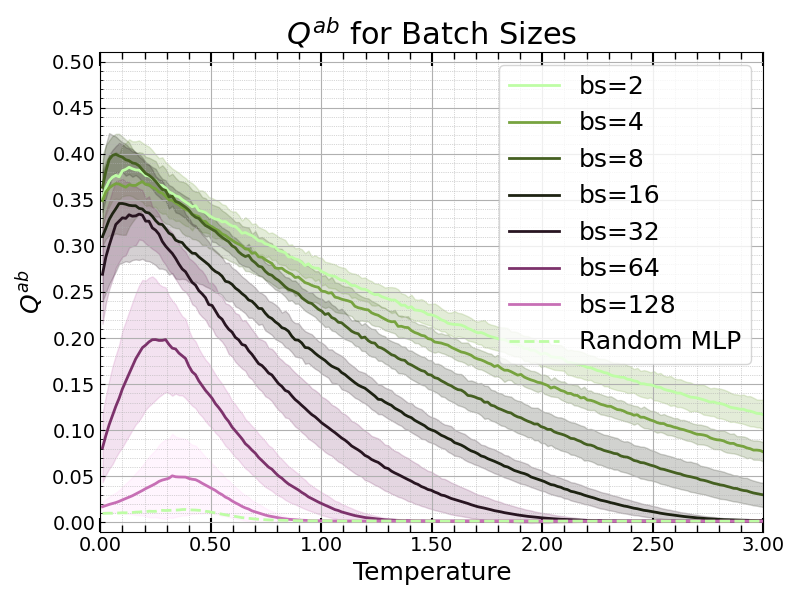} 
    \textnormal{(b)}
\end{minipage}
\caption{$Q^{ab}$ curves of different training conditions: {\bf (a)} learning rates. 
{\bf (b)} batch sizes in SGD}
\label{fig:curve_lr}
\end{figure}
Stochastic gradient descent (SGD) is widely used in training neural networks.
The success of SGD is attributed to the stochasticity, which introduces
regularization and helps the training explore the model space
\cite{Bottou2018,Yang2023}.
The learning rate and the batch size are two important hyper-parameters
that affect the noise term in SGD \cite{Shi2023}.
Such influence is reflected in the $Q^{ab}$ curves in Fig. \ref{fig:curve_lr}.
It is displayed that stronger noises (larger learning rates or smaller batch
sizes) enables the optimization process to escape from 
initial local minima and explore wider regions contains richer metastable structures
\cite{Choromanska2015}.


\begin{figure}
\centering
\begin{minipage}{0.43\textwidth}
    \centering
    \includegraphics[width=\textwidth]{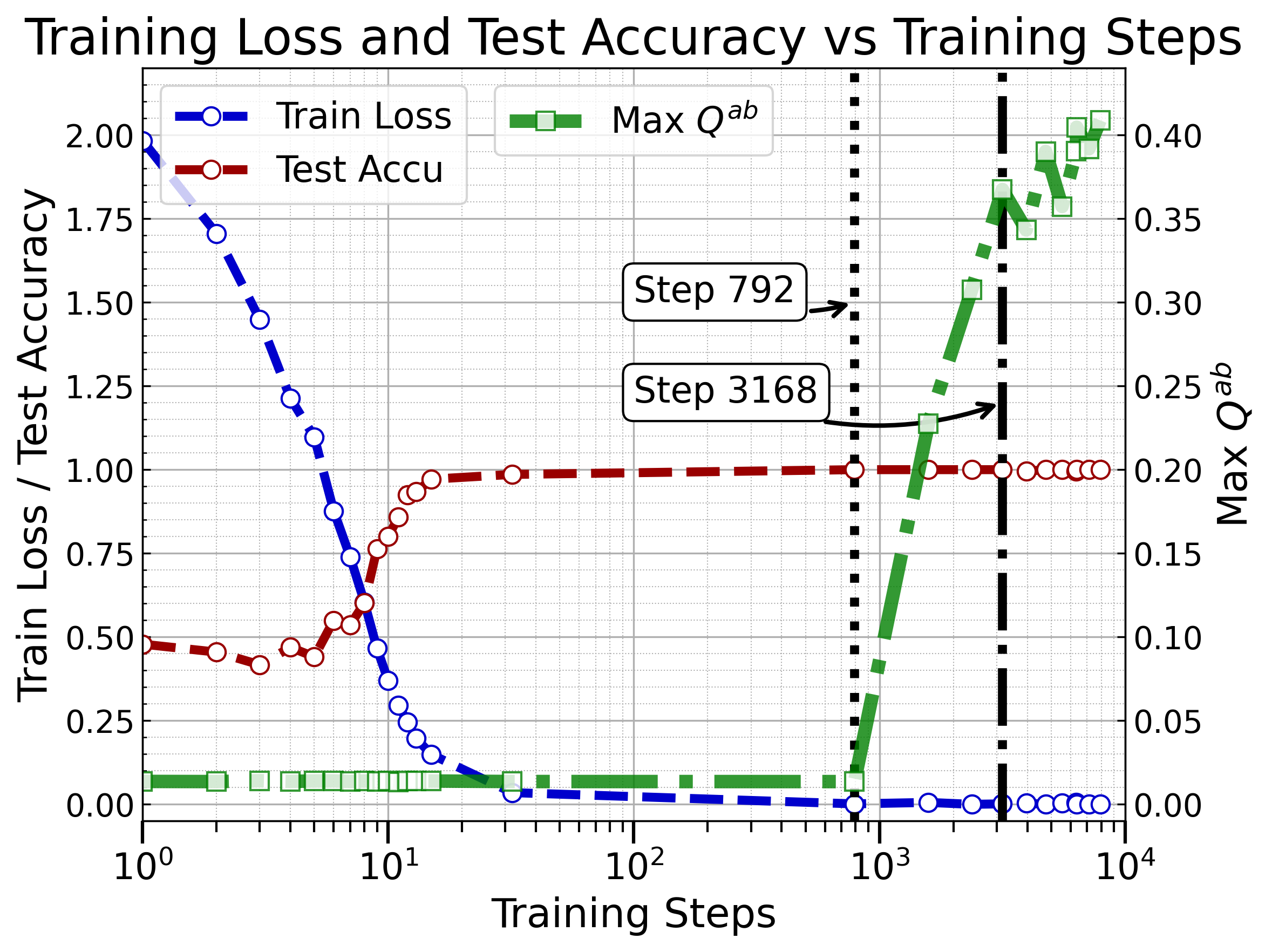} \\
    \textnormal{(a)}
\end{minipage}
\hfill
\begin{minipage}{0.45\textwidth}
    \centering
    \includegraphics[width=0.49\textwidth]{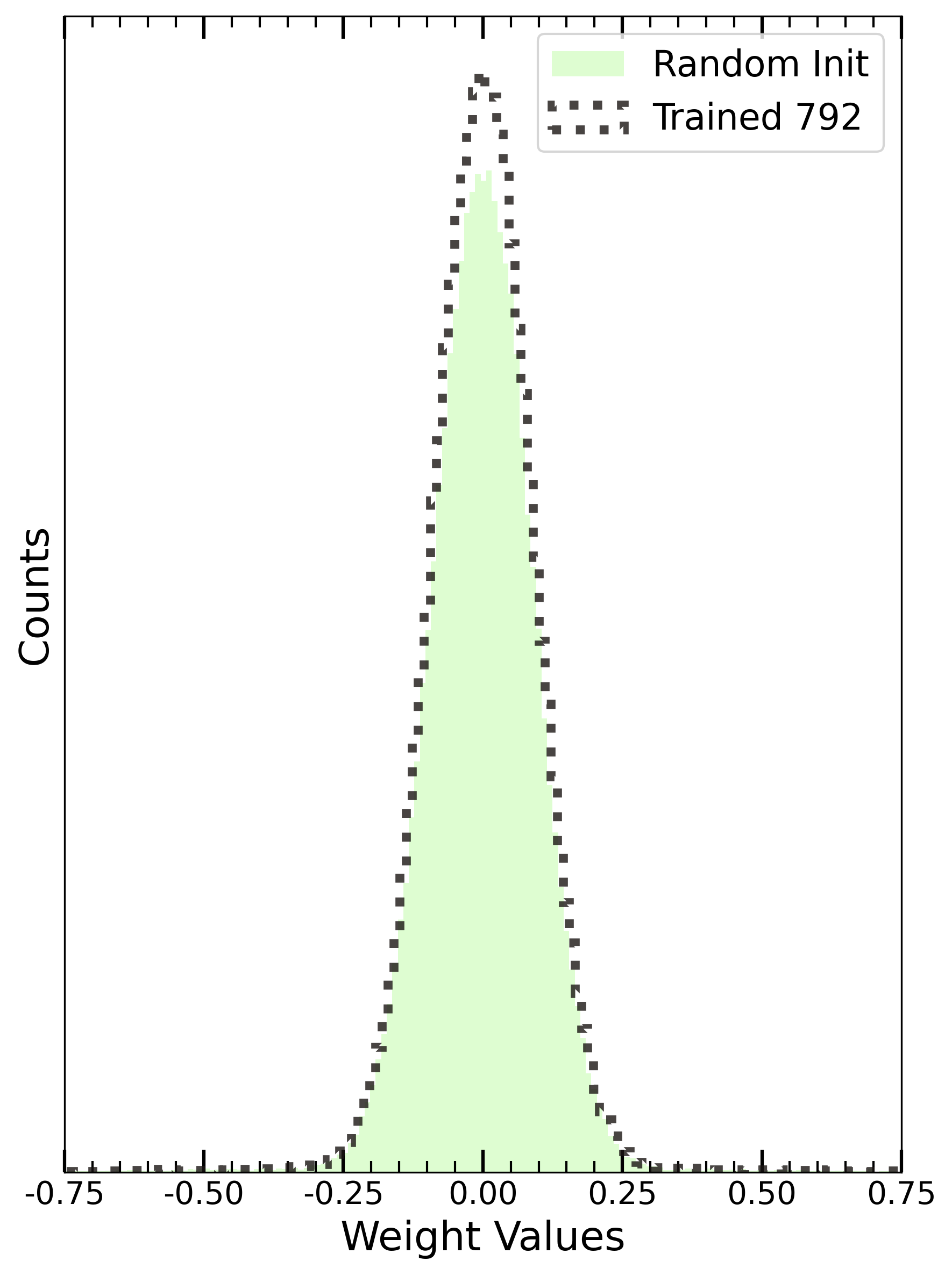} 
    \hfill 
    \includegraphics[width=0.49\textwidth]{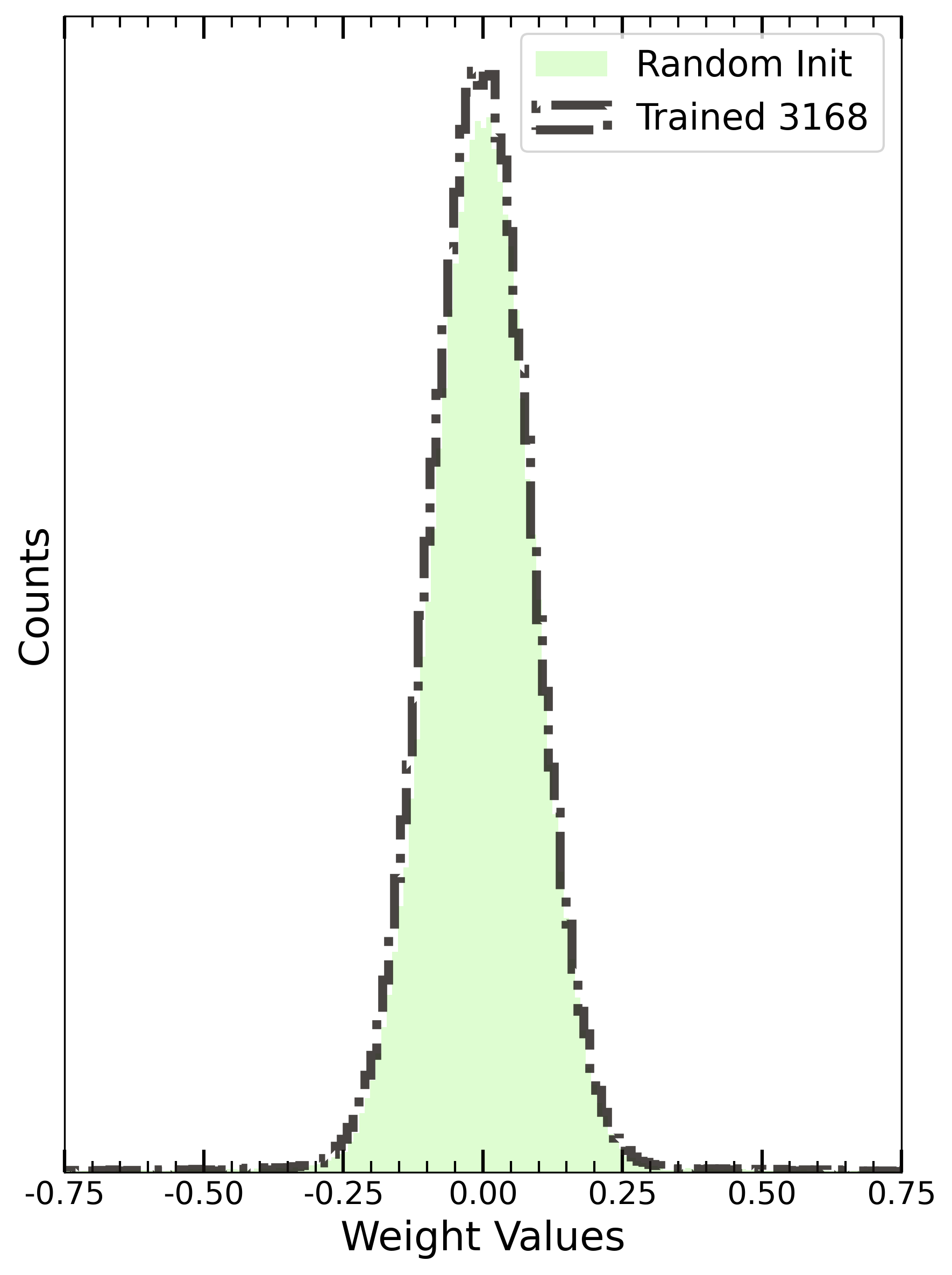} \\
    \centering
    \hfill
    \textnormal{(b$_1$)} 
    \hfill
    $\qquad \qquad$ \textnormal{(b$_2$)} 
    \hfill
    $\ $
\end{minipage}
\caption{Model metrics. 
{\bf (a)} $Q^{ab}$, train loss and test accuracy during training.  
Left y-axis:~the loss and accuracy;
Right y-axis:~the peak $Q^{ab}$ values.
Marked dots indicates the same training step.
{\bf (b)} Weights histograms, before and after $Q^{ab}$
distinction observed (indicated by verticle lines in (a)). 
(b$_1$) {\tt \small step=792, train\_loss$\approx$0.15e-3, test\_accu$>$99.9\%}
(b$_2$) {\tt \small step=3,168, train\_loss$\approx$0.15e-3, test\_accu$>$99.9\%}
}
\label{fig:compare_metrics}
\end{figure}

\paragraph*{Comparison to common metrics}
$Q^{ab}$ curves capture model fitness from a thermodynamic perspective,
which is different from conventional metrics such as training loss, test
accuracy, or parameter statistics.
Fig.~\ref{fig:compare_metrics}(a) compares (i) cross-entropy loss on the
training set, (ii) test set accuracy, and (iii) the peak overlap on the
corresponding $Q^{ab}$ curves.
All quantities are computed from training checkpoints for the {\em default
task}.
As shown in the plots, replica overlap begins to rise {\em after} loss and
accuracy have saturated.
In realistic tasks, phenomena such as double descent and grokking are commonly
observed \cite{Minegishi2025,Davies2023}:
Continuing optimization beyond the point where training loss flattens, test
performance can improve again after a plateau.
In our experiment, continued training results in a shift in spin glass dynamics
manifested by the $Q^{ab}$ curves. Nevertheless, test accuracy does not improve
further in this simple task.

Fig.~\ref{fig:compare_metrics}(b$_{1}$, b$_{2}$) compares weight histograms before
and after significant changes in the $Q^{ab}$ curves (checkpoint details are
provided in the caption).
Shaded regions indicate the initial distribution, while silhouettes outline
weights at two distinct training steps (b${1}$ and b${2}$), respectively.
The histogram plots show that simple statistics fail to capture structural
differences between models, which are revealed by the $Q^{ab}$ curves.

\begin{figure}[t]
\centering
\begin{minipage}{0.45\textwidth}
    \centering
    \includegraphics[width=\textwidth]{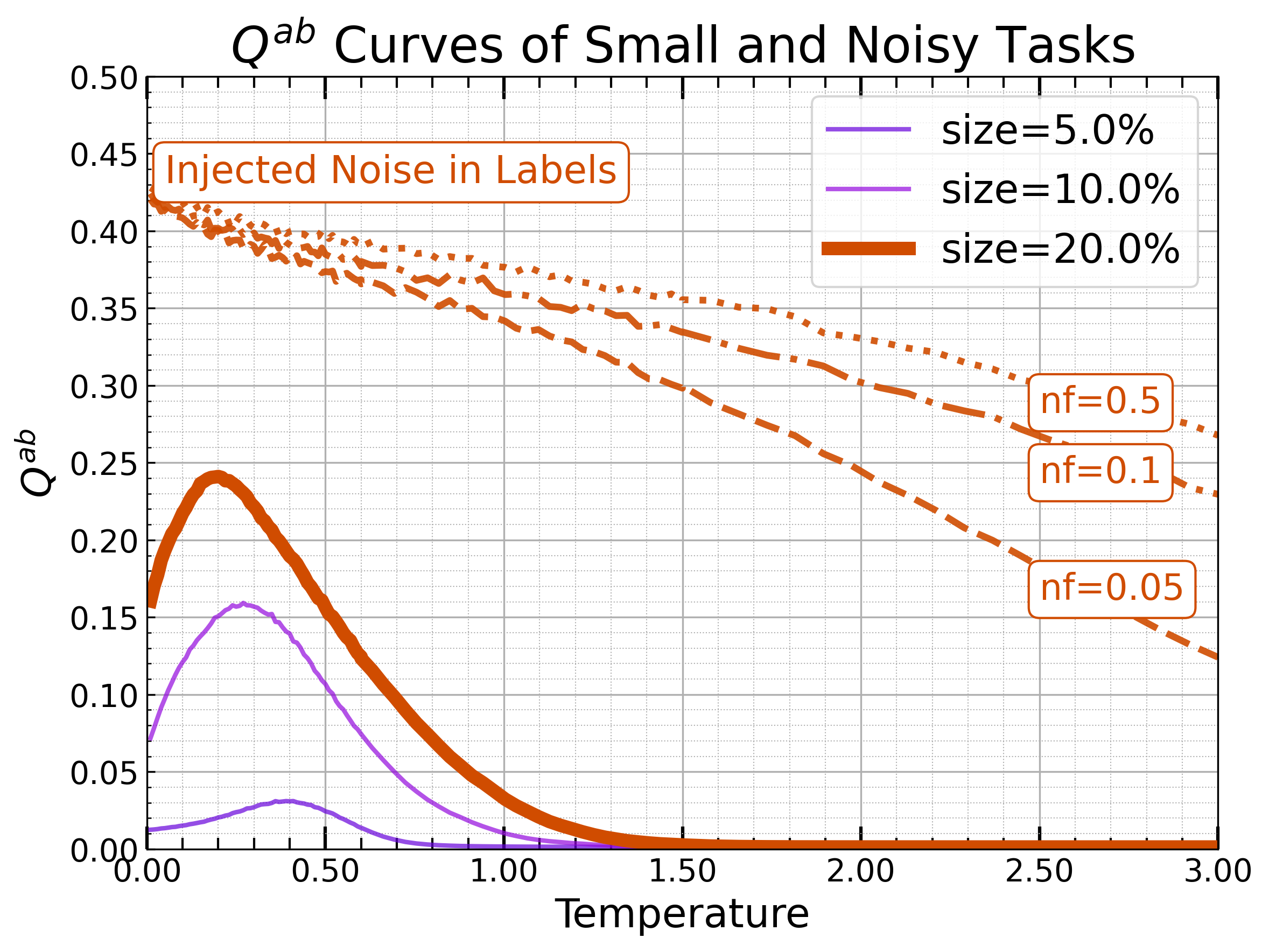} \\
    \textnormal{(a)}
\end{minipage}
\hfill
\begin{minipage}{0.45\textwidth}
    \centering
    \includegraphics[width=\textwidth]{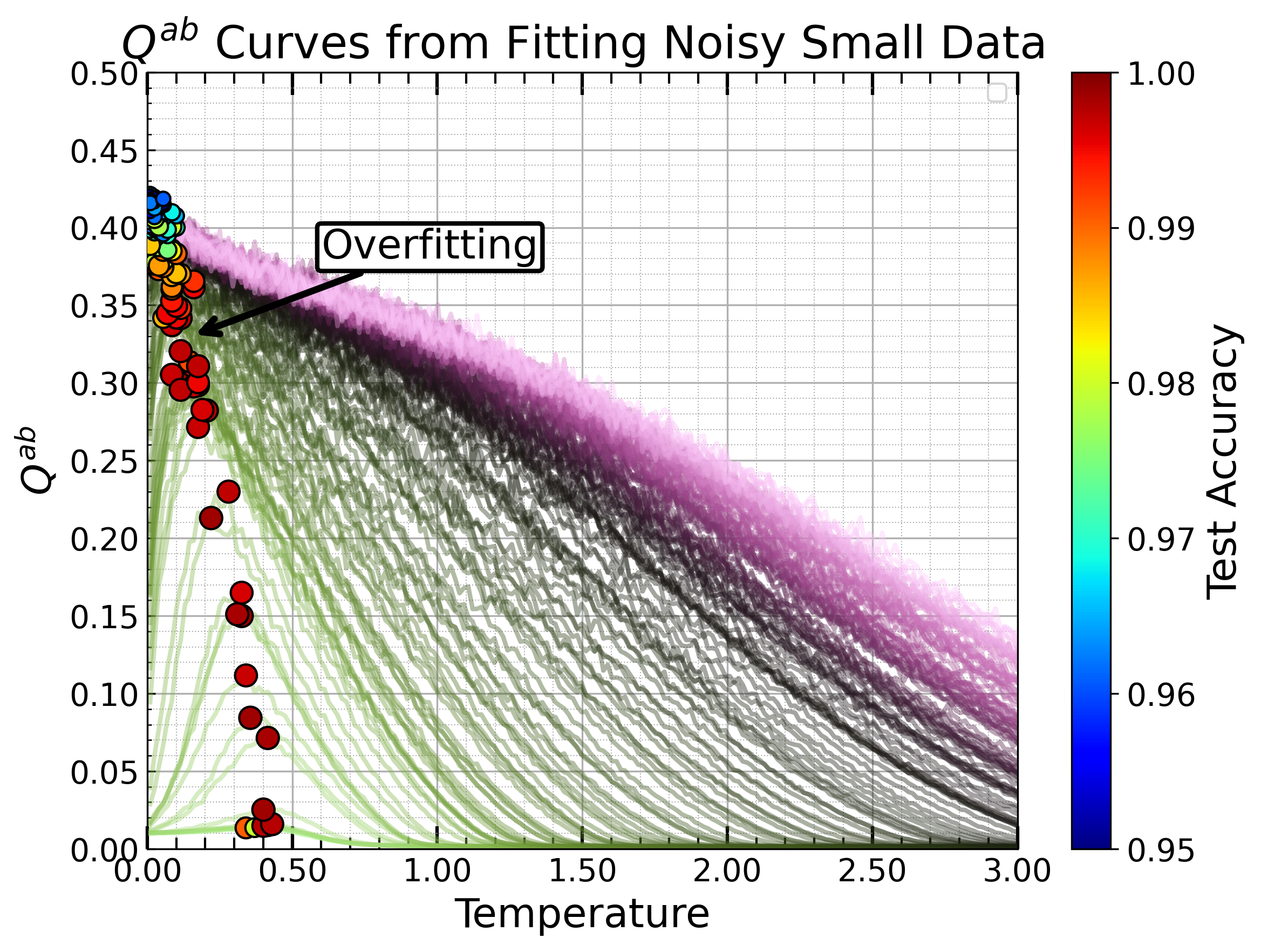} \\
    \textnormal{(b)}
\end{minipage}
\caption{$Q^{ab}$ curves and training conditions.
{\bf (a)} Small datasets and noisy labels.
``{\tt nf}'' denotes the noise factor, i.e., the probability of flipping a
training label.
Noise experiments are conducted on a 20\% subset
Thus, the ``{\tt nf}'' curves are comparable to the bold one on top.
{\bf (b)} Overfitting to noisy datasets. 
Dots indicate the peak of each $Q^{ab}$ curve and are colored by test accuracy.
Overfitting is indicated by the arrow.
\label{fig:mnist_overfit}}
\end{figure}

\subsection{$Q^{ab}$ curves to examine learning abnormalities}
\label{ssec:experi:abnormalities}
Characterizing models and learning tasks via $Q^{ab}$ curves enables examination
of the learning process, such as model pre-conditioning or anomalous data
patterns.

\paragraph*{Data quality and overfitting}
Small or noisy training datasets are common in practice.
However, what counts as “small” or “noisy” is relative to the task and model
capacity.
In this experiment, $Q^{ab}$ curves are used to concretely characterize such
situations.
In the {\em default task}, small subsets (5-20\%) of training data are
used. 
The corresponding $Q^{ab}$ curves are shown in Fig.~\ref{fig:mnist_overfit}(a),
maked ``size''.  
%
Smaller datasets produce $Q^{ab}$ curves that are less distinguishable from
those of random models.
This suggests that less information is encoded in the trained model, and fewer
metastable modes emerge in the spin system.
%
In contrast, when label noise is added to the 20\%-subset (see SM), the $Q^{ab}$
curves exhibit significant values persisting at higher temperatures.
This coincides with a drop in test accuracy (overfitting), indicating that the
model memorized the noise.

\begin{figure}[t]
\centering
\begin{minipage}{0.45\textwidth}
    \centering
    \includegraphics[width=\textwidth]{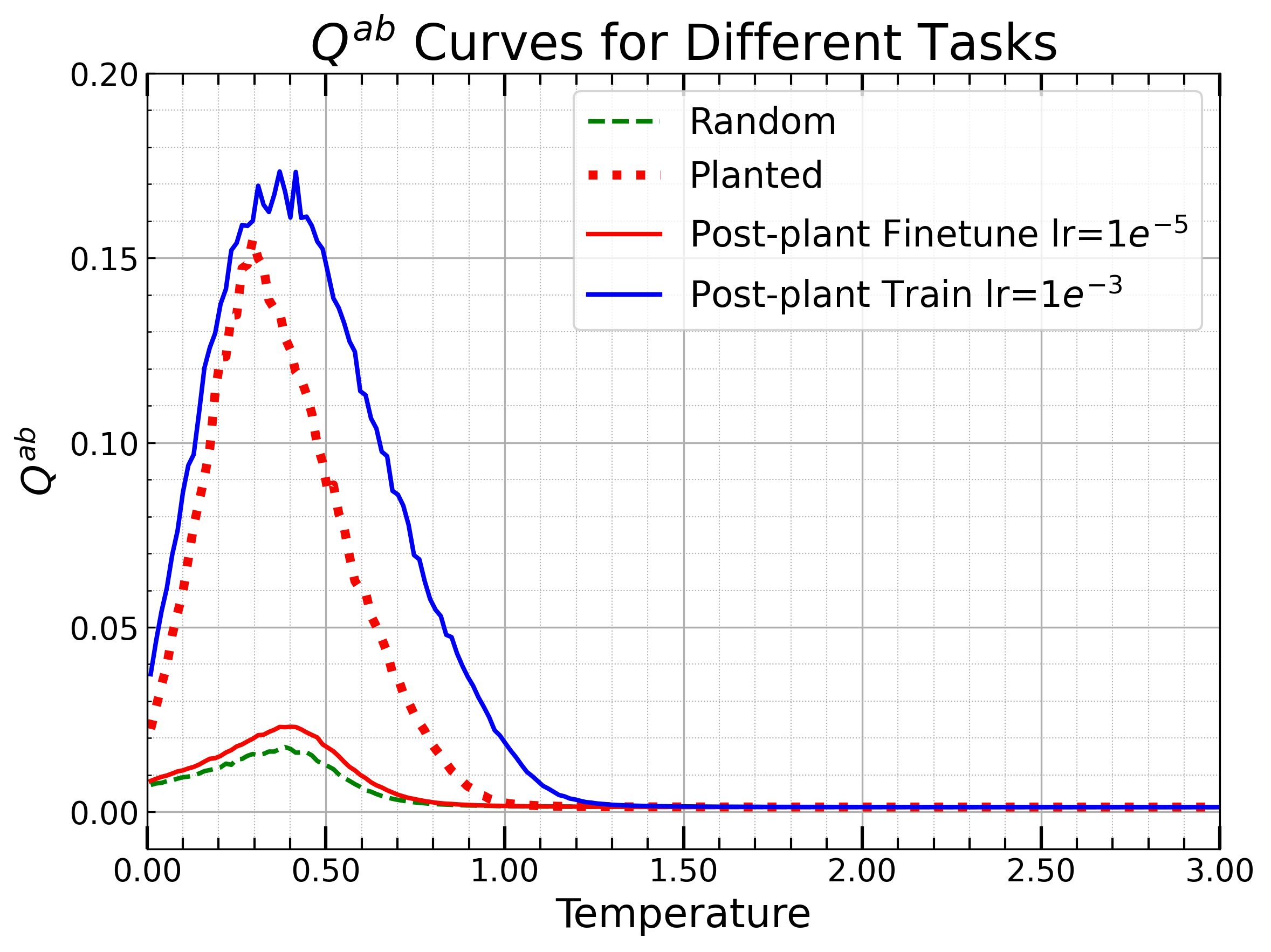} \\
    \textnormal{(a)}
\end{minipage}
\hfill
\begin{minipage}{0.45\textwidth}
    \centering
    \includegraphics[width=\textwidth]{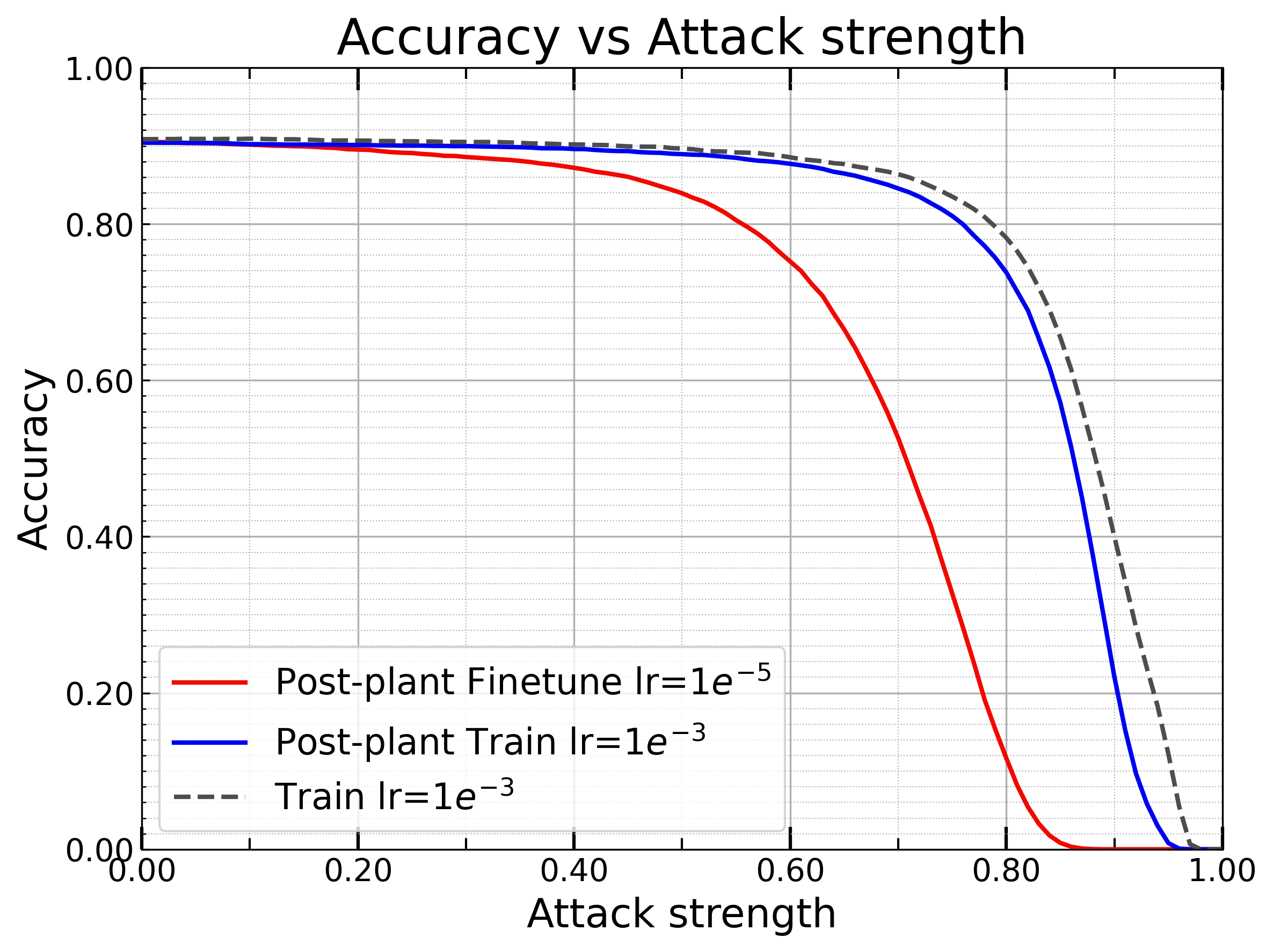} \\
    \textnormal{(b)}
\end{minipage}
\caption{Examine models with planted patterns.
{\bf (a)} $Q^{ab}$ curves of models initialized randomly and with a
a planted pattern. 
The planted model is subsequently trained using two schemes.
Broken lines represent initial models; solid lines represent trained/finetuned
models.
{\bf (b)} Accuracy under attacks using the planted pattern with varying
strengths, for three trained models.
}
\label{fig:cifar_plant}
\end{figure}

\paragraph*{Planted pattern}
%
A scenario with a ``planted pattern'' is tested in this experiment.  
The planted pattern is a random vector of the MLP input ($256$-D)
in the CIFAR-10 task.
The MLP is overfitted so that the classifier produces a fixed response. 
The pre-conditioned parameters are then normalized to match the distribution of
standard random initialization. 
In Fig.~\ref{fig:cifar_plant}(a), the two broken lines show the $Q^{ab}$ curves
of the standard random model and the planted model. 
The distinction is clear and consistent with previous experiments. 
The MLP of the planted model was trained on the standard dataset using learning
rates of $10^{-3}$ (``post-plant train'') and $10^{-5}$ (``post-plant
finetune''), respectively. 
%
Both schemes result in similar training loss and test accuracy.
However, the $Q^{ab}$ curves (solid lines in Fig.~\ref{fig:cifar_plant}(a))
reveal that the models occupy distinct dynamical regimes.

The distinction is further reflected when the trained models are attacked by
injecting the planted pattern with varying strengths. 
The standard model and the post-plant trained model exhibit similar robustness
to the attack. 
In contrast, the post-plant finetuned model shows vulnerability at weaker attack
strengths, as illustrated in Fig.~\ref{fig:cifar_plant}(b). 
This is consistent with the $Q^{ab}$ curve of the finetuned model, which
suggests that the model has not fully escaped the planted pattern, i.e.
the information in the standard training data is not fully
encoded in the corresponding spin system. 

%% file: arxiv_secs/sec2_related.tex
\section{Related Work and Discussion}
\label{sec:related_work}

Importing the ideas from statistical mechanics into the study of neural networks
has a long history.  
Hopfield \cite{Hopfield1982} demonstrated that associative memory
emerges as a collective phenomenon among interacting neurons. The
network retains patterns as fixed points of the dynamics~\eqref{eq:spinprob}.
The capacity of emergent memory has been analyzed using tools from statistical
mechanics
\cite{Amit1985,Gardner1988}. 
E.g., Gardner~\cite{Gardner1988} analyzed the volume of 
weights space for storing a given number of patterns with specified attraction
basin sizes. The size of this volume is formulated as a function of the overlap
between typical samples at the equlibrium, suggesting a close link between the
network's energy landscape and replica overlaps.
%
The success of modern large-scale neural networks has recently raised questions
that challenge classical statistical learning theory~\cite{E2020},
where physics and statistical mechanics provide a powerful framework for studying
large-scale neural networks as complex stochastic systems \cite{Bahri2020}:
Why do large networks generalize \cite{Zhang2017}? 
Why does first-order optimization work well for exploring complex landscapes?~\cite{Li2019}

{\bf Data-independent (prior) ensemble properties} are concerned with the
capacity and expressivity of neural networks.
In the Bayesian framework~\cite{Lee2018,Neal1996}, this corresponds to the model
prior: {\em random neural network ensembles}, i.e., networks with identical
architecture and randomly initialized weights.
A rich body of work has been devoted to studying such ensemble properties.
In~\cite{Poole2016,Raghu2017}, the expressivity of deep networks is studied via
the evolution of input-output correlations across layers, showing that
architectural priors in deep ensembles facilitate learning complex nonlinear
functions.
In~\cite{Schoenholz2017}, a mean-field theory is used to compute how far signals
propagate through layers of a random network ensemble.
The findings apply to both forward and backward passes and help explain
trainability.
Later works~\cite{Zhang2021,Thurn2024} relate trainability to the transition
between ordered and chaotic phases.

{\bf Loss landscape} affects important {\em posterior} properties of learned
models, such as generalization~\cite{Hochreiter1997, Dauphin2014,Zdeborova2016}.
Choromanska et al.~\cite{Choromanska2015} show a connection between the loss of
deep networks and the Hamiltonian of a spherical spin glass. 
Using results from~\cite{Auffinger2013}, it is shown that most critical points
correspond to low loss (energy), and the landscape is relatively flat for large
networks.
Extensive research has been devoted to related topics, e.g., Gaussian
fields~\cite{Bray2007}, Hessian eigenvalue spectra~\cite{Sagun2016},
classification of critical points~\cite{Cheridito2022}, and topological
properties~\cite{Arjevani2024}.
The information-theoretic properties of random neural network ensembles are also
of significant interest, relating to both function priors and
generalization~\cite{Tishby1999}.
Entropy and mutual information between activations in hidden layers are computed
in~\cite{Gabrie2018}.
In~\cite{Kabashima2008}, the typical inference performance of a perceptron is
studied in terms of the size of training samples.

Most existing studies on prior and posterior properties are analytical, focusing
on {\em ensemble} characteristics.
%
%
In contrast, the present work is operational and applies to individual network
instances.
The overlap between replica samples is empirically estimated via Gibbs sampling
to characterize the underlying network.

\twover{
{\bf Dynamics of optimization} is another essential component a learning system.
The optimization algorithm governs the navigation over loss landscapes,
influencing the distribution of learned solutions, which is not necessarily the
same as the one determined by the equilibrium analysis.

For instance, stochastic gradient descent (SGD) is treated as a Bayesian
learning framework \cite{Mandt2017}, where the posterior distribution is
influenced by the learning specifications such as learning rate and the batch
size. The posterior ensemble performance is studied recently in
\cite{Jordahn2025}.  The mechanism underlying SGD has also been studied using
the model of Langevin dynamics \cite{Li2019}.  The noises in SGD are treated
using continuous-time stochastic differential equation (SDE), from which the
Fokker-Planck equation is derived to specify the evolution of the probability
distribution in the parameter space \cite{Shi2023,Chen2025}. It is shown that
the parameter distribution depends on the learning rate in optimization.

Existing statistical mechanics studies on neural networks are mostly concerned
with the equalibrium properties and taking less account of the optimization
dynamics. While the present work reports the empirical $Q^{ab}$ curves that 
corroborate the theoretical analysis. 
}
{
{\bf Dynamics of optimization} governs navigation over loss landscapes and
influences the distribution of learned models, which does not necessarily
coincide with the theoretical equilibrium.
E.g., stochastic gradient descent (SGD) can be interpreted within a
Bayesian learning framework~\cite{Mandt2017}, where the posterior distribution
is influenced by learning specifications such as the learning rate and batch
size.
The posterior ensemble performance is studied recently in~\cite{Jordahn2025}.  
The dynamics underlying SGD have also been analyzed using Langevin
models~\cite{Li2019}.
The noise in SGD is modeled using continuous-time stochastic differential
equations (SDEs), from which the Fokker-Planck equation is derived to describe
the evolution of the parameter-space probability
distribution~\cite{Shi2023,Chen2025}.
It has been shown that the resulting parameter distribution depends on the
learning rate used during optimization.

Existing statistical mechanics studies on neural networks mostly focus on
equilibrium properties, with less attention given to optimization dynamics.
In contrast, the present work reports empirical $Q^{ab}$ curves that corroborate
the theoretical analysis.
}

%% file: arxiv_secs/sec5_con.tex
\subsection*{Discussion and limitations}
\twover{
It is helpful to discuss open questions and incomplete aspects of the work.
The supplementary materials contain more discussions on some topics and
preliminary results.
}{}
{\bf Description of replica symmetry breaking and spin-glass} 
Replica overlaps exhibit rich structure~\cite{Talagrand2011}, but the present
work considers only the average absolute off-diagonal components.
Glassy properties of the system, such as relaxation time, are not explored.
A more complete picture of neural networks may emerge from future
investigations.

\twover{
{\bf Architecture} The proposal is about generic FNNs, but the experiments are
reported on regularly shaped MLPs. Empirical tests on convolutional nets,
transformers and networks with skip connections, or recurrent topologies, are of
interest.  
}
{
{\bf Architecture} The experiments are conducted on regularly shaped MLPs. 
Empirical evaluations on convolutional networks, transformers with skip
connections, or recurrent architectures remain of interest.
Some results on the depth–width trade-off in MLPs are included in the
Supplementary Material.
}

\twover{
{\bf HNN construction} The ``structure-clone'' construction of HNN from FNN is
straightforward and natural. When the model is large, the computational cost of
simulating the spins to obtain the $Q^{ab}$ curves can be prohibitive.  
Methods of subsampling the neurons and obtaining a representative spin-glass
model are of interest.  Moreover, to which extent the continuous neurons
can be represented by the behavior of the binary spins deserves further
investigation.  
}
{
{\bf HNN construction} The ``clone'' construction of HNN from FNN is na\"{i}ve.
However, the computational cost of simulating spins becomes prohibitive for
large models.
Methods for neuron subsampling and constructing representative spin glass models
are of interest.
Moreover, to what extent continuous neurons can be faithfully represented by
binary spin dynamics requires further investigation.
}

Finally, in the planted pattern experiment
(Subsection~\ref{ssec:experi:abnormalities}), some unexpected but intriguing
results were observed.
Directly planting a pattern in the input images did not yield similar $Q^{ab}$
changes or sensitivity in the post-plant fine-tuned model.
A possible explanation is that the convolutional encoder layers transformed the
input signals such that the planted pattern was no longer distinguishable,
warranting further investigation.


\section{Conclusion}
\label{sec:conclusion}

A spin-glass characterization of neural networks is proposed.
Hopfield-type spin systems are constructed from feedforward networks (FNNs).
The phenomenon of replica symmetry breaking (RSB) is used to characterize the
FNNs. Replica samples are generated from the spin-glass model at different
temperatures, and the average overlaps form a $Q^{ab}$ curve.
The $Q^{ab}$ characterization reveals key properties of neural networks,
including their training dynamics and capacity, which are empirically studied.
This work bridges thermodynamic theory with practical neural network analysis by
providing an operational method to characterize individual models via their
spin-glass behavior.
Such characterizations may be of benefit for future 
practical tools for auditing, robustness assessment, and detection of anomalous
behaviors.

%
%
%
%

%% file: arxiv_secs/sec_appendix.tex
\section{Appendix}
\newcommand{\mdef}[2]{{#1} := {#2}}
\newcommand{\wildindex}{{\boldsymbol{\cdot}}}
\newcommand{\rep}[1]{{\ve{\sigma}^{(#1)}}}
\newcommand{\Q}{{\ve{Q}}}
\newcommand{\debugincludegraphics}[2][]{%
  \begingroup
  \setkeys{Gin}{draft=false}%
  \includegraphics[#1]{#2}
  \endgroup
}
\newcommand{\pib}[1]{
    {p_{i, \beta}(\sigma_i=#1; \ve{J}, \ve{\sigma}_{\backslash i})}}

\subsection{Background on statistical mechanics of spin systems}
\subsubsection*{Spin systems and dynamics}

A brief overview of the foundational concepts of statistical mechanics is
included for completeness. For more details, interested readers are referred to
standard textbooks such as \cite{Mezard2009, Engel2001}.

A system of $N$ spins $\ve{\sigma} = \{ \sigma_1, \sigma_2, \dots, \sigma_N\}$
is specified by a Hamiltonian $H(\ve{\sigma})$, which defines the system's
energy landscape.
The Hamiltonian $H$ defined in \eqref{eq:ising} consists of pairwise spin
interactions, analogous to the connections found in typical feedforward neural
networks (FNNs).
For convenience, the Hamiltonian in \eqref{eq:ising} is reproduced below:
\begin{align*}
H(\ve{\sigma}; \ve{J}) & = - \sum_{1\leq i < j \leq N}
    J_{i,j} \sigma_{i} \sigma_{j} 
\end{align*}
It is worth noting that some modern neural networks involve more complex
interactions, such as the attention mechanism introduced in \cite{Vaswani2017}.

The Hamiltonian-defined energy landscape governs the stochastic dynamics of spin
state transitions.
The evolution of the probability distribution is governed by \cite{Glauber1963}:
\begin{align}
\frac{d p_\beta(\ve{\sigma}; \ve{J})}{dt} 
&= \sum_{i=1}^{N} \left[ 
    \omega_i(\ve{\sigma}^{(i)}; \ve{J}) 
    p_\beta(\ve{\sigma}^{(i)}; \ve{J}) 
    - 
    \omega_i(\ve{\sigma}; \ve{J}) 
    p_\beta(\ve{\sigma}; \ve{J}) 
\right]
\label{eq:glauberdynamics}
\end{align}
Here, $\omega_i(\ve{\sigma}; \ve{J})$ denotes the rate at which spin $i$ flips
its state, keeping the remaining spins $\ve{\sigma}_{\backslash i}$ 
fixed. 
The notation $\ve{\sigma}^{(i)}$ represents the configuration obtained by
flipping spin $i$, i.e., $\sigma_i \to -\sigma_i$.
The flipping rate $\omega_i$ is determined by the local field $h_i$ acting on
spin $i$, which depends on the couplings $J_{ij}$ and the states of
neighboring spins, as defined in \eqref{eq:ising}.
\begin{align}
\omega_i(\ve{\sigma}; \ve{J}) = \frac{1}{2} \left[1 - \sigma_i \tanh(\beta
h_i)\right]
\end{align}
This form is consistent with \eqref{eq:spinprob} and underlies the Gibbs
sampling step in Line~\ref{alg:replica:ln:gibbs} of Algorithm~\ref{alg:replica}.
The stationary distribution of the system evolution is the Boltzmann
distribution \eqref{eq:boltzmann}. In the present work, the spin systems are
considered closed, and Boltzmann and Gibbs distributions are equivalent. The
term ``Gibbs distribution'' is used throughout for consistency with the sampling
method.

\subsubsection*{Replica method and replica symmetry}
The following discussion provides a brief introduction to the replica method,
which serves to motivate the technique used in this work.

{\bf Free entropy and observables.}
Consider an observable quantity of interest in a spin system, denoted by
$O(\ve{\sigma})$.
The average value of this observable over the system's ground states is
\begin{align} \label{eq:obsoptimal}
O^*(\ve J) = & \mathbb{E}_{\ve{\sigma} \in \ve{S}^*} [O(\ve{\sigma})]  \\
\ve{S}^* & = \arg\min_{\ve{\sigma}} H(\ve{\sigma}; \ve J)
\end{align}
where the ground states $\ve S^*$ minimize the Hamiltonian. The dependence of
$O$ on $\ve J$ arises because the Hamiltonian is parameterized by $\ve J$, which
in turn determines the ground states $\ve S^*$.
In practice, the expectation over $\ve S^*$ is approximated as the
zero-temperature limit of the expectation under the Gibbs (Boltzmann)
distribution $p_\beta(\ve{\sigma})$ in \eqref{eq:boltzmann}, i.e., as $T \to 0$
or $\beta \to \infty$,
\begin{align}
O^*(\ve J) &= \lim_{\beta \to \infty}  O_\beta (\ve J)\\
O_\beta(\ve J) &=
\sum_{\ve{\sigma}} O(\ve{\sigma}; \ve{J}) p_\beta(\ve{\sigma}; \ve J) \label{eq:obsGibbs}
\end{align}
where the summation runs over all possible spin configurations $\ve{\sigma}$,
and $p_\beta(\ve{\sigma})$ denotes the Gibbs distribution \eqref{eq:boltzmann},
reproduced with the corresponding partition function as 
\newcounter{savedEqCounter}
\setcounter{savedEqCounter}{\value{equation}}
\setcounter{equation}{\numexpr\getrefnumber{eq:boltzmann}-1\relax}
\begin{align}
p_\beta(\ve{\sigma};\ve J) 
&= \frac{1}{Z_\beta} \exp(-\beta H(\ve{\sigma}; \ve J)) \\
Z_\beta(\ve J) 
&= \sum_{\ve{\sigma}} \exp(-\beta H(\ve{\sigma}; \ve J))
\end{align}
\setcounter{equation}{\value{savedEqCounter}}
Substituting \eqref{eq:boltzmann} into the expression for $O_\beta$ yields
\begin{align}
O_\beta(\ve J) & = \frac{1}{Z_\beta(\ve J)} 
\sum_{\ve{\sigma}} 
    O(\ve{\sigma}) \exp(-\beta H(\ve{\sigma}; \ve{J})) \label{eq:obsGibbs2} 
\end{align}
The expectation $O_\beta(\ve{J})$ can alternatively be derived by introducing an
{\em augmented partition function},
\begin{align}
\tilde{Z}_\beta(\ve J, \alpha) = \sum_{\ve{\sigma}} 
    \exp\big[-\beta H(\ve{\sigma};\ve{J}) +
    \alpha O(\ve{\sigma})\big] \label{eq:Ztilde}
\end{align}
Taking the derivative of $\log \tilde{Z}_\beta(\ve J, \alpha)$ with respect to
$\alpha$ at $\alpha = 0$ recovers the expectation in \eqref{eq:obsGibbs2}:
\begin{align}
    \frac{\partial }{\partial \alpha} \log \tilde{Z}_\beta(\ve J,\alpha) \mid_{\alpha=0} 
    &= 
    \frac{1}{\tilde Z_\beta(\ve J, 0)} \sum_{\ve{\sigma}} \big[
        \exp(-\beta H(\ve{\sigma})) O(\ve{\sigma}) 
    \big]
\end{align}
In the remainder of this section, the classical partition function is considered
without specifying any observable $O(\cdot)$; practical observables are assumed
to be incorporated into an effective Hamiltonian.
The free entropy $\log Z_\beta$—or equivalently, the free energy
$-\frac{1}{\beta} \log Z_\beta$—is the quantity of interest, as it characterizes
the macroscopic behavior of the system
\cite{Mezard1987,Kabashima2008,Gabrie2018, Engel2001}.



{\bf Quenched disorder.}
Given a fixed parameter set $\ve{J}$, evaluating the partition function
$Z_\beta(\ve{J})$ involves summing over all $2^N$ spin configurations, rendering
$\log Z_\beta(\ve{J})$ intractable.
%
In many practical scenarios, interest lies in the typical behavior of systems
governed by a distribution over parameters $\ve{J}$.
This requires computing the average free entropy, as reproduced from
\eqref{eq:logZavg}:
\setcounter{savedEqCounter}{\value{equation}}
\setcounter{equation}{\numexpr\getrefnumber{eq:logZavg}-1\relax}
\begin{align}
\qchavg{\log Z_\beta(\ve J)} & = \int_{\ve{J}} 
\log Z_\beta(\ve J) \prod\nolimits_{i,j} P(J_{i,j}) dJ_{i,j} 
\end{align} \setcounter{equation}{\value{savedEqCounter}}
This average is physically meaningful and also relevant in machine learning
contexts—for example, in characterizing the typical performance of models
trained under specified conditions, where the behavior of a single instance is
assumed representative of the ensemble due to self-averaging.
The parameters $\ve{J}$ evolve on a much slower timescale than the system's
thermodynamic dynamics, a condition referred to as {\em quenched disorder}.

%
%

{\bf Replica method.} The quenched average $\qchavg{\log Z_\beta(\ve J)}$ is
computed using the identity
\begin{align}
\qchavg{\log Z_\beta} & = \lim_{n\to 0} \frac{\log \qchavg{Z_\beta^n}}{n} \\
&= \lim_{n\to 0} \frac{\qchavg{Z_\beta^n} - 1}{n} \label{eq:replica:quenched}
\end{align}
where $Z_\beta^n$ denotes the partition function of $n$ replicated systems, defined as
\begin{align}
Z_\beta^n(\ve J) &= 
\sum_{\rep{1}, \rep{2}, \dots, \rep{n}}
\exp\big\{ -\beta \sum_{a=1}^n \mathcal{H}(\rep{a}; \ve{J}) \big\}
\end{align}
Here, $\ve{\sigma}^{(a)}$ denotes the $a$-th replica, consisting of $N$ spin
variables.
The quenched average of the replicated partition function is given by
\begin{align}
& \int \prod_{i,j} dJ_{i,j} P(J_{i,j}) Z_\beta^n \label{eq:replica:int} \\
&=\int \prod_{i,j} dJ_{i,j} P(J_{i,j}) \sum_{\ve{\sigma}^{(1)}, \ve{\sigma}^{(2)}, \dots, \ve{\sigma}^{(n)}}
\exp(-\beta \sum_{a=1}^n \mathcal{H}(\ve{\sigma}^{(a)}; \ve{J}))  
\end{align}
where $P(J_{i,j})$ is the probability density function of the quenched disorder
$\{J_{i,j}\}$. 
The replica trick removes the logarithm from the quenched average and enables
analytical progress by exchanging the order of integration and summation:
\begin{align}
\qchavg{Z_\beta^n} & = 
\sum_{\ve{\sigma}^{(1)}, \ve{\sigma}^{(2)}, \dots, \ve{\sigma}^{(n)}}
\underbrace{
\int \prod_{i,j} dJ_{i,j} P(J_{i,j}) 
\exp (-\beta \sum_{a=1}^n \mathcal{H}(\ve{\sigma}^{(a)}; \ve{J}))
}_{A} \label{eq:replica:avg:xchg} \\
A(\rep{1\dots n}) &= 
\left\langle 
\left\langle 
    \exp (-\beta \sum_{a=1}^n \mathcal{H}(\ve{\sigma}^{(a)}; \ve{J}))
\right\rangle 
\right\rangle  \label{eq:replica:avg:Areplica} 
\end{align}
Note that (i) in \eqref{eq:replica:avg:Areplica}, the dependence of the quenched
average $\qchavg{e^{-\beta H}}$ on the specific replica configuration
$\{\ve{\sigma}^{(a)}\}_{a=1}^n$ is made explicit through the functional form of
$A(\rep{1\dots n})$,
and (ii) on the left-hand side of \eqref{eq:replica:avg:xchg}, the dependence of
$Z_\beta^n$ on $\ve{J}$ is omitted, since $\ve{J}$ is integrated out in the
quenched average.

When both the Hamiltonian and the observable consist of simple terms involving
only a few spins, the integral $A$ in \eqref{eq:replica:avg:xchg} and
\eqref{eq:replica:avg:Areplica} can be reformulated as a function of the
overlaps between replica configurations.
\begin{align}
A(\rep{1\dots n}) = \exp \left\{ - n N F\big(\Q(\rep{1\dots n})\big) \right\}
\end{align}
where $\Q$ is the $n \times n$ overlap matrix for the replica configurations
$\rep{1\dots n}$, with elements
\begin{align}
\Q(\rep{1\dots n})_{[a, b]} = \frac{1}{N} \sum_{i=1}^{N} \sigma_i^{(a)} \sigma_i^{(b)}
\end{align}
and $F(\Q)$ is a function characterizing the joint energy and entropy associated
with $\Q$.
Using the notion of $\Q(\rep{1\dots n})$, the sum in \eqref{eq:replica:avg:xchg}
can be approximated by an integral over overlap matrices $\Q$
\begin{align}
\qchavg{Z_\beta^n} &= \sum_{\rep{1\dots n}} A(\rep{1\dots n}) \nonumber \\
&= \int_\Q d\Q \exp \big( - n N F(\Q) \big) \label{eq:replica:avg:Q} 
\end{align}
where $d\Q$ denotes the associated measure.
The function $F(\Q)$ encodes both the energetic contribution, inherited from the
generalized Hamiltonian, and the entropic contribution, which accounts for the
volume of replica configurations consistent with the overlap matrix $\Q$.

In the thermodynamic limit $N \to \infty$, the integral in \eqref{eq:replica:avg:Q} is dominated by its saddle point $\Q^*$, leading to the approximation
\begin{align}
\lim_{N\to \infty} \qchavg{Z_\beta^n} = \exp \big( - n N F(\Q^*) \big),
\end{align}
where the saddle point $\Q^*$ is given by
\begin{align}
\Q^* = \arg\min_\Q F(\Q).
\end{align}

{\bf Replica overlap and system characteristics.}
Loosely speaking, the structure of $\Q^*$ reflects the organization of the Gibbs
distribution at a given temperature.
For example, if the energy landscape has a unique minimizer $\ve{\sigma}^{(0)}$,
the dominant configurations tend to correlate similarly with
$\ve{\sigma}^{(0)}$.
In this case, the replicas exhibit two types of overlap: (i) self-overlap when
$a = b$, and (ii) mutual overlap when $a \neq b$ \cite{Talagrand2011}.
Such a simple $\Q$ matrix structure is referred to as the {\em replica symmetry
ansatz}, where all diagonal entries share one value and all off-diagonal entries
share another.

The ansatz breaks down when the Gibbs distribution exhibits a complex structure,
such as multiple minima or metastable states.
In such cases, the overlap matrix $\Q$ no longer has the simple two-level form.
This phenomenon is known as {\em replica symmetry breaking} (RSB), and commonly
occurs in spin glass systems at low temperatures.

In this work, the connection between RSB and qualitative changes in the Gibbs
distribution is used to analyze existing neural networks.
Replica dynamics are simulated using a Hopfield network (HNN) constructed from a
given feedforward network (FNN).
The procedure relies on the self-averaging property of the replica potential
with respect to the quenched disorder.
Accordingly, the quenched average is approximated using a single realization of
the network parameters.

\subsection{Technical Details of the Method}
This subsection provides technical details on constructing Hopfield networks
(HNNs) from feedforward networks (FNNs), as well as the sampling procedure for
the associated spin system. Definitions and conceptual motivations are discussed
in Section~\ref{sec:method} of the main text.


\begin{figure}
\begin{centering}
\includegraphics[width=10cm]{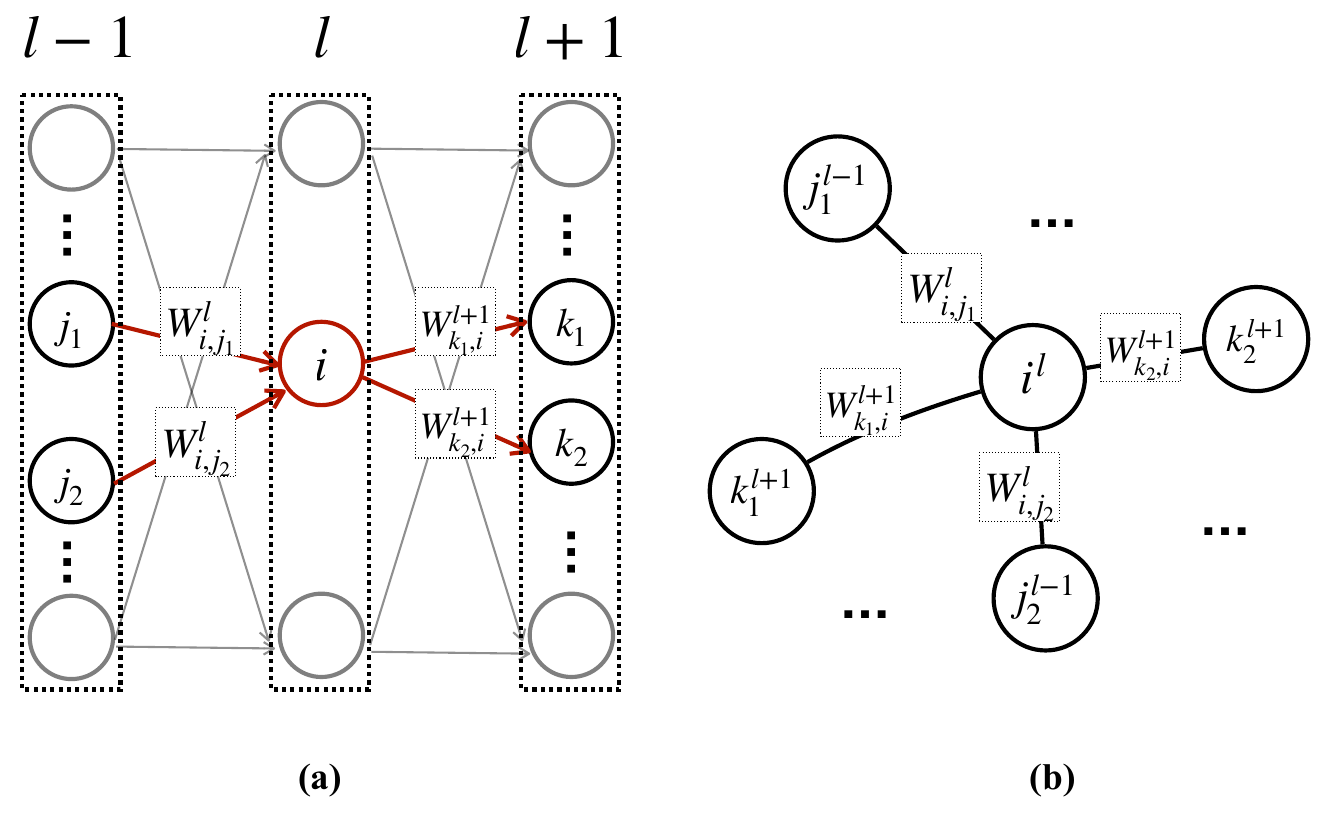}
\par
\end{centering}
\caption{Correspondence between the computational models of a feedforward neural
network (FNN) and a Hopfield neural network (HNN). 
\label{fig:nn-illu}
{\bf (a)} A neuron $i$ in layer $l$ of the FNN (bold red circle), with its
synaptic connections (bold red arrows) to neurons in the adjacent layers $l{-}1$
and $l{+}1$. 
{\bf (b)} The corresponding neuron in the HNN, labeled as $i^l$ (superscript
indicates the original FNN layer), and its neighborhood in the HNN (bold
circles). All connections are made bidirectional.  In the resulting HNN, neurons
such as $k_\wildindex^\wildindex$ and $j_\wildindex^\wildindex$ belong to
$\nbr{i}$, and the weight parameters $W_\wildindex^\wildindex$ become symmetric
coupling strengths $J_{i,j}$.}
\end{figure}

\subsubsection*{From FNNs to HNNs: Structural Mapping}
The construction of the Hopfield spin system (HNN) from a feedforward neural
network (FNN) follows a natural correspondence.
This correspondence is illustrated in Fig.~\ref{fig:nn-illu}, which is exact,
intuitive, and computationally convenient.
The remaining details are provided for completeness and may be skipped by
readers familiar with neural network models, or those who prefer to consult the
accompanying computer program for an exact description.

A FNN is specified by a set of weight matrices:
\begin{align}
\mdef{\ve{W}}{\{\ve{W}^{l}\}_{l=1}^{L-1}}
\end{align}
Each $\ve{W}^l$ is an $n_l \times n_{l-1}$ matrix, where $n_l$ denotes the
number of neurons in layer $l \in \{0, \dots, L-1\}$.
The network consists of $L$ layers, and the input layer is indexed by $l = 0$.
For $l > 0$, the neurons in layer $l$ are computed as \eqref{eq:fnn}, reproduced
as:
\setcounter{savedEqCounter}{\value{equation}}
\setcounter{equation}{\numexpr\getrefnumber{eq:fnn}-1\relax}
\begin{align}
x^l_i = \phi \big(\sum\nolimits_{j=1}^{n_{l-1}} W_{i,j}^{l} x_{j}^{l-1}\big)
\end{align}
\setcounter{equation}{\value{savedEqCounter}}
Thus, for any neuron {\color{red} $i$} in layer $1 \leq l \leq L{-}1$, its activation depends on:  
\begin{enumerate}[label=(\roman*)]
    \item neurons in the previous layer $l-1$, via the weights $W^{l}_{{\color{red} i},j}$;
    \item neurons in the next layer $l+1$, via the weights $W^{l+1}_{k,{\color{red} i}}$.
\end{enumerate}
In the corresponding Hopfield network (HNN), each neuron $i$ in layer $l$ of the
FNN is mapped to a spin variable $\sigma_{i_{\mathrm{HNN}}}$.
The flattened index $i_{\mathrm{HNN}}$ is defined as
\begin{align}
i_{\mathrm{HNN}} = \mathsf{HNNIndex}(i, l) := i + \sum_{l'=0}^{l-1} n_{l'}
\end{align}

The neighborhood of $i_{\mathrm{HNN}}$ is given by
\begin{align}
\nbr{i_{\mathrm{HNN}}} =
\left\{
    \mathsf{HNNIndex}(j, l{-}1)
\right\}_{j = 1}^{n_{l-1}}
\cup
\left\{
    \mathsf{HNNIndex}(k, l{+}1)
\right\}_{k = 1}^{n_{l+1}}
\end{align}
where $j$ and $k$ index the neurons in the $(l{-}1)$-th and $(l{+}1)$-th layers
of the original FNN, respectively.
Note that for $l = 0$, the input layer has no preceding layer, and for $l =
L{-}1$, the output layer has no subsequent layer.

The coupling matrix $\ve{J}$ of the HNN is constructed from the
feedforward weights $\ve{W}$ by symmetrizing the local connections between
adjacent layers. For any pair of neurons $i$ in layer $l$ and $j$ in layer
$l{-}1$, the corresponding HNN indices are
\begin{align}
i_{\mathrm{HNN}} = \mathsf{HNNIndex}(i, l), \quad
j_{\mathrm{HNN}} = \mathsf{HNNIndex}(j, l{-}1)
\end{align}
The symmetric coupling strength is defined as
\begin{align}
J_{i_{\mathrm{HNN}}, j_{\mathrm{HNN}}} 
= J_{j_{\mathrm{HNN}}, i_{\mathrm{HNN}}} 
:= \frac{1}{2} \left( W^l_{i,j} + W^{l}_{i,j} \right) = W^l_{i,j}
\end{align}
since the FNN weight matrix $\ve{W}^l$ defines a directed connection from layer
$l{-}1$ to layer $l$.

Similarly, the backward connection from $i$ in layer $l$ to $k$ in layer $l{+}1$
contributes
\begin{align}
J_{i_{\mathrm{HNN}}, k_{\mathrm{HNN}}} := W^{l+1}_{k,i}
\end{align}
Hence, the full symmetric coupling matrix $\ve{J}$ is defined by:
\begin{align}
J_{i_{\mathrm{HNN}}, j_{\mathrm{HNN}}} := 
\begin{cases}
W^{l}_{i,j} & \text{if } i \in \text{layer } l, \ j \in \text{layer } l{-}1 \\
W^{l+1}_{j,i} & \text{if } i \in \text{layer } l, \ j \in \text{layer } l{+}1 \\
0 & \text{otherwise}
\end{cases}
\end{align}
The resulting matrix $\ve{J}$ is sparse, symmetric, and encodes the layer-wise
topology of the original FNN in the HNN representation.

\subsubsection*{Generalized Gibbs sampling} 
\begin{figure}[h]
\centering
\begin{minipage}{0.49\textwidth}
    \centering
    \includegraphics[width=\textwidth]{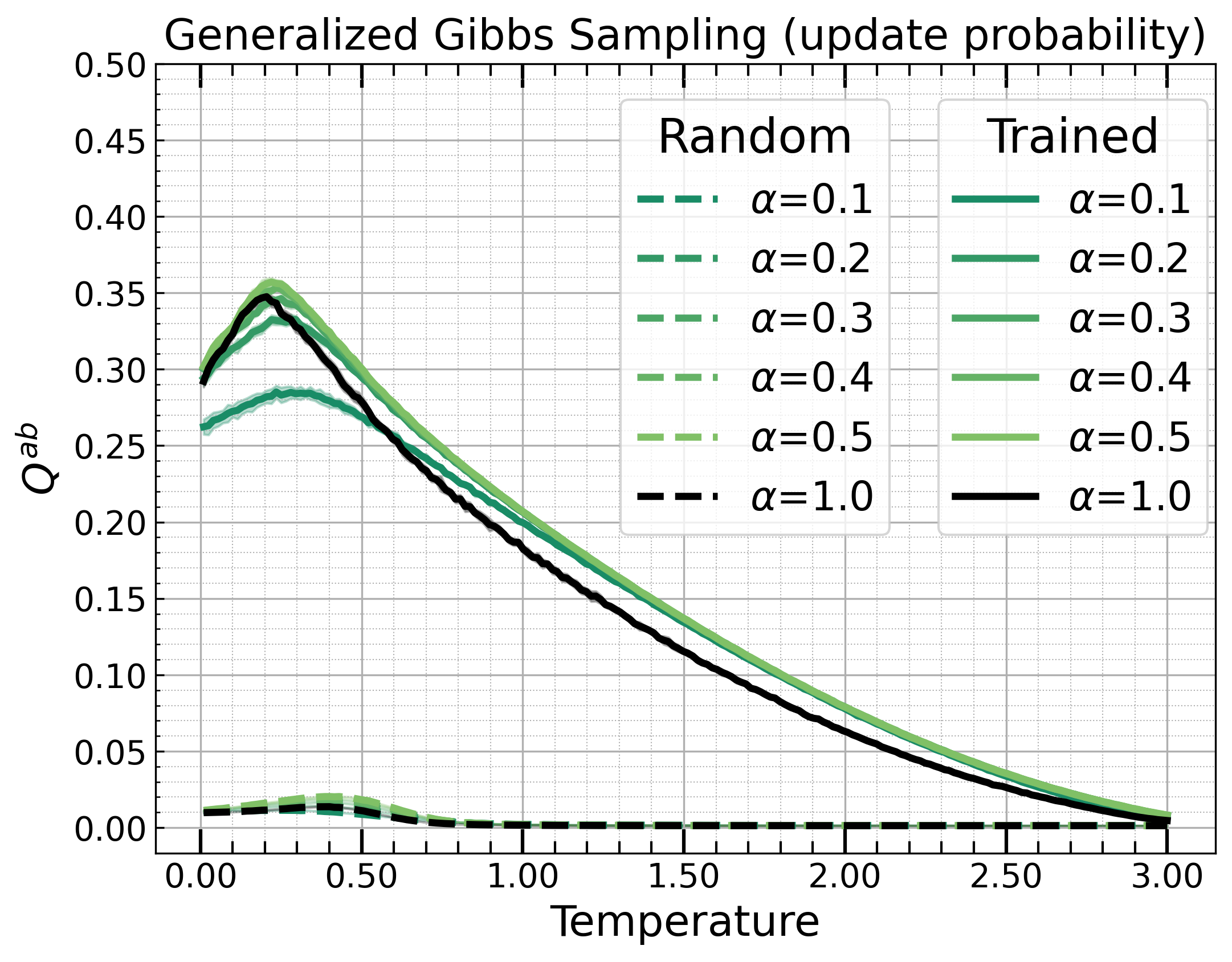}\\
    \textnormal{(a)}
\end{minipage}
\hfill
\begin{minipage}{0.49\textwidth}
    \centering
    \includegraphics[width=\textwidth]{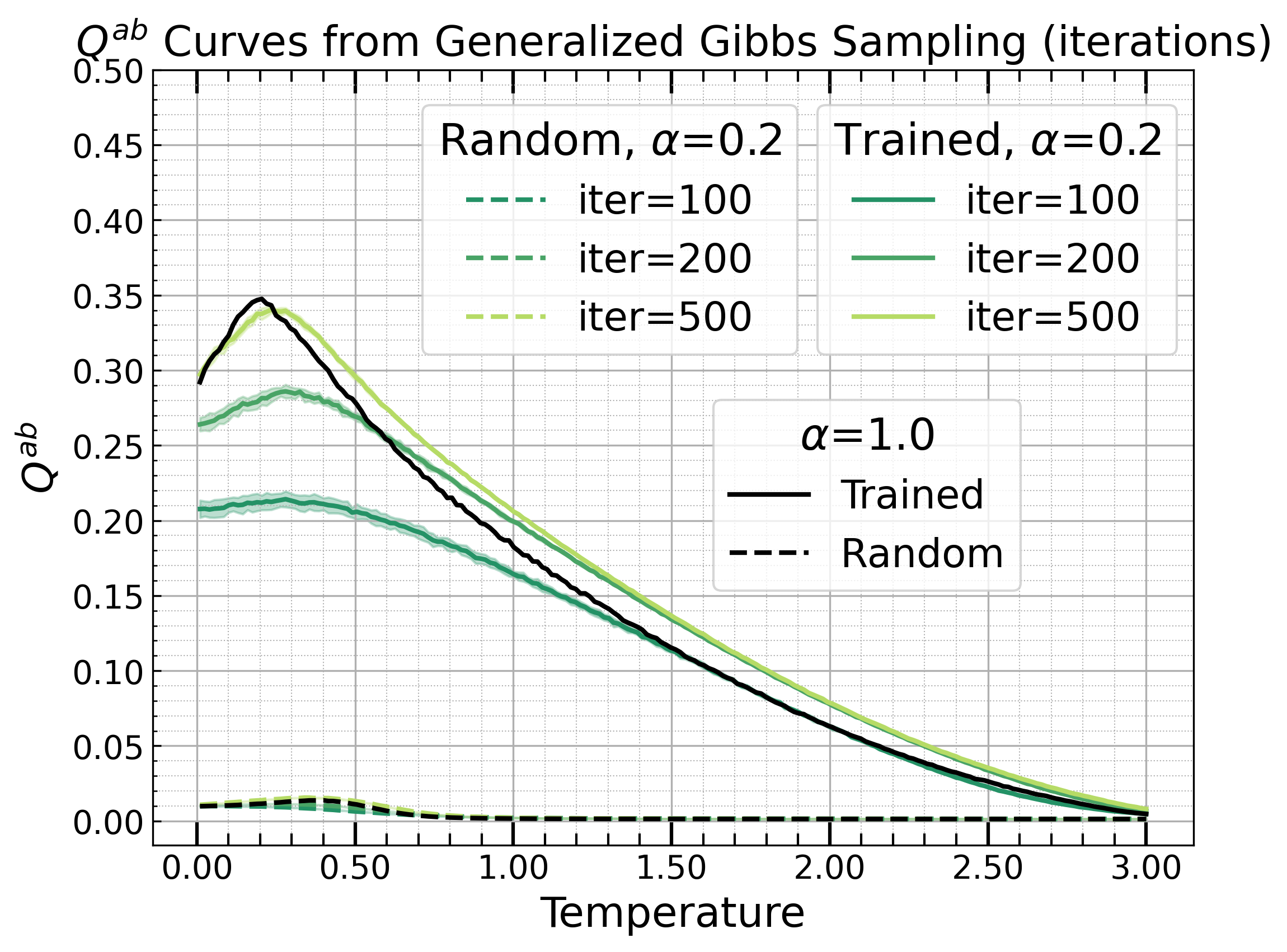}\\
    \textnormal{(b)}
\end{minipage}
\caption{$Q^{ab}$ curves under generalized Gibbs sampling.
This figure evaluates the validity of the grouped spin update scheme (see
Algorithm~\ref{alg:gibbs}) on a randomly initialized spin system and a spin
system derived from a trained neural network model.
{\bf (a)} Curves of different spin update rates, at 200 burn-in iterations.
{\bf (b)} Curves of different number of iterations, at $\alpha=0.2$ spin update
rate.  Curves produced with $\alpha=1.0$ and {\tt iterations=200} are
are included as references.
\label{fig:curve_ggibbs}
}
\end{figure}

The spin update dynamics defined in \eqref{eq:spinprob} operate by sequentially
updating individual spins. In principle, this process can be implemented using
fully parallel hardware-accelerated sampling. However, such implementations
often require low-level optimization tailored to specific FNN architectures and
hardware platforms.
To simplify implementation and preserve generality, this work adopts a grouped
update scheme, in which spins are updated in blocks. Specifically, in the HNN
representation, spin groups are naturally defined by the layer structure of the
original FNN.

\begin{algorithm}[h]
\caption{Generalized Gibbs sampling}
\label{alg:gibbs}
\SetAlgoNoEnd
\LinesNumbered
\KwIn{Initial spin configuration $\ve \sigma$, 
    inverse temperature $\beta$, 
    number of iterations $N_{\rm iter}$,
    spin update probability $\alpha$}
\KwOut{Gibbs sample of spin states $\ve \sigma$ (in-place update)}
\For{$t$ = 1 \KwTo $N_{\rm iter}$}{
    \For{$l=1$ \KwTo $L$}  
    { \label{alg:gibbs:ln:layer} 
        \For{$j \in \{\mathsf{HNNIndex}(\wildindex, l)\}$}{
            $H_t^j \leftarrow \texttt{LocalField}(\ve \sigma; j)$ \tcp*[f]{Compute via \eqref{eq:spinH}} \\
            \If{$\textrm{UniformRand}\big([0,1]\big) < \alpha$}{ \label{alg:gibbs:ln:alpha}
                $\sigma_j \leftarrow \texttt{GibbsSample}(H^j_t, \beta)$
                \tcp*[f]{Update using \eqref{eq:spinprob}}
            }
        }
    } 
}
\Return $\ve \sigma$\\
\end{algorithm}
The sampling procedure is detailed in Algorithm~\ref{alg:gibbs}.
The inner loop (line~\ref{alg:gibbs:ln:layer}) updates all spins associated with
a single FNN layer, and can be efficiently implemented using tensorized
operations in PyTorch~\cite{Paszke2019}.
Line~\ref{alg:gibbs:ln:alpha} implements a soft layer-wise update scheme, where
each spin is updated independently with probability $\alpha$.
When $\alpha$ is chosen to be of order $O(N_{\mathrm{layer}}^{-1})$, where
$N_{\mathrm{layer}}$ denotes the number of spins in a typical layer,
Algorithm~\ref{alg:gibbs} effectively approximates the standard Gibbs sampling
process, where spins are updated one at a time in random order.

Fig.~\ref{fig:curve_ggibbs} presents the $Q^{ab}$ curves computed using the
generalized Gibbs sampling procedure.
The two sets of models—random and trained on the {\em default task}—are
setup identically to those described in Subsection~\ref{subsec:experi:1}.
In Fig.~\ref{fig:curve_ggibbs}(a), the overlaps are computed using Gibbs samples
obtained with varying spin update probabilities $\alpha \in [0.1, 1.0]$, where
$\alpha = 1.0$ corresponds to full layer-wise updates.
All curves are computed after 200 burn-in iterations.
The sampling variance introduced by different values of $\alpha$ is insignificant
compared to the difference between random and trained models, and its influence
is consistent across model types.

Fig.~\ref{fig:curve_ggibbs}(b) shows that with sufficient burn-in iterations,
even small $\alpha$ values yield $Q^{ab}$ curves that closely match those
obtained with full layer-wise updates ($\alpha = 1.0$).
Moreover, the regions where the $Q^{ab}$ curves vary due to $\alpha$ are
distinct from the regions that separate trained and random models.
This supports the validity of using $Q^{ab}$ curves for model comparison,
provided that the sampling protocol is applied consistently across models.

All subsequent experiments use full layer-wise updates with 200 burn-in
iterations as the default sampling protocol.
This choice is justified by three considerations:
(i) the $Q^{ab}$ curves exhibit consistent qualitative behavior across different
sampling settings,
(ii) the fine structure of the overlap matrix $\ve Q$ is not directly analyzed,
and
(iii) the implementation is significantly simplified.

\subsubsection*{Tasks and models in experiments}

The {\bf default task}, introduced at the beginning of
Section~\ref{sec:experiment}, is a binary classification of digits ``0'' and
``1'' from the MNIST dataset~\cite{Lecun1998}. (An extended task with more
classes is discussed separately in a later experiment.)
As a preprocessing step, the $28\times 28$ pixel images are flattened into
784-dimensional vectors, followed by principal component analysis (PCA) for
feature representation.
The multi-layer perceptrons (MLPs) used for the {\em default task} have
architecture {\tt 10-256-256-256-2}, corresponding to a 10-dimensional PCA input
and 2-dimensional output logits.
Models are trained using the Adam optimizer~\cite{Kingma2014} with a learning
rate of $0.001$ and batch size of $16$.

\begin{figure}
\centering
\begin{minipage}{0.49\textwidth}
    \centering
    \includegraphics[width=\textwidth]{figs/task_complexity/qab_tasks_input_dim_10.png} \\
    \textnormal{(a)}
\end{minipage}
\hfill
\begin{minipage}{0.49\textwidth}
    \centering
    \includegraphics[width=\textwidth]{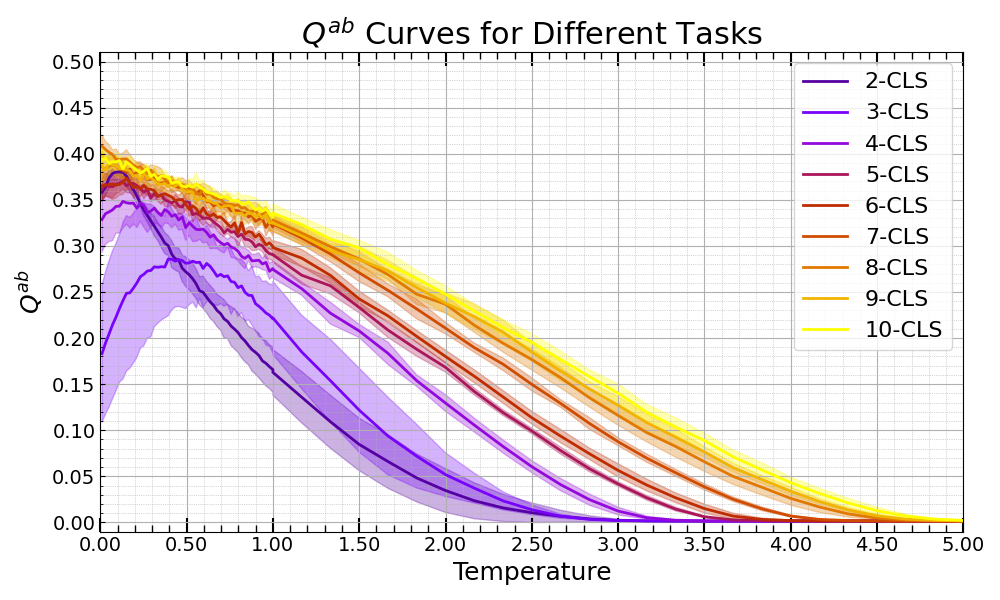} \\
    \textnormal{(b)}
\end{minipage}
\caption{$Q^{ab}$ curves for classification tasks with $C = 2$ to $10$ classes
on the MNIST dataset, comparing the effect of input dimensionality.  
{\bf (a)} Curves for models with 10-dimensional input (reproduced from
Fig.~\ref{fig:curve_fitness_epoch_task}(b)).  
{\bf (b)} Curves for models with 32-dimensional input.  
\label{fig:curve_input_dim}}
\end{figure}
An additional experiment on the {\em default task} is conducted to examine how
input representation affects the shape of the $Q^{ab}$ curves.
The input dimension is increased from the original 10 to 32. The results for
the 10-dimensional setting (from Figure~\ref{fig:curve_fitness_epoch_task}(b))
are reproduced in Figure~\ref{fig:curve_input_dim}(a) for comparison.
Figure~\ref{fig:curve_input_dim}(b) shows the corresponding $Q^{ab}$ curves
using the 32-dimensional input.
Among trained models, most tasks with $C \geq 3$ classes exhibit lower spin
overlap.
This phenomenon suggests questions for further investigation.
A possible explanation is that higher input dimensionality enables the model to
separate classes more easily, thereby reducing the likelihood of convergence to
narrow regions of the parameter space.
As a result, the overlaps are reduced—i.e., the $Q^{ab}$ curves become more
similar to those of the random model than to the trained model with
10-dimensional input.

Two additional tasks are described in Subsection~\ref{subsec:experi:1}, with the
following settings.
For the {\bf CIFAR-10 classification task}, the input encoder is a compact
ResNet-style convolutional neural network (CNN) with three stages of residual
blocks. Each stage uses $3 \times 3$ kernels, applies downsampling via strided
convolutions, and doubles the number of channels: $16 \to 32 \to 64$.
A global average pooling layer reduces the final feature map to a 64-dimensional
vector.
The encoder is pretrained using a classification head applied to the
64-dimensional feature vector.
The feature encoder is frozen during the training of the MLP body.
An MLP with architecture {\tt 64-256-256-256-10} is applied, producing 10 output logits.
All remaining procedures follow those of the {\em default task}.
The middle three layers (each with 256 neurons) are used to construct the
corresponding HNN, on which the $Q^{ab}$ curves are computed.

The {\bf Mini-Shakespeare character-level language modeling task} is adopted
from~\cite{Karpathy2020}.
The dataset consists of excerpts from the works of Shakespeare.
The task is to predict the next character given the previous context.
The text is tokenized into 65 distinct characters (letters, digits, and
punctuation), resulting in approximately one million tokens.
The input encoder is a compact transformer~\cite{Vaswani2017} with 4 blocks,
each using 4 attention heads and a 128-dimensional hidden size.
As in the CIFAR-10 task, the transformer is pretrained for the generation
objective, and its feature encoder is kept fixed thereafter.
An MLP with structure {\tt 128-256-256-256-65} is applied, and the middle three
hidden layers (each with 256 units) are used to construct the HNN for computing
the $Q^{ab}$ curves.


Here are more details on the visualization in Fig.~\ref{fig:vis_overlap}, 
which illustrates Gibbs samples and replica overlap in two {\bf toy spin-grid
systems}:
(i) a fully connected system with Gaussian-distributed couplings
(SK-type~\cite{Sherrington1975}); and (ii) a spatially localized system with
grid-based couplings. Both systems consist of $N = 64 \times 64 = 4,096$ spins.

The SK-type model (``Toy Spin 1'') has non-zero coupling $J_{ij}$ drawn from a
normal distribution for all $i \ne j$. 
In the localized model (``Toy Spin 2''), spins are arranged on a $64 \times 64$
two-dimensional grid. The coupling matrix $J_{ij}$ is sparse and symmetric, with
non-zero entries only between spatial neighbors. Specifically, each spin $i$ is
coupled to its four nearest neighbors: the spins directly above, below, to the
left, and to the right on the grid.

Letting $i = \texttt{row} \times 64 + \texttt{col}$, spin $i$ at position
$(\texttt{row}, \texttt{col})$ is coupled to spin $j$ at the following grid
locations:
$$
(\texttt{row}, \texttt{col} \pm 1) \quad \text{(horizontal neighbors)}, \qquad
(\texttt{row} \pm 1, \texttt{col}) \quad \text{(vertical neighbors)},
$$
whenever the corresponding $j$ falls within bounds.
In the main-text example spin $i = 116$ (at grid position $(1, 52)$) is coupled
to its neighbors $j \in \{52, 115, 117, 178\}$, corresponding to the positions
$(0, 52)$, $(1, 51)$, $(1, 53)$, and $(2, 52)$.

The non-zero entries $J_{ij}$ are drawn from Gaussian distributions with
distinct means depending on the neighbor direction: 
for example, in the illustration shown in Fig.~\ref{fig:vis_overlap}, the
vertical couplings (up/down) are sampled with mean $\mu_v = 1.0$, while
horizontal couplings (left/right) have mean $\mu_h = 0.1$; all couplings use
variance 
$\sigma^2 = 0.25 = \frac{1}{\text{\#.neighbours=4}}$. This
anisotropic structure encourages stronger vertical alignment in the resulting
Gibbs samples.

The non-zero entries $J_{ij}$ are drawn from Gaussian distributions with
direction-dependent means. As used in the configuration shown in
Fig.~\ref{fig:vis_overlap}, vertical couplings (up/down) are drawn from
$\mathcal{N}(\mu_v = 1.0,\ \sigma^2 = 0.25)$, while horizontal couplings
(left/right) are drawn from $\mathcal{N}(\mu_h = 0.1,\ \sigma^2 = 0.25)$. The
variance corresponds to the reciprocal of the number of neighbors ($1/4$). 
This anisotropic structure encourages stronger vertical alignment in the
resulting Gibbs samples.
The complete matrix $J$ is symmetrized as $J \leftarrow (J + J^T)/2$.

\begin{figure}[h]
\centering
\setlength{\tabcolsep}{2pt}
\renewcommand{\arraystretch}{1.0}
\begin{tabular}{cccccc}
    \includegraphics[width=0.14\textwidth]{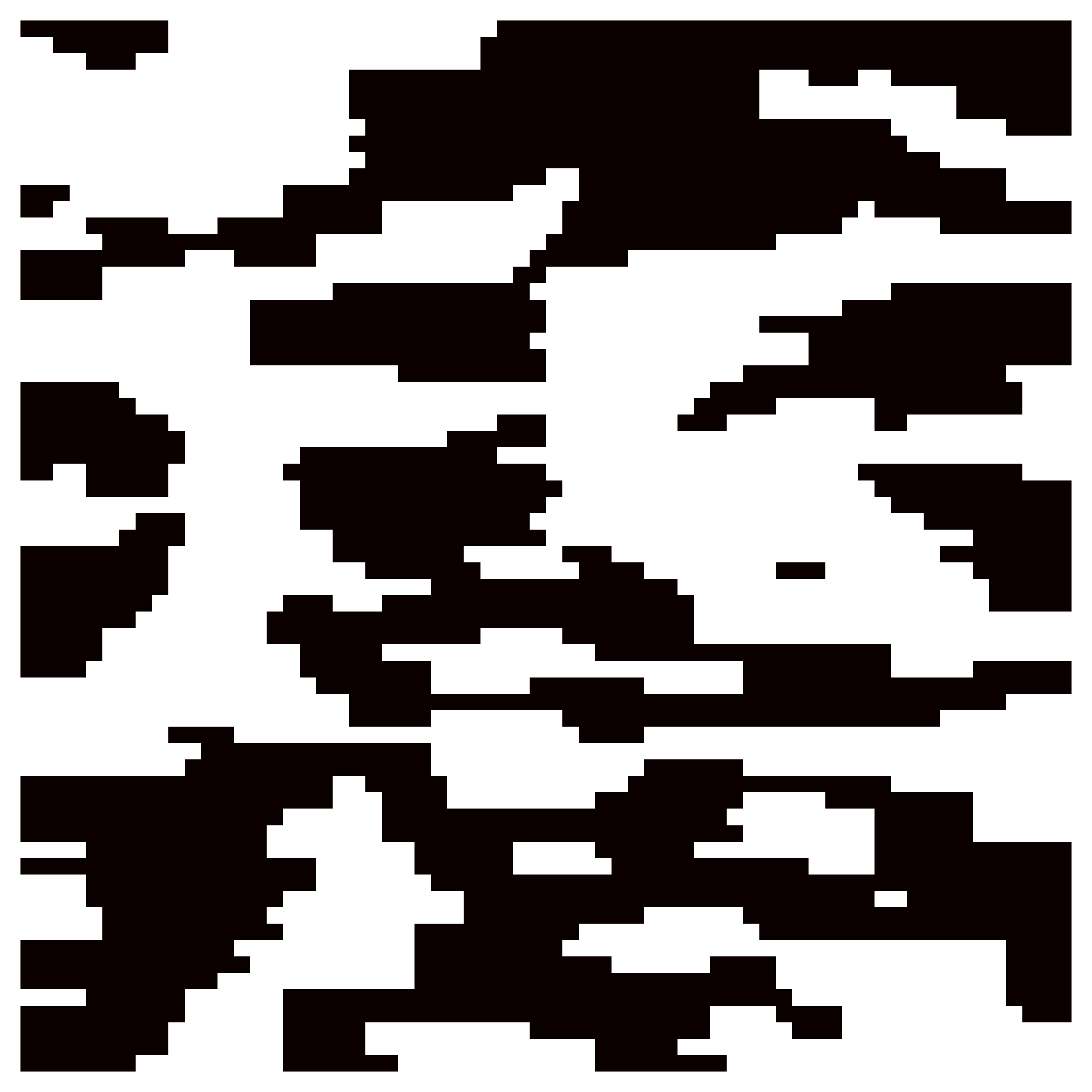} &
    \includegraphics[width=0.14\textwidth]{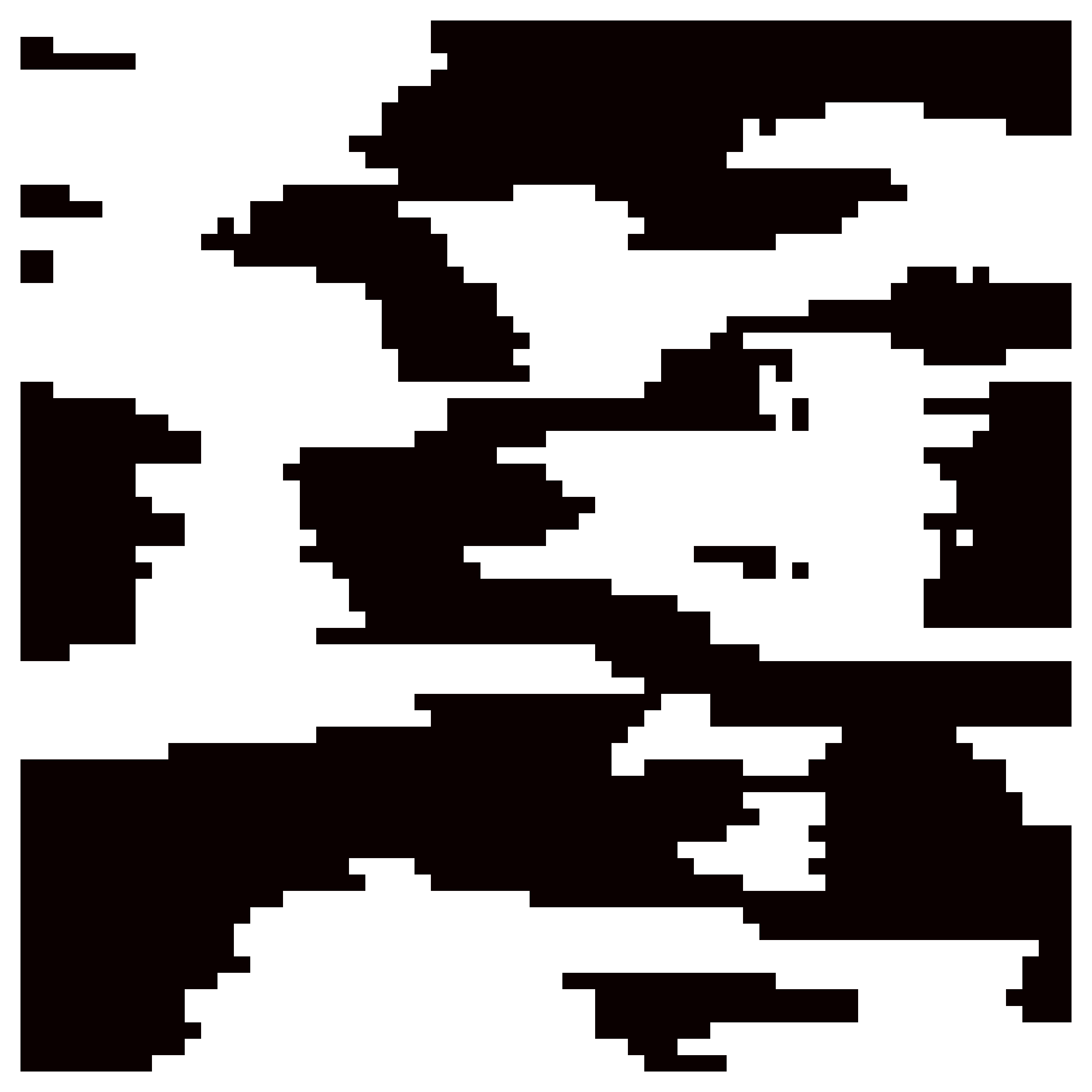} &
    \includegraphics[width=0.14\textwidth]{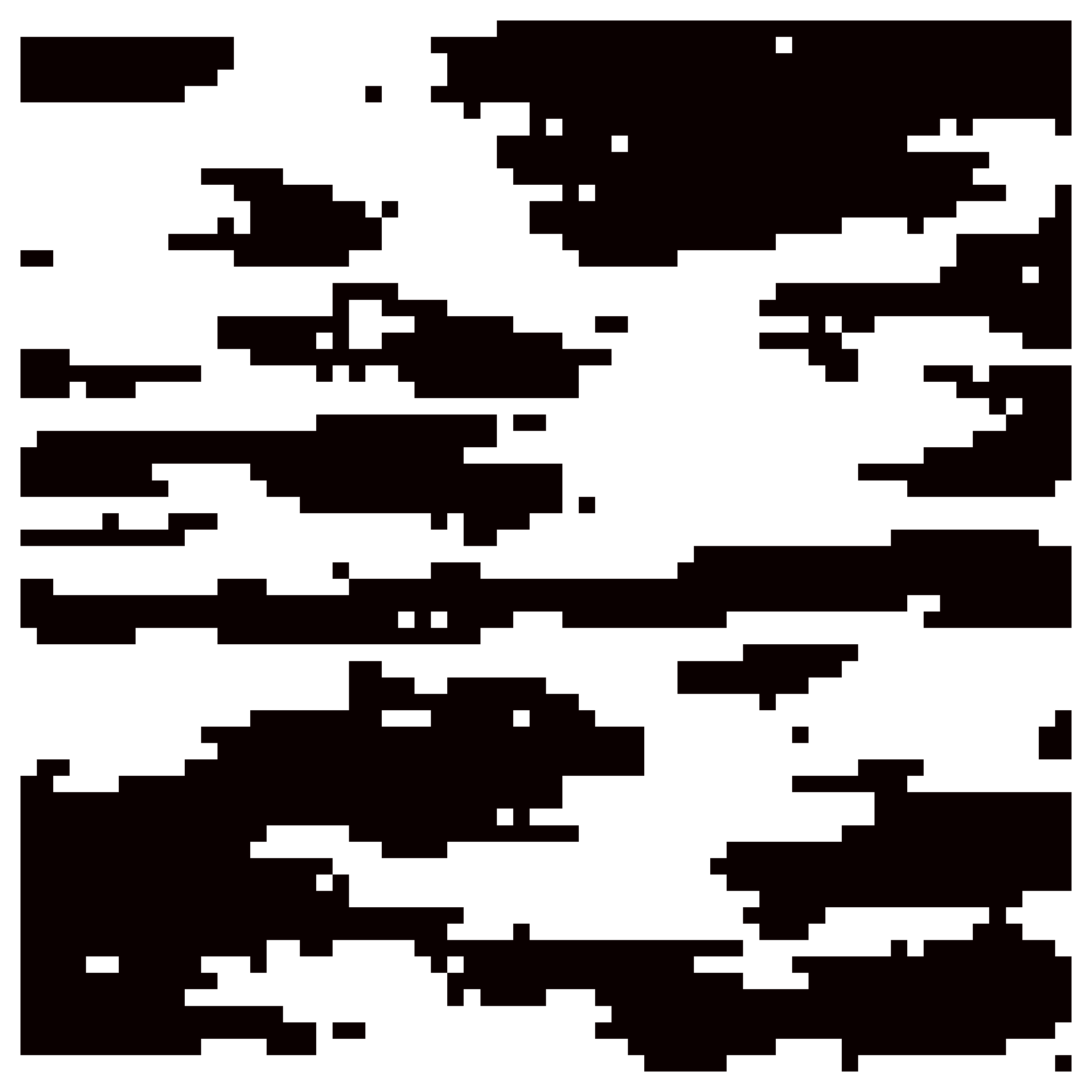} &
    \includegraphics[width=0.14\textwidth]{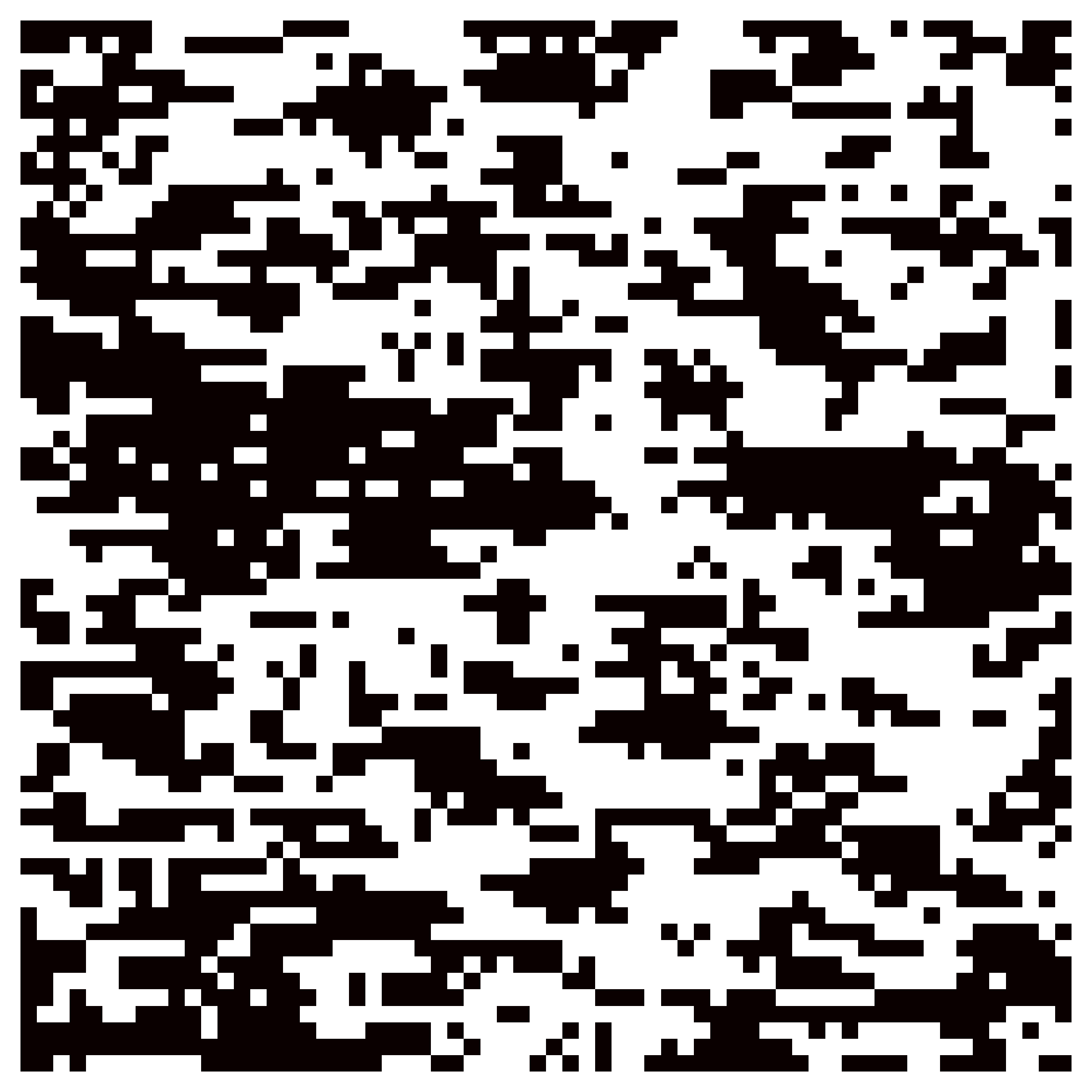} &
    \includegraphics[width=0.14\textwidth]{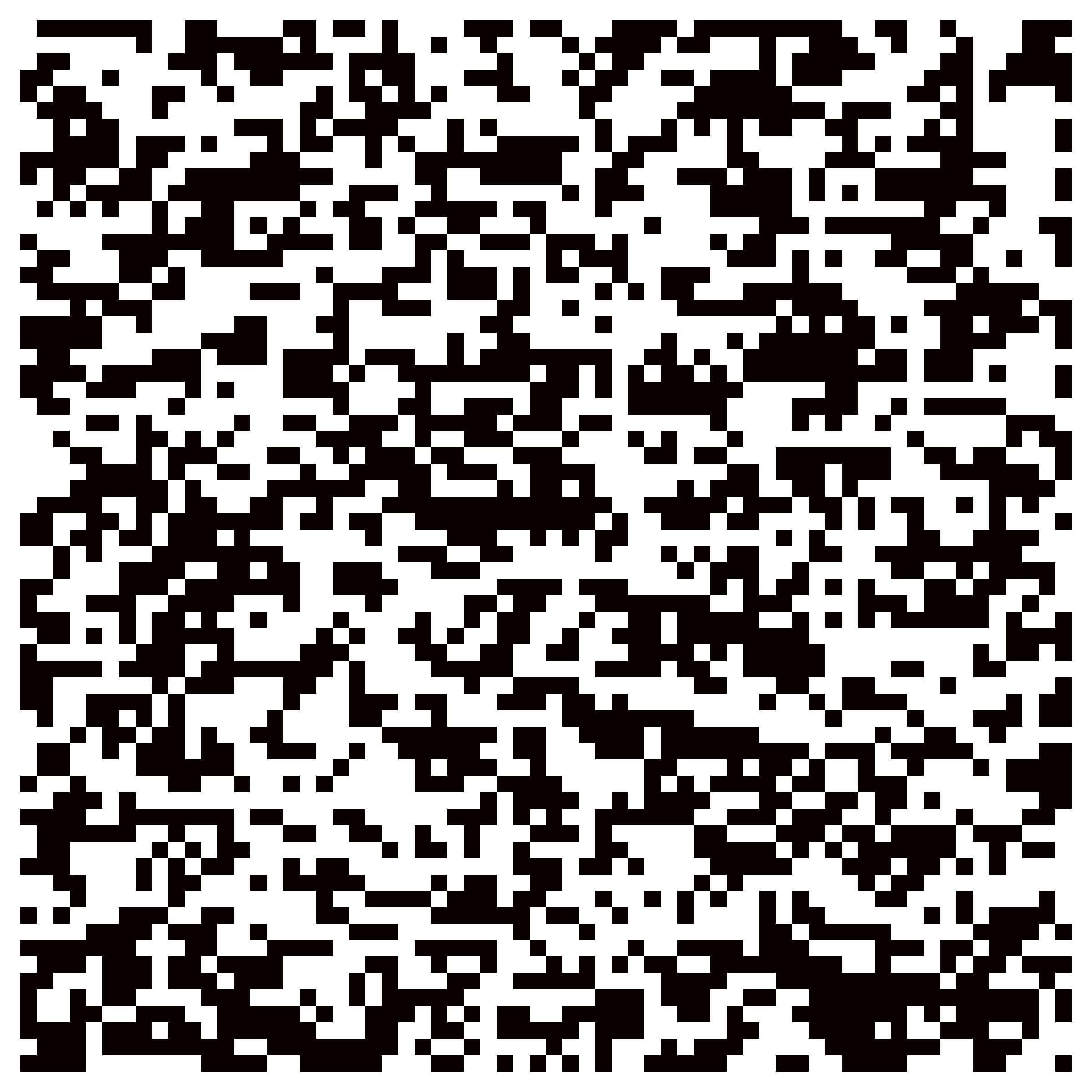} &
    \includegraphics[width=0.14\textwidth]{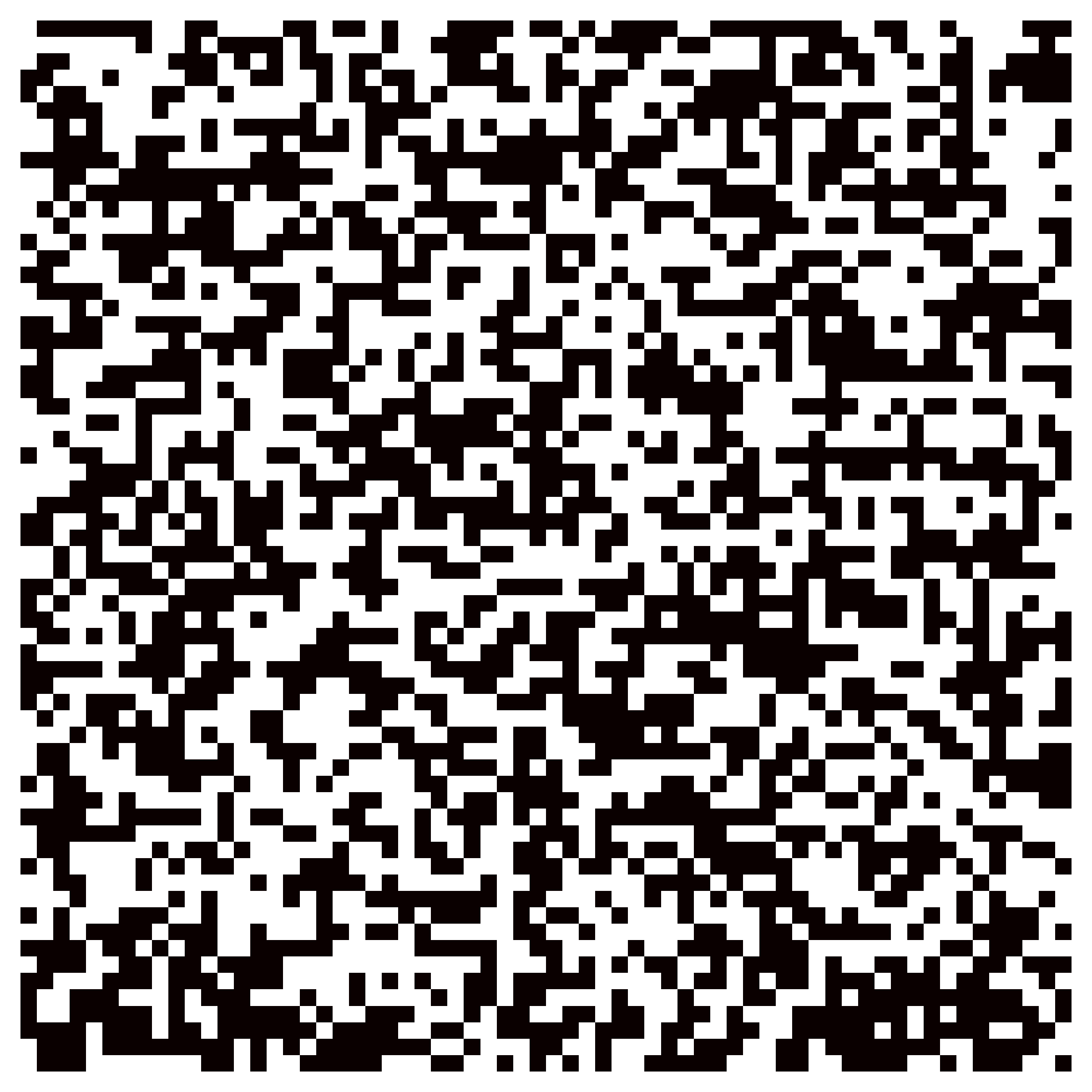} \\
    \includegraphics[width=0.14\textwidth]{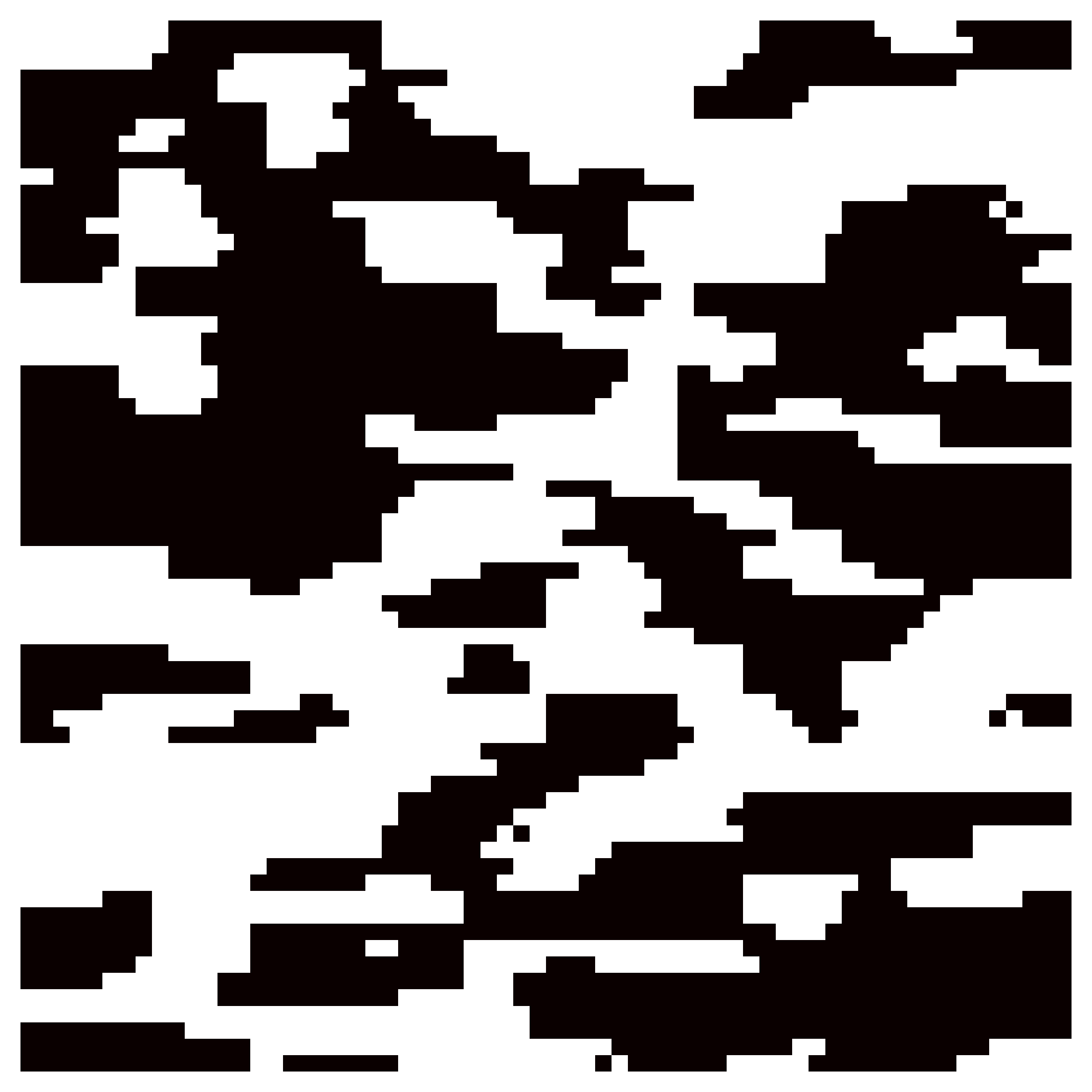} &
    \includegraphics[width=0.14\textwidth]{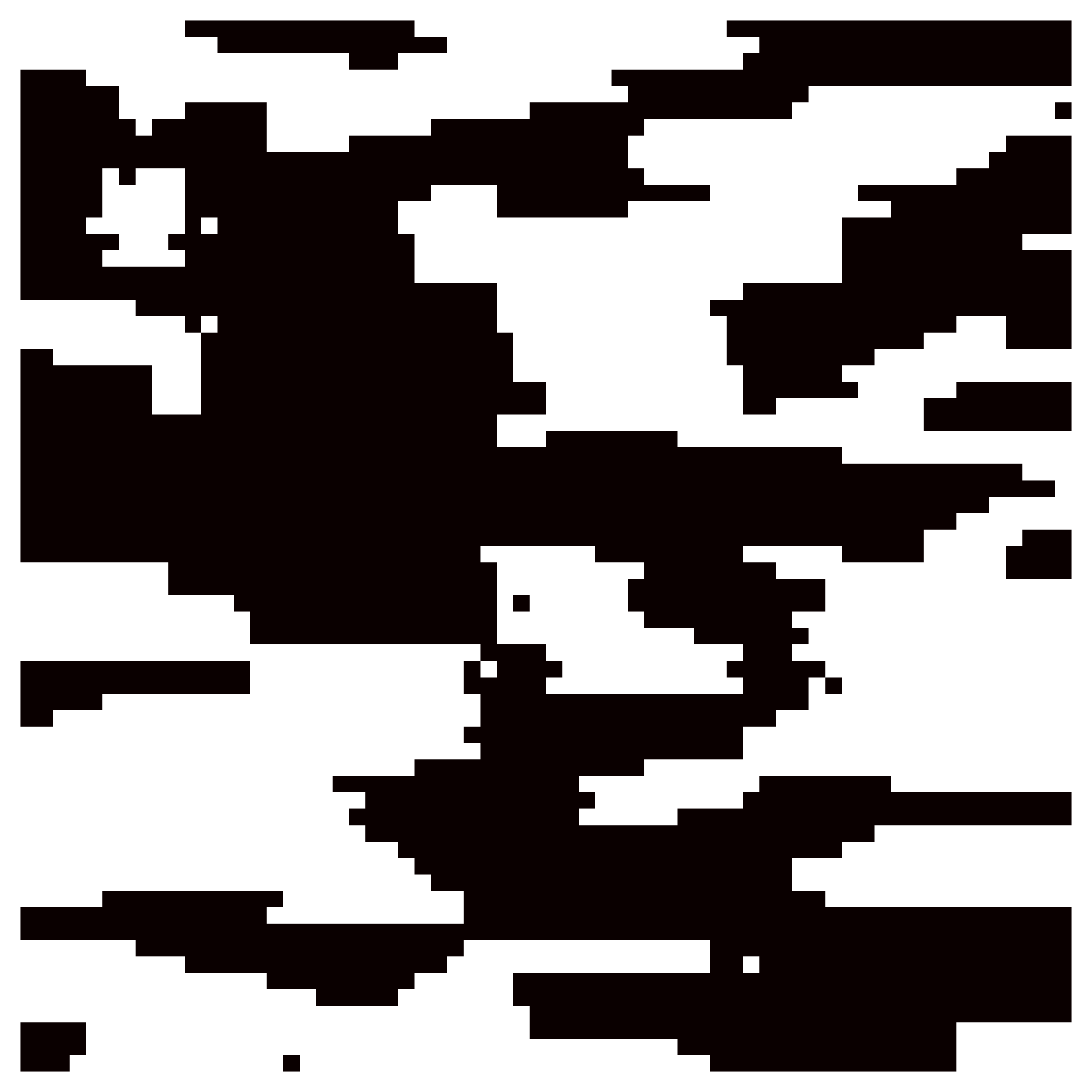} &
    \includegraphics[width=0.14\textwidth]{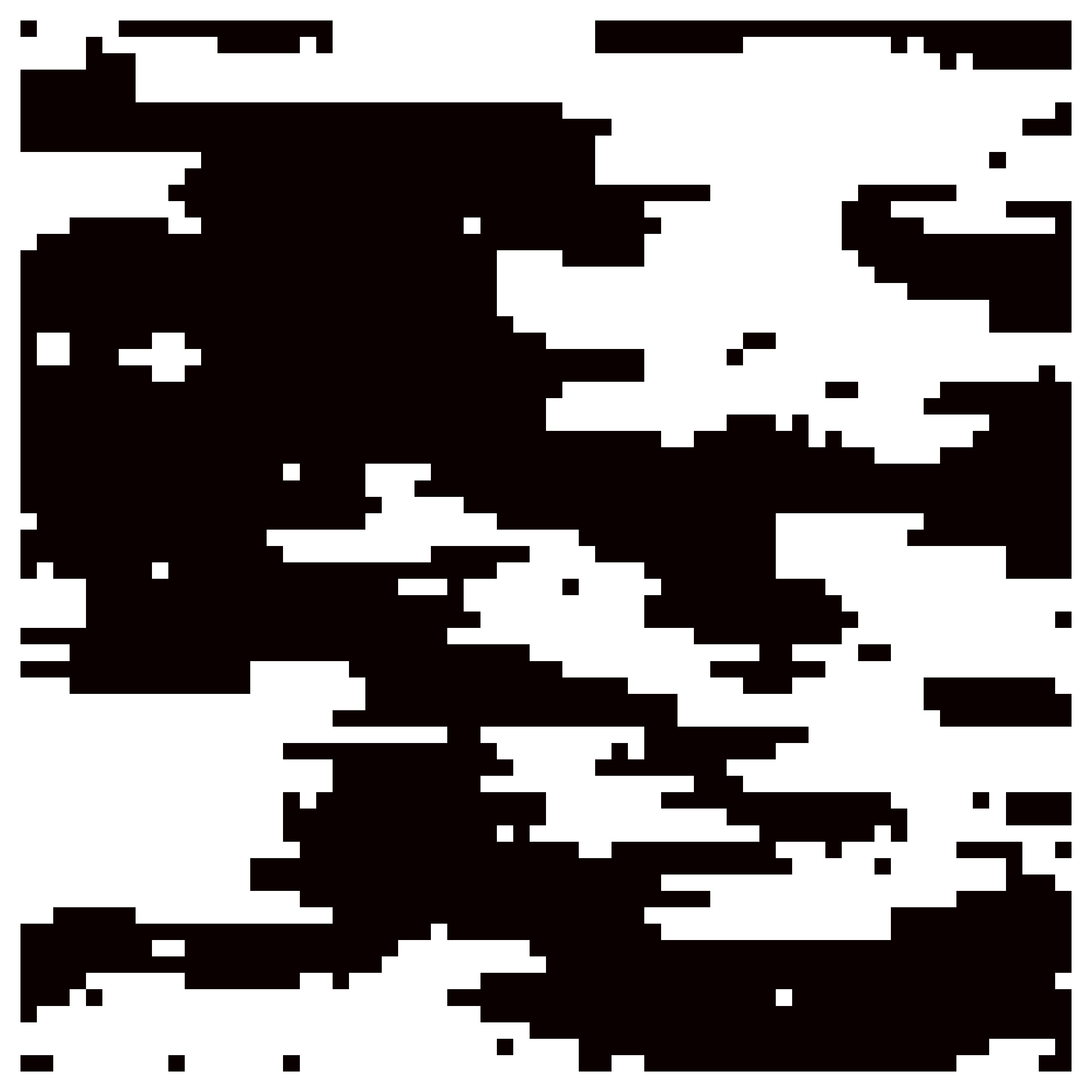} &
    \includegraphics[width=0.14\textwidth]{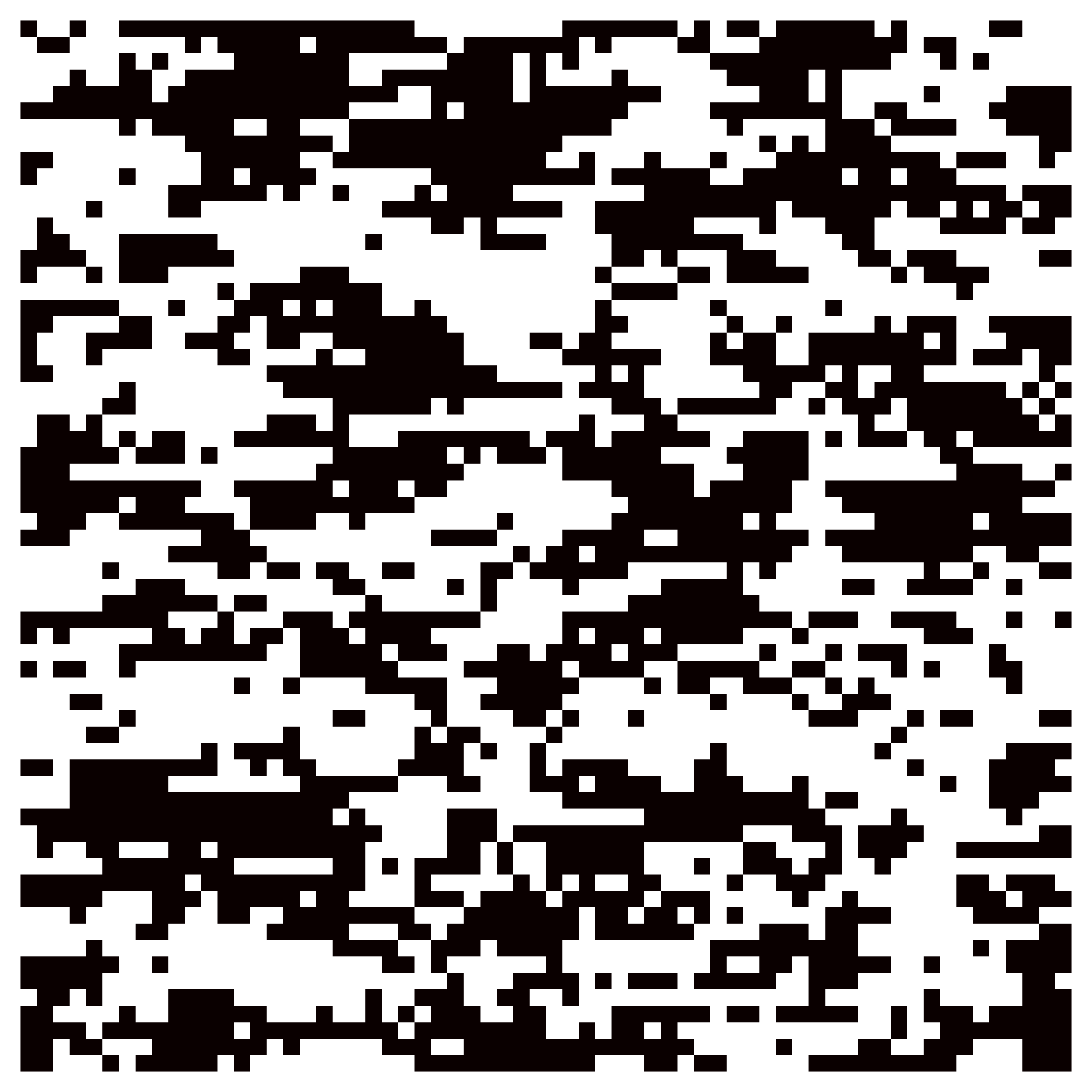} &
    \includegraphics[width=0.14\textwidth]{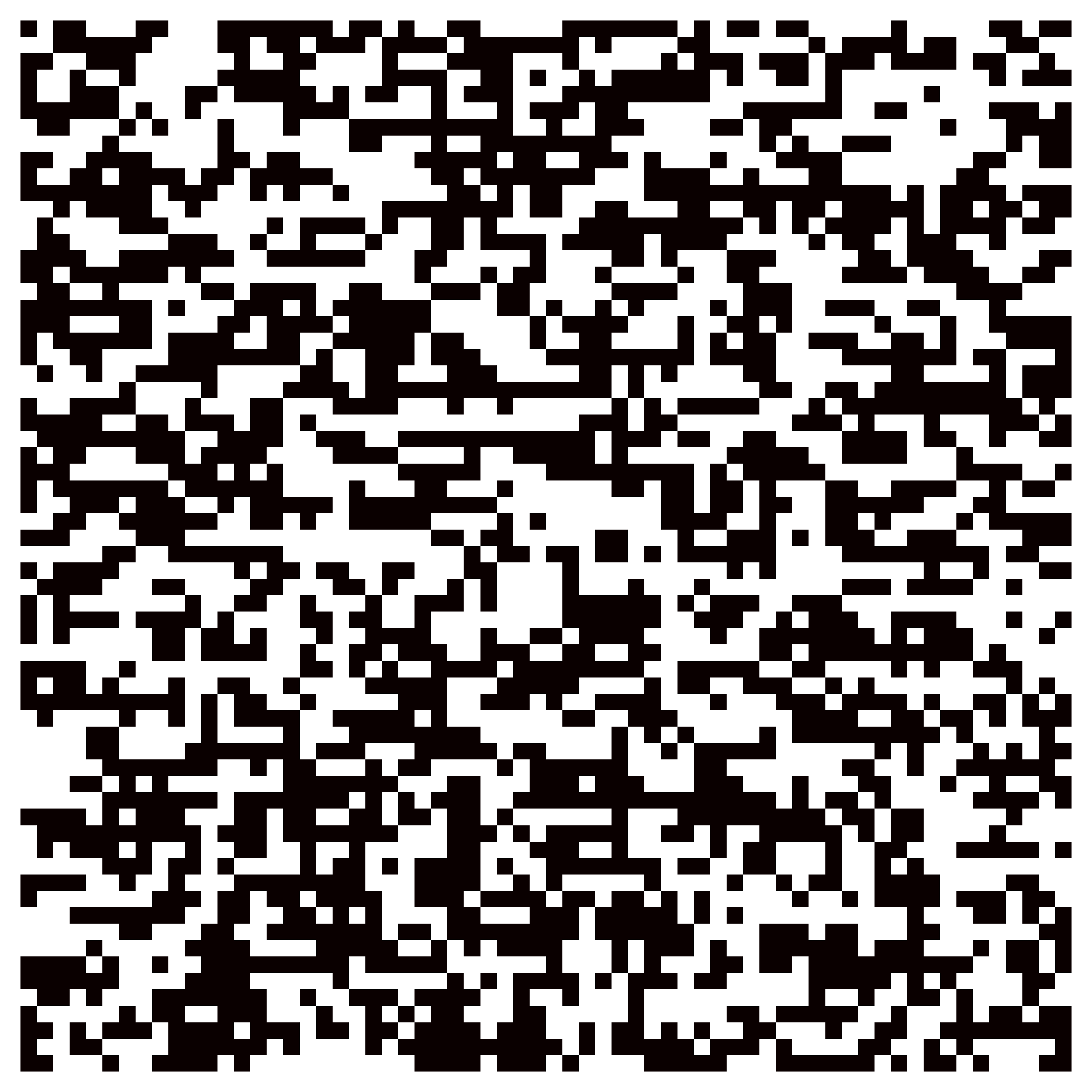} &
    \includegraphics[width=0.14\textwidth]{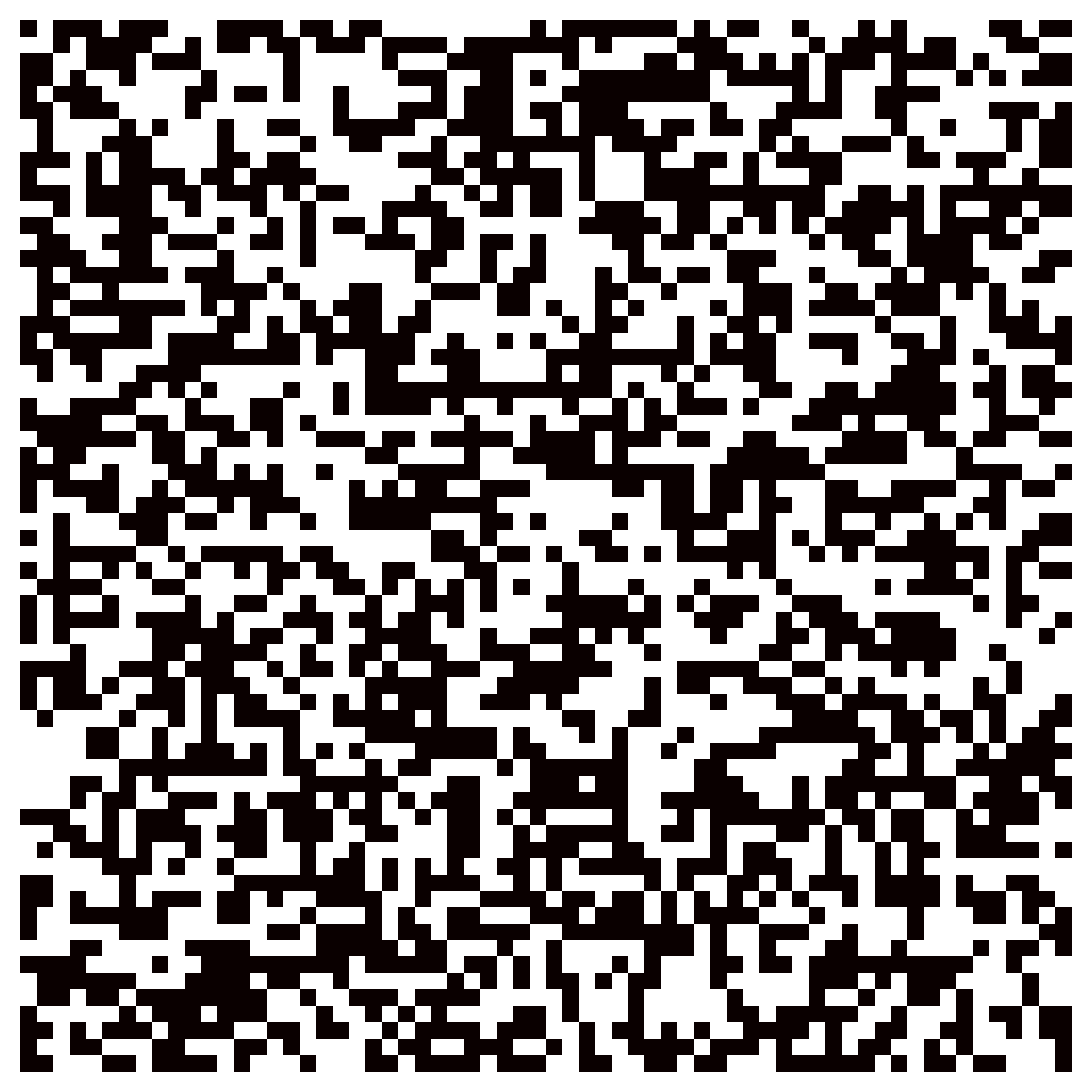} \\
    \includegraphics[width=0.14\textwidth]{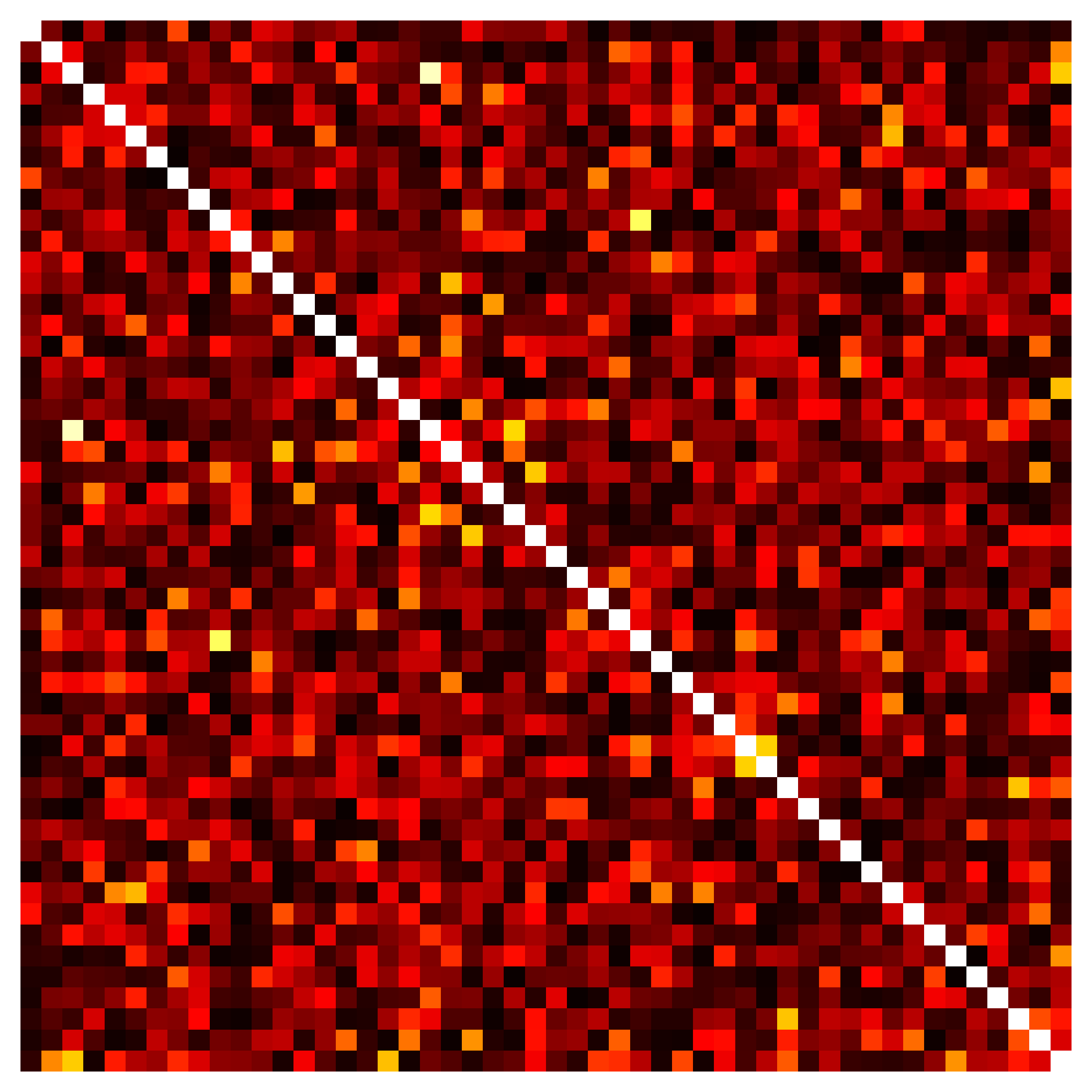} &
    \includegraphics[width=0.14\textwidth]{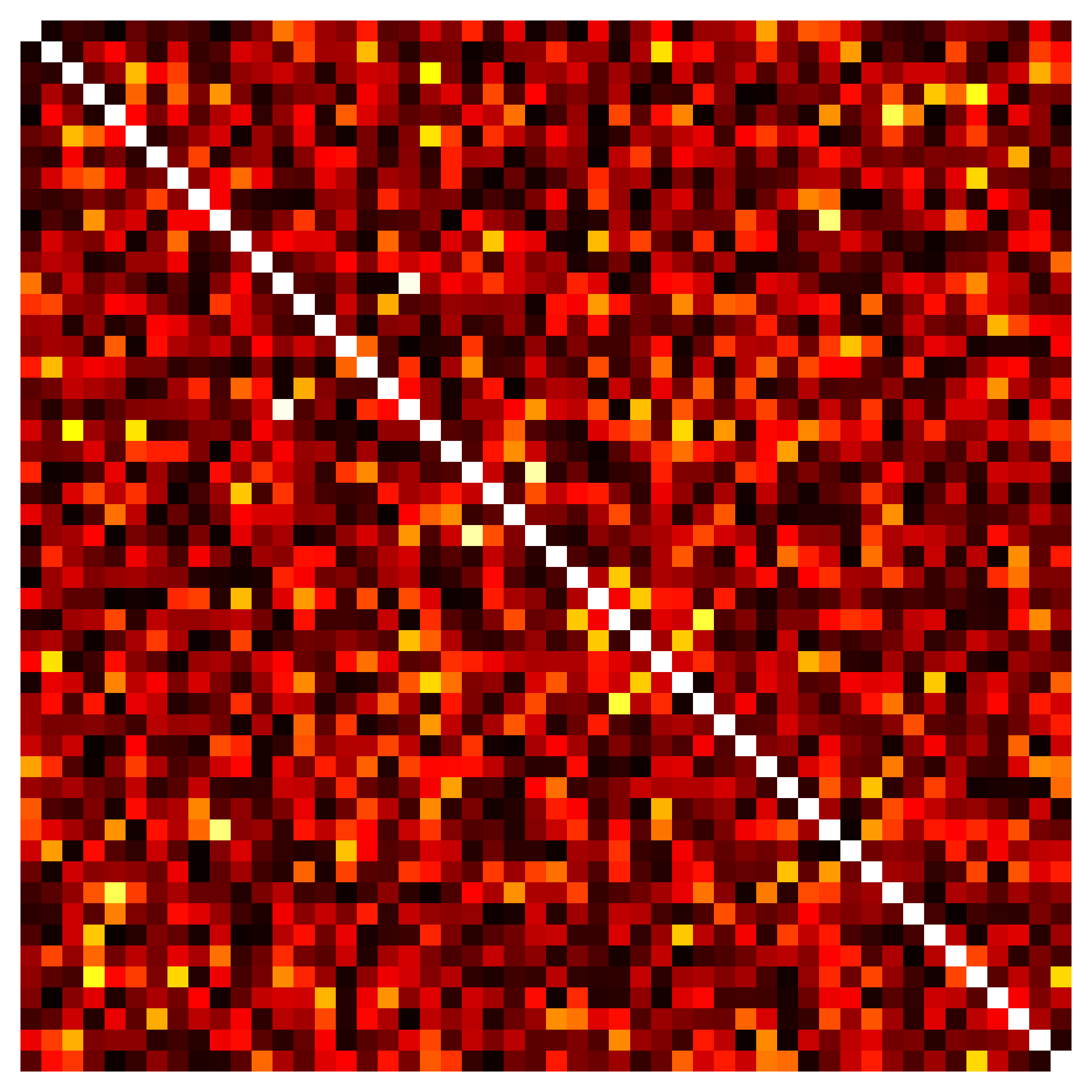} &
    \includegraphics[width=0.14\textwidth]{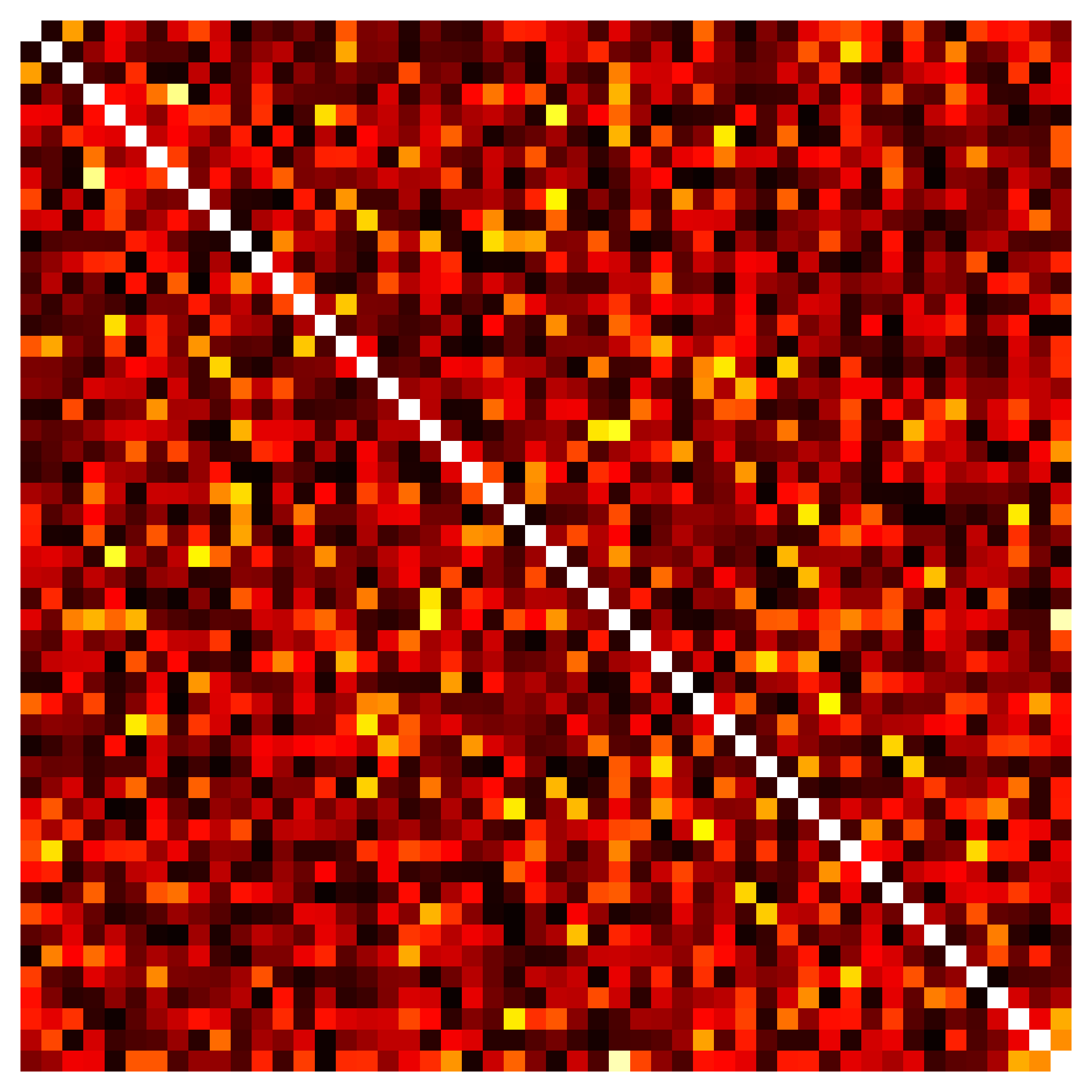} &
    \includegraphics[width=0.14\textwidth]{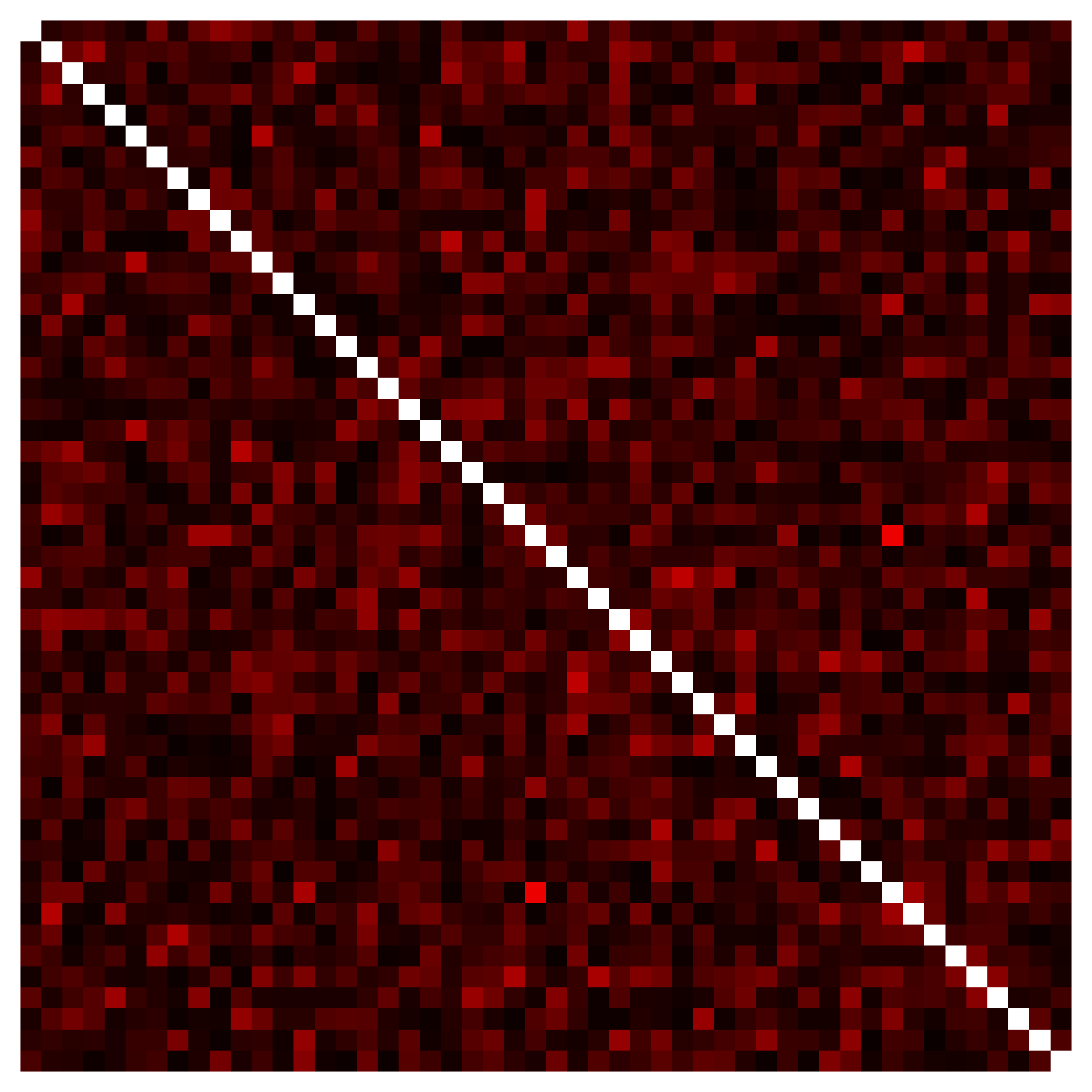} &
    \includegraphics[width=0.14\textwidth]{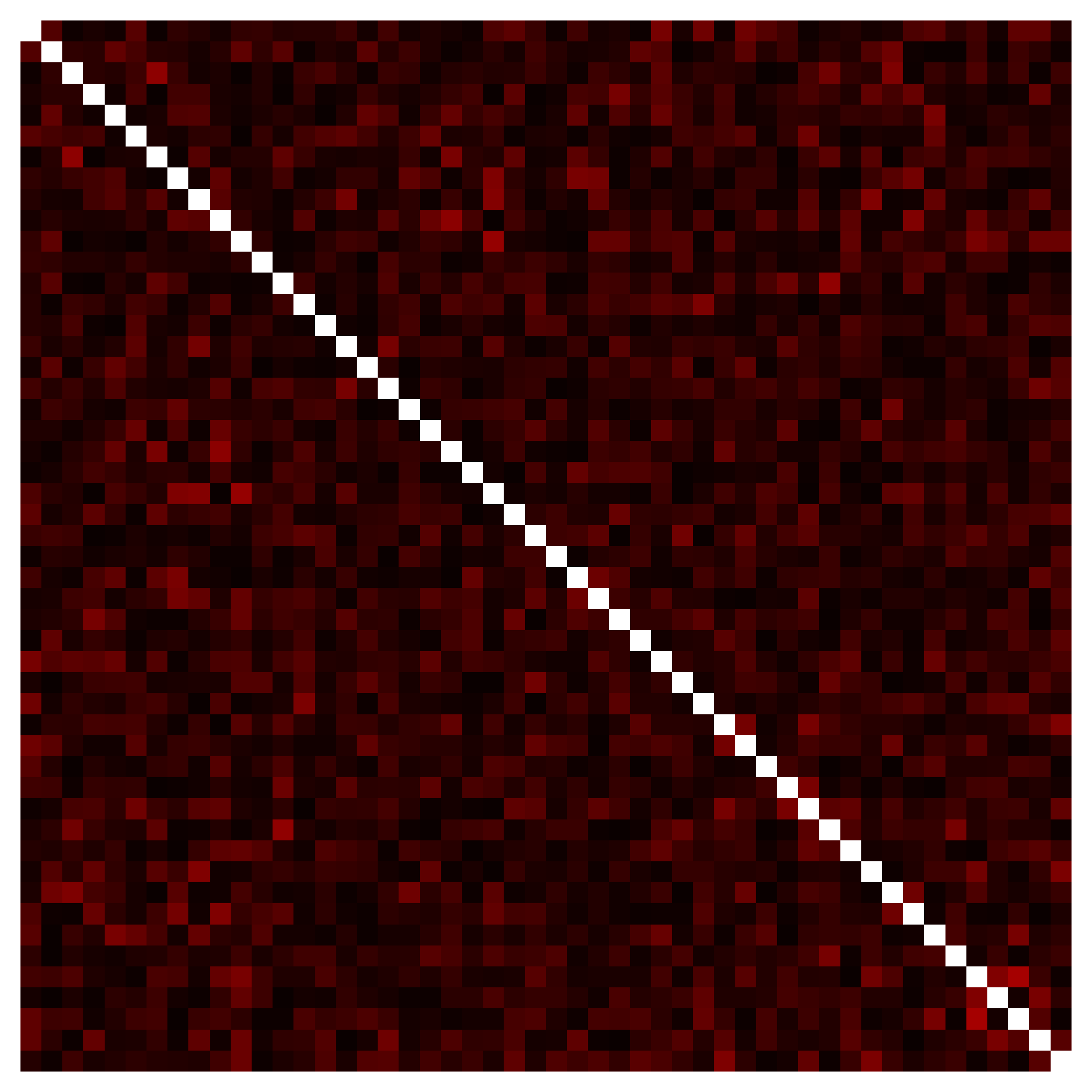} &
    \includegraphics[width=0.14\textwidth]{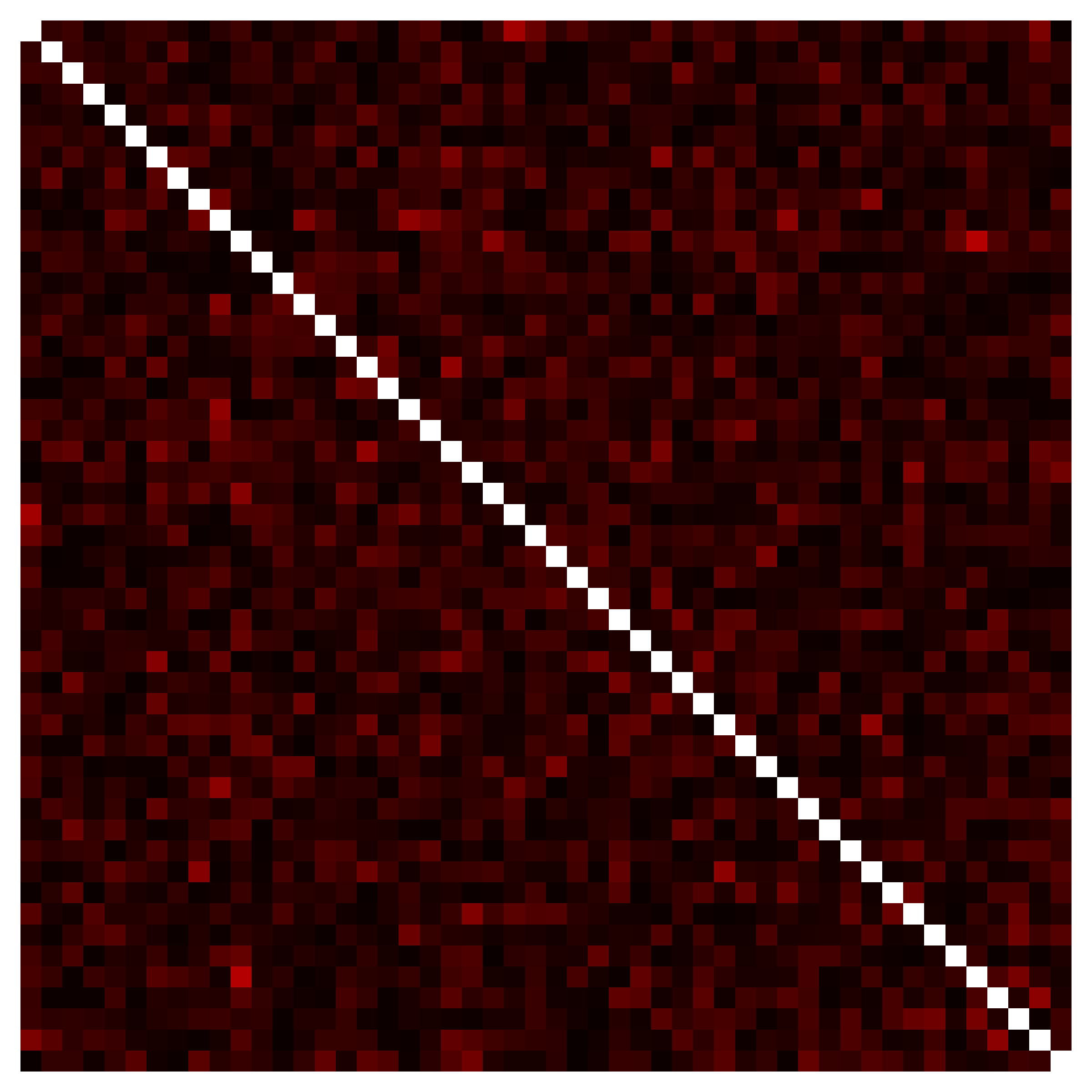} \\
    $T = 0.10$ & $T = 0.20$ & $T = 0.40$ & 
    $T = 0.80$ & $T = 1.60$ & $T = 3.20$  
\end{tabular}
\caption{Spin configurations of the localized model (Toy Spin 2) at selected
temperatures. At low temperatures, strong local structure emerges due to
anisotropic couplings. As temperature increases, the configurations become
progressively disordered.}
\label{fig:toy_spin2_samples}
\end{figure}

Fig.~\ref{fig:toy_spin2_samples} presents additional samples from the localized
model (Toy Spin 2). The figure shows that local structure emergence similar to
the behavior in Fig.~\ref{fig:vis_overlap}.
In this setting, the horizontal and vertical coupling means are set to $\mu_h =
1.0$ and $\mu_v = 0.2$, respectively.
The code used to generate and visualize these samples is included in the
supplementary materials.


\subsection{Additional experiment results}
\subsubsection*{Model architecture}
\begin{figure}
\centering
\begin{minipage}{0.49\textwidth}
    \centering
    \includegraphics[width=\textwidth]{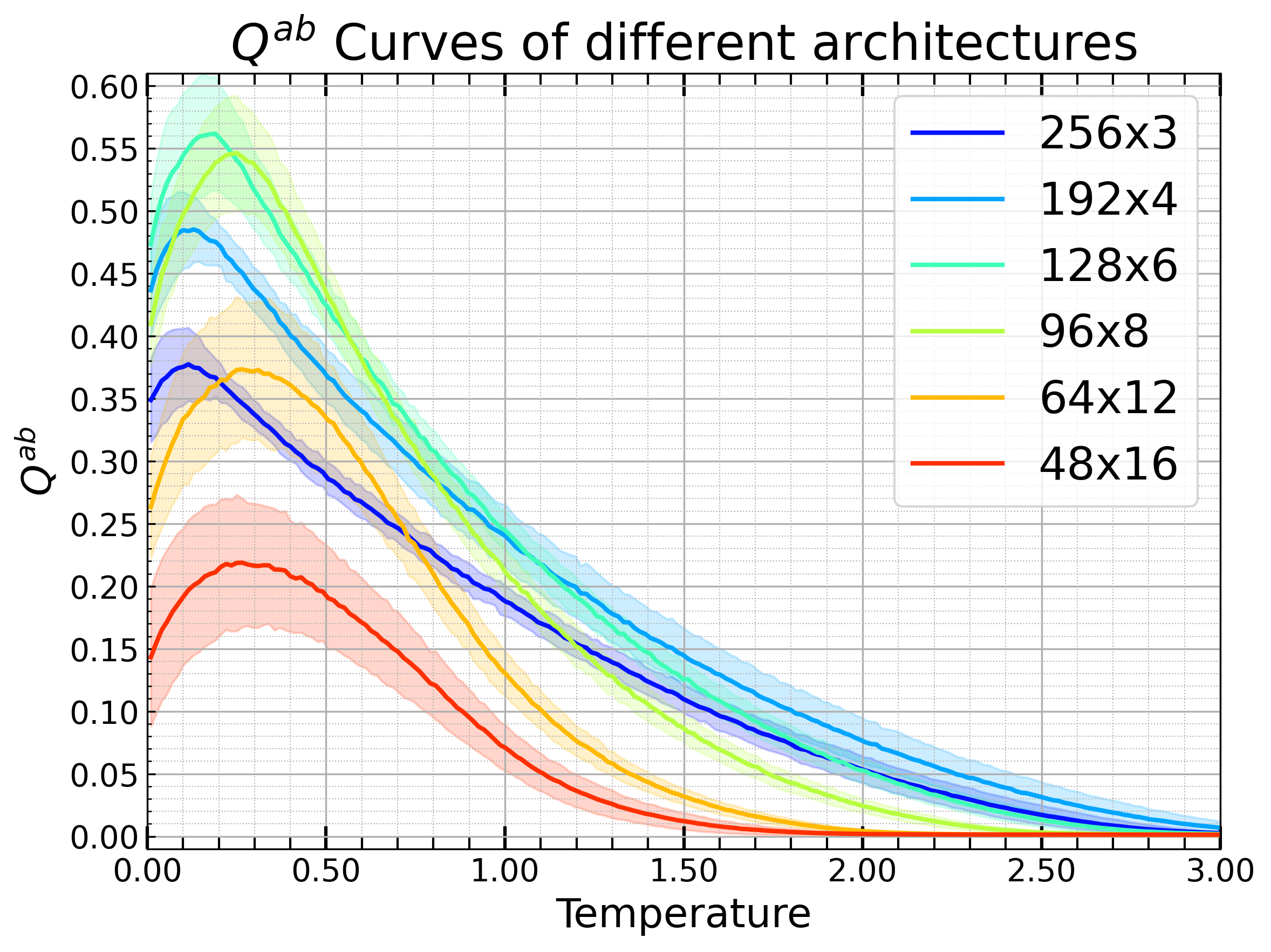}
    \textnormal{(a)}
\end{minipage}
\hfill
\begin{minipage}{0.49\textwidth}
    \centering
    \includegraphics[width=\textwidth]{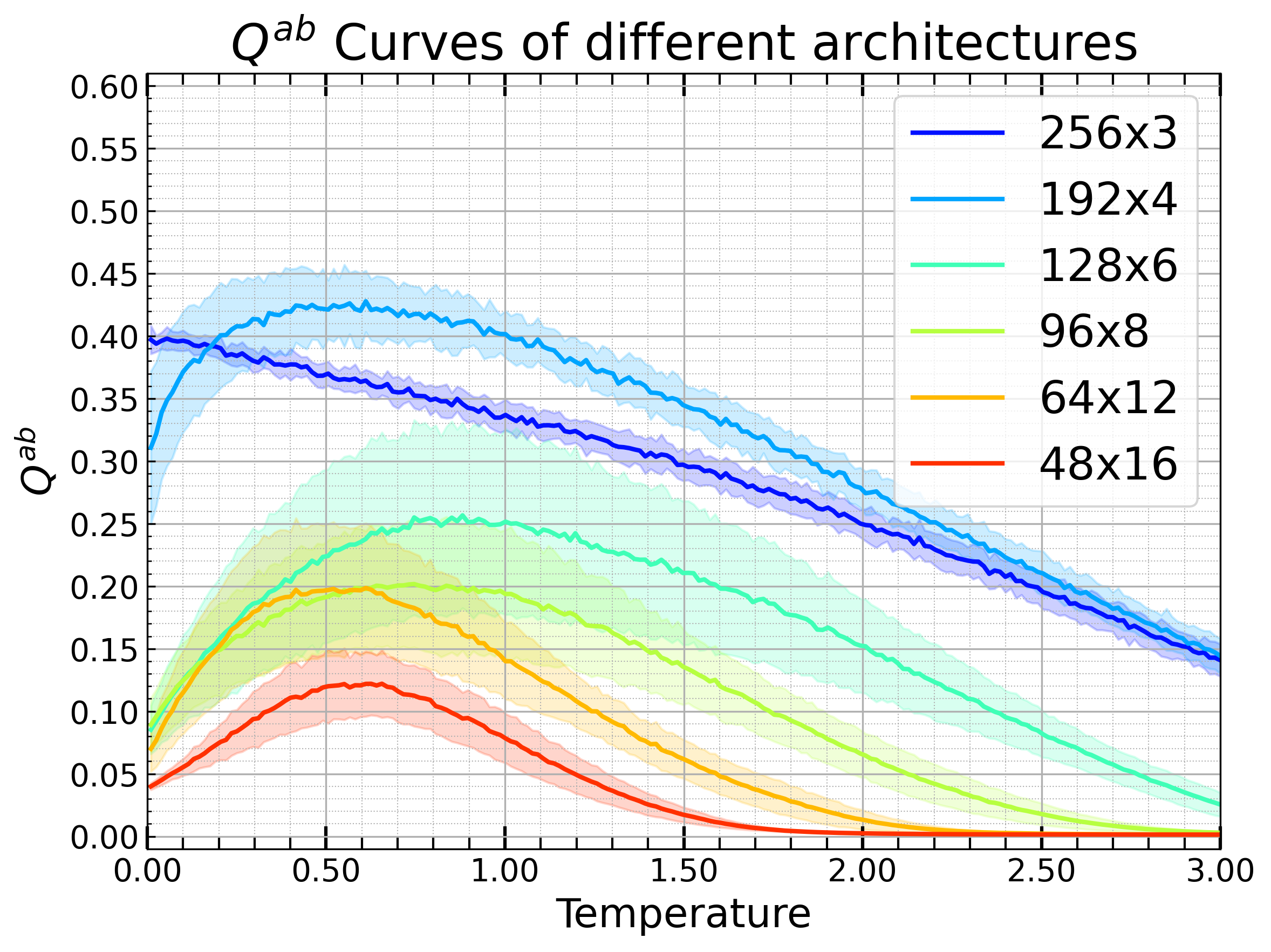}
    \textnormal{(b)}
\end{minipage}
\caption{$Q^{ab}$ curves of different model architectures. The figure compares
trained models of $768$ neurons organized in a range of different architectures. 
The tests are on two task settings:
{\bf (a)}: the {\em default task}, 10 input dimensions and 2 target classes {\bf
(b)}: 32 input dimensions and 10 target classes.
\label{fig:curve_arch}
}
\end{figure}

The layer structure of a FNN determines the connectivity topology of the
corresponding HNN. These differences in connectivity manifest in the $Q^{ab}$
curves.
In most experiments, the MLP body follows a {\tt 256-256-256} architecture,
comprising a total of 768 hidden neurons.
To examine the effect of depth and width, the 768 hidden units are organized
into a range of architectures with
varying depths, while keeping the total number of neurons constant. The tested
structures include:  
{\small
\[
[256, 256, 256],\quad [192] \times 4,\quad [128] \times 6,\quad [96] \times 8,\quad [64] \times 12,\quad [48] \times 16
\]
}
Here, $[d] \times k$ denotes an MLP with $k$ identical layers of width $d$.

Fig.~\ref{fig:curve_arch} shows that, for a fixed task, different architectures
result in distinct model states as reflected in the corresponding $Q^{ab}$ curves.
Higher overlap between spin replicas is observed in deeper architectures with
moderate layer widths.
This suggests that overlap is maximized when model depth and width are in a
balanced configuration.
Moreover, the architecture that maximizes $Q^{ab}$ varies across tasks of
different complexity.
Detailed investigation of these structure-task relationships needs to be
investigated in future research.
It is also possible that the full relationship between task structure and
network architecture is more complex than what can be captured by $Q^{ab}$
curves alone, and may require more refined characterizations of the Gibbs
distribution.


\subsubsection*{Planting patterns}

\begin{figure}[h]
\centering
\begin{minipage}{0.49\textwidth}
    \centering
    \debugincludegraphics[width=\textwidth]{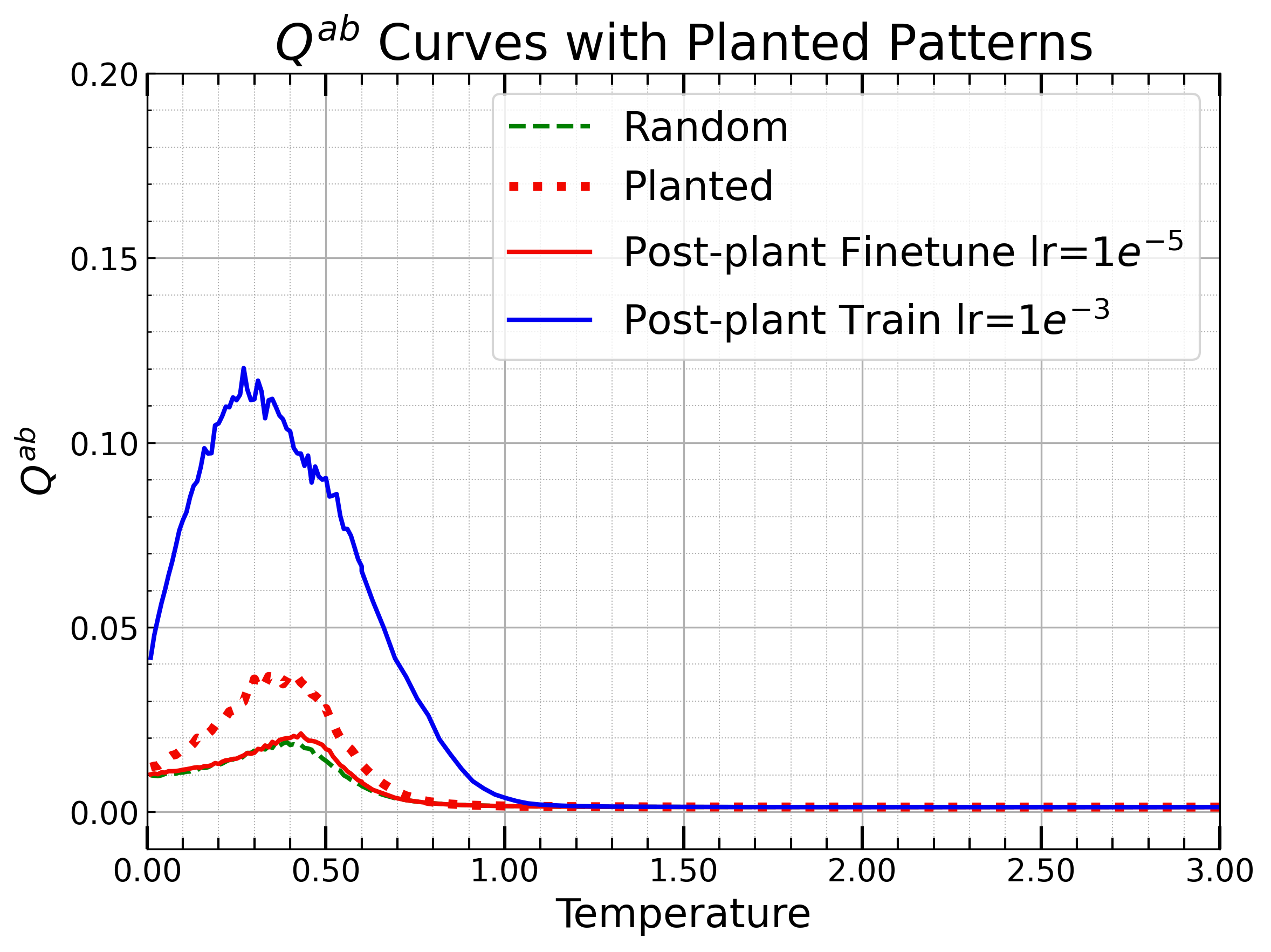} \\
    \textnormal{(a)}
\end{minipage}
\hfill
\begin{minipage}{0.49\textwidth}
    \centering
    \debugincludegraphics[width=\textwidth]{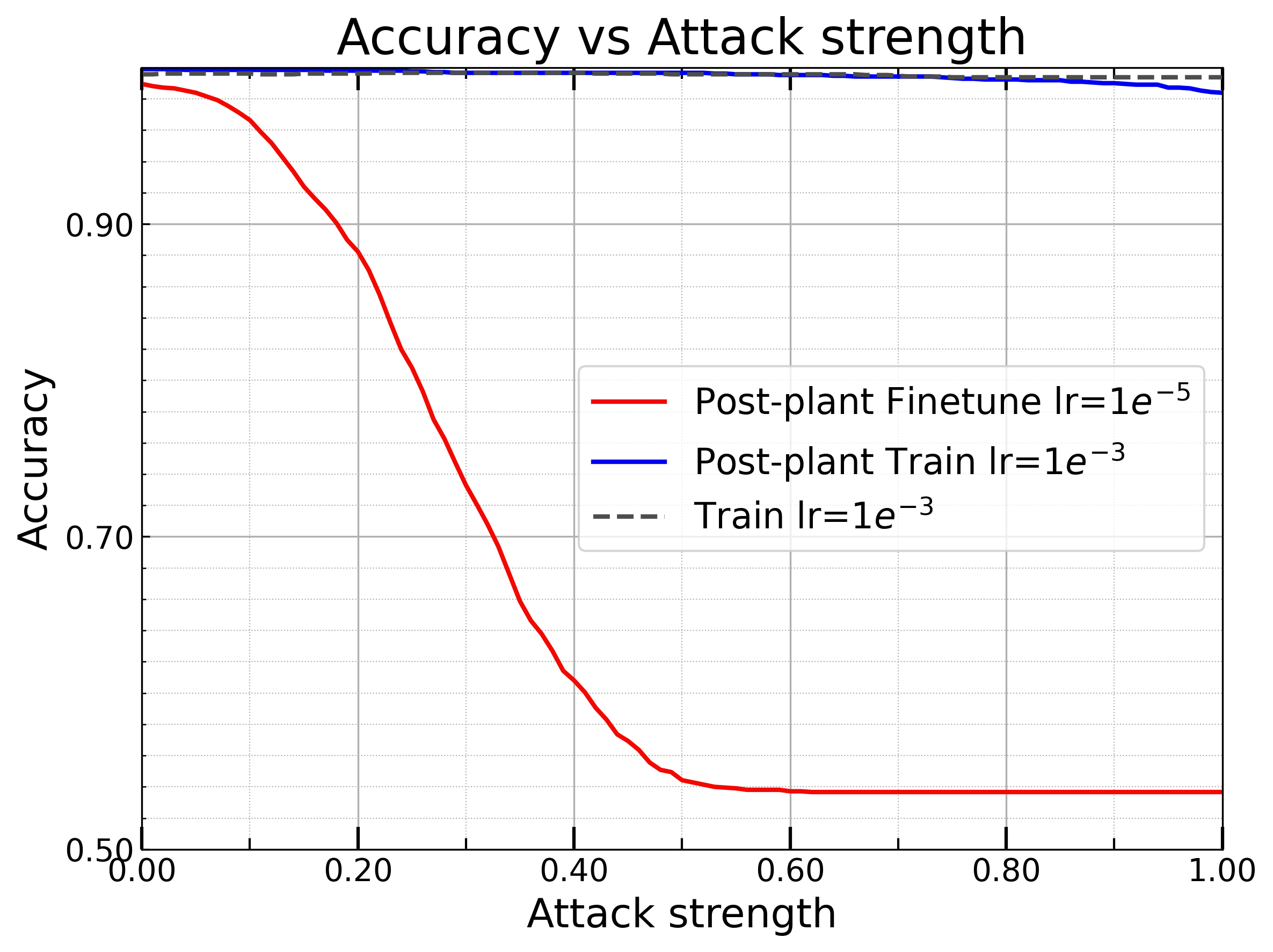} \\
    \textnormal{(b)}
\end{minipage}
\caption{Examine models with planted patterns on the {\em default task}.
{\bf (a)} $Q^{ab}$ curves of models initialized randomly and with a planted
pattern. 
The planted-pattern model is subsequently trained using two learning-rate schedules.
Dashed lines indicate the initial (planted) models; solid lines indicate the trained or
finetuned versions.
{\bf (b)} Test accuracy under input perturbations aligned with the planted
pattern, across three trained models.
The model trained from scratch, and the one finetuned with a learning rate of
$10^{-3}$, remain robust across perturbation levels (curves plotted at the top).
The model with a planted pattern finetuned using a learning rate of
$10^{-5}$ is vulnerable to such perturbations.
}
\label{fig:mnist_plant}
\end{figure}

Subsection~\ref{ssec:experi:abnormalities} presents an experiment, 
where $Q^{ab}$ curves are used to examine a model with a ``planted pattern'': the
model appears random but contains a planted pattern.
The evolution of the model's $Q^{ab}$ curves during training reveals behavior
that deviates from models with typical random initialization.
Initially, the $Q^{ab}$ curve differs significantly from that of typical
randomly initialized models.
When the training proceeds with a small learning rate, the model's $Q^{ab}$
curves become increasingly similar to those of the standard {\em random} models.
This effect disappears when a larger learning rate is used.

A similar effect is observed in the {\em default task} on the MNIST subset, as
shown in Figure~\ref{fig:mnist_plant}.
The phenomenon is more evident in this toy setting, where the input distribution
exhibits simple structure.
When the input signal is sufficiently corrupted, a model with a planted pattern
(without adequate post-training) fails, with test accuracy approaching
the random baseline of $0.5$.

Unexpected negative results were encountered in early implementations of the
planted-pattern experiment on the CIFAR-10 dataset (see main text).
When the attack signal was implemented directly in the input space using the
same {\em visual pattern}—e.g., by overlaying a red square with varying
intensities—the planted model did not show increased vulnerability compared to a
standard model.
However, when the attack signal was extracted from the encoder's internal
representation and used to perturb the input feature vector of the planted MLP
directly, the attack became effective, and vulnerability was observed, similar
to the results shown in subplot~(b) of Figure~\ref{fig:mnist_plant} and
Figure~\ref{fig:cifar_plant}.

{\bf Remark:} Two additional bibliography items are cited in the
Appendix~\cite{Kingma2014, Vaswani2017}. A separate reference list is included
for the Appendix only. As a result, reference numbers may differ between the
main text and the Appendix.
